\documentclass[aps,pra,reprint]{revtex4-2}

\usepackage{amsmath}
\usepackage{amssymb}
\usepackage{bbm}
\usepackage{graphicx}
\usepackage{xcolor}
\usepackage{hyperref}
\usepackage[normalem]{ulem}

\DeclareMathOperator{\Tr}{Tr}

\begin{document}

\title{Adiabatic time evolution of highly excited states}

\author{Hadi Yarloo}
\author{Hua-Chen Zhang}
\author{Anne E. B. Nielsen}
\affiliation{Department of Physics and Astronomy, Aarhus University, Ny Munkegade 120, DK-8000 Aarhus C}

\begin{abstract}
Adiabatic time evolution of quantum systems is a widely used tool with applications ranging from state preparation through simplifications of computations and topological transformations to optimization and quantum computing. Adiabatic time evolution generally works well for gapped ground states, but not for thermal states in the middle of the spectrum that lack a protecting energy gap. Here we show that quantum many-body scars -- a particular type of highly excited states -- are suitable for adiabatic time evolution despite the absence of a protecting energy gap. Considering two rather different models, namely a one-dimensional model constructed from tensor networks and a two-dimensional fractional quantum Hall model with anyons, we find that the quantum scars perform similarly to gapped ground states with respect to adiabatic dynamics when the required final adiabatic fidelity is around 0.99. The maximum speed at which the scar state of the one-dimensional model can be adiabatically transformed decreases as a power law with system size, as opposed to exponentially for both generic thermal and disorder-driven localized states. At constant and very low ramp speed, we find that the deviation of the fidelity from unity scales linearly with ramp speed for scar states, but quadratically for gapped ground states. The gapped ground states hence perform better when the required adiabatic fidelities are very high, such as 0.9999 and above. We identify two mechanisms for leakage out of the scar state and use them to explain our results. While manipulating a single, isolated ground state is common in quantum applications, adiabatic evolution of scar states provides the flexibility to manipulate an entire tower of ground-state-like states simultaneously in a single system.
\end{abstract}

\maketitle

\section{Introduction}

According to the adiabatic theorem \cite{born1928,schwinger1937,kato1950}, a quantum system that is initially prepared in an eigenstate of the Hamiltonian stays in that eigenstate if the parameters of the Hamiltonian are varied sufficiently slowly. This theorem has a profound place in physics because of its numerous applications, and its significance can hardly be overemphasized. From a theoretical perspective, it tremendously simplifies computations for slowly varying systems. Instead of solving a dynamical problem in a huge Hilbert space, it suffices to keep track of a single eigenstate of the Hamiltonian. Experimentally, the adiabatic theorem provides a key for preparing ground states \cite{aspuru-guzik2005}. Instead of cooling the system, which is often challenging, one can initialize the system in a simple ground state of a simple Hamiltonian and from there adiabatically prepare the desired ground state by slowly varying the parameters of the model. Computationally, the adiabatic theorem provides the basis for both adiabatic quantum computing \cite{farhi2000,albash2018} and solving challenging optimization problems through quantum annealing \cite{Annealing}. Yet another application is to reap the advantages of topological systems, where topological invariants emerge from cyclic adiabatic processes \cite{thouless1983}.

From a practical point of view, it is important that sufficiently slowly is not too slowly, as both speed and maintaining quantum coherence are important for most applications. Several studies have been carried out to quantify how slow processes need to be to achieve adiabaticity (see, e.g., Refs.\ \cite{Marzlin:2004,Tong:2005,Tong:2007,Altland:2008,Du:2008,Amin:2009,comparat2009,Tong:2010,Bachmann:2017,Lychkovskiy:2017,bachmann:2018}). The resulting measures depend on the energy gap to states adjacent in energy and also on matrix elements of the time derivative of the Hamiltonian, the so-called generalized force operator. Adiabatic time evolution generally works well for ground states of gapped Hamiltonians, as the gap ensures that one can move relatively fast. For typical thermal states in the middle of a spectrum, however, there are many states nearby in energy, and adiabaticity is very hard to achieve even for quite small systems.

Thermodynamics is a highly successful theory, and in a quantum context thermodynamics is understood through the eigenstate thermalization hypothesis \cite{gogolin2016,mori2018}. We therefore generally expect that highly excited states of generic quantum systems are thermal, i.e.\ follow the eigenstate thermalization hypothesis. Nevertheless, it has turned out that highly excited states can be nonthermal. Here, we are particularly interested in quantum many-body scars \cite{Shiraishi2017,bernien2017,Turner:2018,Turner:2018_PRB,moudgalya2018a,moudgalya2018b,Schecter2019,serbyn2021,moudgalya2022,Aramthottil2022,chandran2023}. A quantum many-body scar is a nonthermal, highly excited state with low entanglement entropy that is embedded in a thermal spectrum of a strongly correlated quantum many-body system. If one has a tower of quantum scars separated equidistantly in energy, they can give rise to revivals that provide a way to directly observe the nonthermal behavior and also have potential applications in quantum sensing and metrology \cite{dooley2021,dooley2023}. Quantum many-body scars have been observed in experiments with Rydberg atoms in optical tweezers \cite{bernien2017,Turner:2018,Turner:2018_PRB} and on superconducting processors \cite{zhang2023many}.

Ground states are special in many ways. They can be prepared by cooling \cite{diedrich1989,bergenfeldt2009}, they typically have low entanglement entropy \cite{srednicki1993,eisert2010}, they can be manipulated adiabatically in gapped systems, and they are often the target for applications involving quantum effects. Quantum many-body scars have obvious similarities to ground states, as they are nonthermal and have low entanglement entropy. This poses the intriguing question to what extent quantum many-body scars can be utilized for the same purposes as quantum many-body ground states. Here, we answer a crucial element of that question by showing that one can achieve adiabatic time evolution of quantum scar states on reasonable timescales that grow only as a power law with system size.

Intuitively, one may expect that adiabatic time evolution is not possible for quantum scar states. They sit in the middle of a thermal spectrum with a large number of other states nearby in energy. It is important to note, however, that the nearby states have rather different properties compared to the scar states. In particular, they have much higher entanglement entropy. For local Hamiltonians, it seems difficult to dynamically generate a large amount of entanglement. Such an ``\textit{entropy gap}'' (instead of an energy gap) may reduce the amount of transfer to these states. As there is a large number of such states, however, it is not clear how well adiabatic time evolution will work for scar states.

The main result of this paper is to show through numerical investigations of two rather different models that adiabatic time evolution works about equally well for scar states and ground states, unless extremely high adiabatic fidelities are required. We identify two mechanisms for leakage out of the scar state. One mechanism is leakage to thermal states that cross the scar state, and for this type of leakage the deviation of the fidelity from unity increases linearly with the ramp speed for small ramp speeds. The other mechanism is leakage to states of a similar nature at a different energy, and this type of leakage is similar to the leakage that happens when the initial state is a gapped ground state. Both mechanisms are, however, inefficient, either because of the different nature of the states involved or because of the energy difference. As a result, the ramp speed at which an adiabatic fidelity of $0.99$ is achieved in the considered examples is about the same when the initial state is a scar state or the ground state. For one of the models, we study several different system sizes, and we find that the fastest speed at which the parameters in the Hamiltonian can change while still remaining close to adiabatic shows a slow power law decrease with system size when the initial state is a scar or ground state. This is to be contrasted with exponential decrease for thermal states. We also discuss the effect of perturbations for this model. We identify a family of perturbations for which the scar state remains an exact eigenstate, and for this family the physics is robust, even for strong perturbations. For other types of perturbations, we find that the physics is unaltered if the perturbations are weak enough that the scar state of the perturbed model has a high fidelity with the exact scar state of the unperturbed model.

In the following, we first consider a quantum scar model constructed from a matrix product state (MPS) \cite{perez-garcia2007}. The model depends on a parameter that can be varied while maintaining the scar properties of the model. This allows us to study the time evolution, when varying the parameter slowly in time. We also investigate a scar model that has a Laughlin state with two Abelian anyons as its scar state. In this case the positions of the anyons provide suitable parameters to vary. Similar conclusions are obtained for both models.

Our results suggest that one can, indeed, manipulate quantum many-body scars in the same ways as one can manipulate quantum ground states. This opens up interesting possibilities, as one can, in this way, have a whole tower of highly excited states that are like ground states. Instead of manipulating only a single state, one can hence manipulate several states in parallel and also allow for transitions between different scar states.

\section{Scar model from matrix product states} \label{sec:scar_MPS}

To investigate adiabatic dynamics of scar states, the model must fulfil certain requirements. First, the Hamiltonian should depend on a parameter $s$ that can be varied in time. Second, the model should be a quantum scar model for all the considered values of $s$. This means in particular that there should be at least one highly excited state with low entanglement embedded in an otherwise thermal spectrum. Third, the scar state should change as a function of $s$ as the system will otherwise trivially stay in the scar state when $s$ is varied. Fourth, it is preferable if a simple change of the Hamiltonian moves the scar state from a highly excited state to the ground state. The latter is not necessary, but convenient for a more direct comparison of the dynamics when the initial state is a scar state or the ground state.

Frustration-free systems provide an interesting starting point for obtaining such models. In particular, consider a Hamiltonian $H(s)=\sum_i c_i h_i(s)$, where $c_i$ are real numbers and $h_i(s)$ are Hermitian and positive semi-definite operators that all annihilate a particular quantum state $|\Phi_0(s)\rangle$, i.e.\ $h_i(s)|\Phi_0(s)\rangle=0$ for all $i$. If we choose $c_i=+1$ for all $i$, $|\Phi_0(s)\rangle$ is a ground state of the model, because it has energy zero, which is the lowest possible energy. If, however, we choose some of the $c_i$ to be negative numbers, $H$ is no longer positive semi-definite, and there will generally be both positive and negative energies in the spectrum. As $|\Phi_0(s)\rangle$ still has zero energy, it will now be in the middle of the spectrum. One then needs to check numerically, if the rest of the spectrum is thermal, and whether $|\Phi_0(s)\rangle$ has low entanglement entropy compared to the thermal states.

In this section, we utilize the framework of MPS scars, proposed by Moudgalya \textit{et al.\ }\cite{Moudgalya:2020}, to obtain such models and analyze their properties. This construction has several advantages. First, the construction leads to a Hamiltonian with only local terms. Second, the scar states have low entanglement entropy by construction. Third, it is easy to insert a parameter $s$ that both maintains the scar model and changes the wave functions of the scar states. Fourth, it is possible to construct models with a tower of scar states. Fifth, as the scar states are represented by MPSs, they can be prepared efficiently with quantum circuits \cite{Schon2005,Malz2023}. In the following, we consider a model with one scar state, as this is already sufficient to capture the physics of adiabatic dynamics of scar states. The generalization to a tower of scar states will be discussed in Sec.\ \ref{sec:tower}.

\subsection{Model} \label{sec:model_AKLT}

The MPS framework for constructing scar models with a single scar state works quite generally, as long as we start from an MPS with a sufficiently low bond dimension. Here, we choose the MPS to be a particular deformation of the Affleck-Kennedy-Lieb-Tasaki (AKLT) state \cite{Affleck:1987,Schollwock:2011,Orus:2014}, as this model can be modified to host a tower of scar states (see Sec.\ \ref{sec:tower}). Consider a one-dimensional chain consisting of $N$ spin-$1$ particles. The local Hilbert space on site $j$ is spanned by the three vectors $|m_j\rangle$, where $m_j\in\{+,0,-\}$ labels the $z$ component of the $j$th spin-$1$. We define the MPS
\begin{multline}\label{eq:MPS_manifold}
	|\Phi_0(s)\rangle =\\ \mu_N\sum_{m_1,\ldots,m_N}{\ \Tr \left[ A^{m_1}_1(s)  \cdots  A^{m_N}_N(s)\right] |m_1, \cdots m_N\rangle},
\end{multline}
where $\mu_N$ is a normalization constant and the $2\times2$ matrices
\begin{eqnarray} \label{eq:gen_AKLT_MPS}
	A_j^{\pm} (s) &=& \pm \left(1-\dfrac{s}{2}\right) \sqrt{\frac{2}{3}} \sigma^\pm,\\
	A_j^{0} (s) &=& -\left(1+\dfrac{s}{2}\right)\frac{1}{\sqrt{3}} \sigma^z
\end{eqnarray}
are expressed in terms of the Pauli matrices
\begin{equation}
\sigma^+=\left(\begin{tabular}{cc}0 & 1 \\ 0 & 0\end{tabular}\right),
\quad
\sigma^-=\left(\begin{tabular}{cc}0 & 0 \\ 1 & 0\end{tabular}\right),
\quad
\sigma^z=\left(\begin{tabular}{cc}1 & 0 \\ 0 & -1\end{tabular}\right).
\end{equation}
The real, free parameter $s$ has an implicit time dependence $s\equiv s(t)$, and below we study the dynamics when $s$ varies from $0$ to $1$. When $s=0$, the state \eqref{eq:MPS_manifold} reduces to the AKLT state. The AKLT state is the ground state of the AKLT chain, which is a prototype of frustration-free systems. Deforming away from the AKLT point can substantially modify $|\Phi_0(s)\rangle$, while keeping its low-entangled nature intact.

Using matrix product state methods \cite{fannes1992finitely,nachtergaele1996spectral,wolf2006quantum} as in \cite{Moudgalya:2020}, we derive a set of local operators $h_i(s)$, $i\in\{1,2,\ldots,N\}$, that annihilate the MPS,
\begin{equation}
h_i(s) |\Phi_0(s)\rangle = 0,\quad\forall i.
\end{equation}
The operator $h_i(s)$ acts on the spins at sites $i$ and $i+1$, where site $N+1$ is identified with site $1$. They take the form
\begin{equation}\label{eq:AKLT_hi}
h_{i}(s) = \mathcal{J}_0  |K_0 (s)\rangle\langle K_0 (s) | + \sum_{m=\pm1,\pm2} \mathcal{J}_m |J_{2,m}\rangle\langle J_{2,m}|,
\end{equation}
where $|J_{2,m}\rangle$ denote the total angular momentum eigenstates of two coupled spin-$1$'s with total spin $S=2$ and total magnetization $m\in\{0,\pm1,\pm2\}$, i.e.\
\begin{equation}\label{eq:Jm}
	|J_{2,\pm 2}\rangle = |\pm\ \pm\rangle, \quad |J_{2,\pm 1}\rangle = \frac{1}{\sqrt{2}}\left(|\pm \ 0\rangle + |0\ \pm\rangle\right).
\end{equation}
The only $s$-dependent contribution comes from
\begin{equation}\label{eq:K0}
|K_0 (s)\rangle = \sqrt{\lambda_s}(|+-\rangle+|-+\rangle) + \sqrt{1-2\lambda_s}|00\rangle,
\end{equation}
where
\begin{equation}
\lambda_s=\frac{\left(s + 2\right)^{4}}{{4 \left(s - 2\right)^{4} + 2 \left(s + 2\right)^{4}}}.
\end{equation}
Note that the $h_i(s)$ annihilate the MPS \eqref{eq:MPS_manifold} for general choices of the five parameters $\{\mathcal{J}_m\}$. We shall here take $\{\mathcal{J}_m\}$ to be real to ensure that $h_i(s)$ is Hermitian.
	
If we choose the parameters $\{\mathcal{J}_m\}$ to be non-negative, real numbers, the state $|\Phi_0(s)\rangle$ is a ground state of the frustration-free Hamiltonian
\begin{equation}\label{eq:Hp}
H^{+}(s)= \sum_i h_i(s),\quad  \mathcal{J}_m = \mathcal{J}^+_m \geq 0,
\end{equation}
independent of the choice of $s\in[0,1]$. In the computations below, for which we use $H^+(s)$, we avoid simply putting all $\{\mathcal{J}_m\}$ equal to $+1$ as this leads to an $\mathrm{SU}(2)$ symmetry at $s=0$.
Instead, we choose each $\mathcal{J}_m$ to be a random number in the interval $[0,1]$, namely,
\begin{align}
    \mathcal{J}^+_{-2} &=0.97545513805816, \nonumber\\
    \mathcal{J}^+_{-1}&=0.84883205409987, \nonumber\\
    \mathcal{J}^+_{0}&=0.40823209824201, \nonumber\\
    \mathcal{J}^+_{+1}&=0.32544692707096, \nonumber\\
    \mathcal{J}^+_{+2}&=0.55079799361114. \nonumber
\end{align}
We make this choice only once and keep it throughout. $H^+(s)$ constructed in this way remains invariant under translation by one lattice spacing, spatial reflection $\mathcal{I}$, and the $\mathrm{U}(1)$ symmetry corresponding to the conservation of the total magnetization $S^z_{\textrm{tot}}=\sum_i S^z_i$, where $S_i^z$ is the $z$ component of the spin-1 on site $i$. The embedded MPS \eqref{eq:MPS_manifold} lies in the symmetry sector of $S^z_{\textrm{tot}}=0$, zero quasi-momentum $k=0$, and even $\mathcal{I}$, labeled by the set $(S^z_{\textrm{tot}},k,\mathcal{I})=(0,0,+1)$. We hence restrict all numerical computations below to that symmetry sector. We find numerically that the ground state is non-degenerate for all $s$ within the considered range.

We shall also consider the Hamiltonian
\begin{equation}\label{eq:Hm}
H^{-}(s)= \sum_i h_i(s),\quad  \mathcal{J}_m=\mathcal{J}^-_m \equiv (-1)^m,
\end{equation}
for which we choose $\mathcal{J}_m=(-1)^m$. Note that $H^{-}(s)$ is not positive semi-definite, and the MPS \eqref{eq:MPS_manifold} is now a scar state rather than the ground state. With this choice of parameters, $H^-(s)$ possesses an additional spin-inversion symmetry $\mathcal{Z}$ within the $S^z_{\textrm{tot}}=0$ and $k=0$ sector. The embedded scar state $|\Phi_0(s)\rangle$ lies in the $(S^z_{\textrm{tot}},k,\mathcal{I},\mathcal{Z})=(0,0,+1,+1)$ symmetry sector, and we hence restrict our analysis to that symmetry sector throughout.

\begin{figure*}
	\includegraphics[width=1\linewidth]{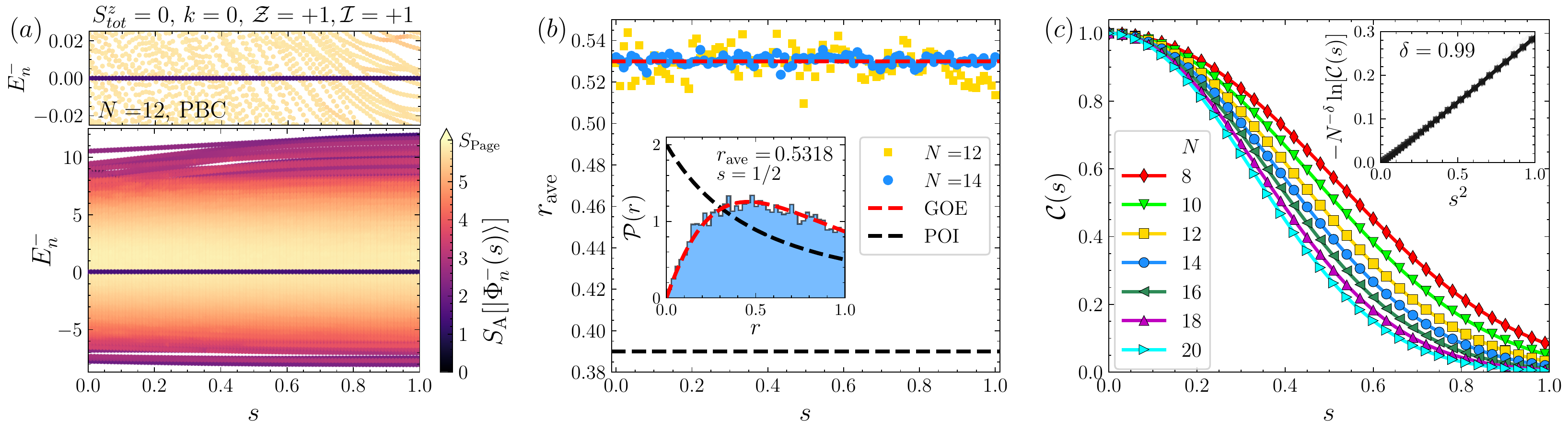}
	\caption{\textbf{Scarring properties of the MPS Hamiltonian $H^-(s)$.} (a) Half-chain entanglement entropy of the eigenstates of $H^-(s)$ within the symmetry sector $(S^z_{\textrm{tot}},k,\mathcal{I},\mathcal{Z})=(0,0,+1,+1)$ for $N=12$ sites and periodic boundary conditions. The scar state produces the dark line at zero energy, while the other states in the middle of the spectrum are seen to have entropies close to the Page value indicating thermal behavior. The zoomed panel shows several crossings between thermal states and the scar state.
    (b) The average of the level spacing ratio $r_{\textrm{ave}}$ within the symmetry sector $(S^z_{\textrm{tot}},k,\mathcal{I},\mathcal{Z})=(0,0,+1,+1)$ is seen to approach the GOE value for increasing system size for the entire range of $s$. The inset shows the level spacing statistics for $s=1/2$ and $N=14$.
    (c) The scar state changes as a function of $s$, and we here quantify this change by plotting the instantaneous fidelity in Eq.\ \eqref{eq:inst_fid}. The data collapse for different system sizes $N=10,100,200,\cdots,1000$ in the inset shows that $\ln [\mathcal{C}(s)]\approx -0.28\cdot N^\delta s^2$ with $\delta\sim 1$, as expected for finitely-correlated MPSs. The results in (a) and (b) have been computed using exact diagonalization, and the results in (c) have been computed by utilizing the MPS structure of the state $|\Phi_0(s)\rangle$.}
 \label{fig:AKLT_static}
\end{figure*}

We find numerically that the scar state is nondegenerate except for accidental degeneracies with a thermal state at specific values of $s$ as seen in the zoomed panel of Fig.\ \ref{fig:AKLT_static}(a). Since the scar state is an exact eigenstate at zero energy for all $s$, it cannot hybridize with other states with similar energies. As a result, the other states can cross the scar state in the spectrum as a function of $s$, and this produces the accidental degeneracies. Note also that this means that the number of states in the spectrum with energy less than the scar state is generally not preserved when $s$ varies.

Let us finally introduce some notation that we will use in the following. In addition to the state $|\Phi_0(s)\rangle$, the Hamiltonian $H^{\pm}(s)$ has $\mathcal{D}_{\pm}-1$ other eigenstates $|\Phi^{\pm}_n(s)\rangle$ with energies $E_n^{\pm}(s)$ within the relevant symmetry sector that are obtained by solving the stationary Schr\"odinger equation
\begin{equation}
H^{\pm}(s)|\Phi^{\pm}_n(s)\rangle
=E^{\pm}_n(s)|\Phi^{\pm}_n(s)\rangle.
\end{equation}
Here, $\mathcal{D}_{\pm}$ is the dimension of the Hilbert space describing the relevant symmetry sector, and $n\in\{1,2,\ldots,\mathcal{D}_{\pm}-1\}$ labels the states in order of increasing energy.

\subsection{Scarring properties of $H^-(s)$}

Before investigating the dynamics, we show numerically that
$H^-(s)$ is a scar model for the considered parameters. We do this by evaluating the entanglement entropy and the average level spacing ratio. We also quantify the change of the scar state as a function of the parameter $s$.

\subsubsection{Entanglement entropy}

We first evaluate the half-chain von Neumann entanglement entropy, which is defined as
\begin{equation}\label{eq:vNS}
S_{A}(s) = -\Tr_{A}  \left\{ u(s) \ln[u(s)] \right\},
\end{equation}
where
\begin{equation}
u(s)=\Tr_{\bar{A}} (|\Phi^-_n(s)\rangle\langle \Phi^-_n(s) |) \nonumber
\end{equation}
is the reduced density matrix of subsystem $A$ consisting of $N/2$ contiguous sites and $\bar{A}$ is the complement of $A$. In Fig.\ \ref{fig:AKLT_static}(a), we plot the half-chain entanglement entropy for a chain of length $N=12$ within the symmetry sector containing the scar state $|\Phi_0(s)\rangle$. The scar state is seen as a dark line at zero energy due to its low entanglement entropy. The other states in the middle of the spectrum are seen to have entanglement entropies close to the Page value $S_{\textrm{Page}}=\frac{N}{2} \ln 3-\frac{1}{2}$ \cite{page1993}, in agreement with the predictions of the eigenstate thermalization hypothesis \cite{Alessio:2014_rev}. The result hence indicates that the states in the middle of the spectrum are thermal, except for the scar state.

\subsubsection{Level spacing statistics}

To further confirm that the spectrum of $H^-(s)$ is thermal for the considered values of $s$, we also examine the level spacing ratio, defined by
\begin{equation}\label{eq:level_spacings}
r_n = \frac{\min(\delta_n,\delta_{n+1})}{\max(\delta_n,\delta_{n+1})},
\end{equation}
where $\delta_n=\tilde{E}_{n+1}^{-}-\tilde{E}_n^{-}$ is the energy difference between two adjacent energy levels in the spectrum within the considered symmetry sector. We put a tilde on the energies here, because for this computation we label all $\mathcal{D}_{-}$ states from $1$ to $\mathcal{D}_{-}$ in order of increasing energy, including also the scar state. For ergodic systems with time-reversal symmetry, the distribution of $r_{n}$ is generally expected to obey the Gaussian orthogonal ensemble (GOE) with the mean value $r_\mathrm{GOE}\approx 0.536$ \cite{Poilblanc:1993,Atas:2013}. In contrast, in an integrable or many-body localized system, the levels do not repel each other, leading to Poisson (POI) statistics with $r_\mathrm{POI} \approx 0.386$. Figure \ref{fig:AKLT_static}(b) presents the instantaneous level spacing ratio averaged over all energy differences within the considered symmetry sector,
\begin{equation}\label{eq:mean_level_spacing_ratios}
r_{\mathrm{ave}} = \frac{1}{\mathcal{D}_{-}-1} \sum_{n=1}^{\mathcal{D}_{-}-1} r_n.
\end{equation}
It is seen that $r_{\mathrm{ave}}$ approaches the GOE value with increasing system size for the entire range of $s$, and the inset of Fig.\ \ref{fig:AKLT_static}(b) shows that the corresponding level spacing distribution also fits the GOE statistics well. This indicates that the spectrum is thermal.

\subsubsection{Change of the scar state with $s$} \label{subsubsec:CE}

To quantify the change in the structure of the embedded scar state as a function of $s$, we plot the instantaneous fidelity
\begin{equation}\label{eq:inst_fid}
\mathcal{C}(s) = |\langle \Phi_0(0) | \Phi_0(s) \rangle|^2
\end{equation}
in Fig.\ \ref{fig:AKLT_static}(c). It is seen that $\mathcal{C}(s)$ is a monotonically decreasing function of $s$ and decays exponentially fast with system size. Moreover, data for different system sizes collapse onto a single curve (see the inset), showing the asymptotic form $-\ln[\mathcal{C}(s)]\approx C_N s^2$ at large $N$ with the exponent $C_N\approx 0.28\cdot N^{0.99}$. This implies that the global structure of the scar state is substantially modified by a weak deformation, i.e., adiabatically connected scar states are nearly orthogonal in the limit of large $N$, despite the change in $s$ being small.

This is a manifestation of the so-called orthogonality catastrophe \cite{Anderson:1967}; a genuine many-body phenomenon by which infinitesimal local perturbations evolve the system to an orthogonal state in the thermodynamic limit. While the extended states of a metallic system exhibit power-law orthogonality catastrophe with the \textit{catastrophe exponent} $C_N\sim \ln(N)$ \cite{Anderson:1967}, finitely-correlated states subject to local, homogeneous transformations in systems with spatial dimension $D$ are expected to show an exponential orthogonality catastrophe with $C_N\sim N$, in agreement with the above estimate for the MPS $|\Phi_0(s)\rangle$. A similar exponential (but statistical) orthogonality catastrophe has been observed in Anderson localized or many-body localized eigenstates which are subjected to strictly local, single-site driving \cite{Khemani:2015,Deng:2015}.

\subsection{Adiabatic time evolution} \label{sec:dynamics_AKLT}

As already mentioned in Sec.\ \ref{sec:model_AKLT}, one can see the parameter $s\equiv s(t)$ as an adiabatically modulated (ramp) parameter through which a nondegenerate scar state can be modified. This feature makes the MPS embedded Hamiltonian \eqref{eq:Hm} a suitable playground to explore adiabatic response of quantum many-body scars. We are interested in whether a system initially prepared in an exact scar state can adiabatically follow the ramp and remain in its instantaneous eigenstate, or the lack of a protecting gap for highly excited states will lead to ``\textit{quantum leakage}'' \cite{Michailidis:2020} to the thermal subspace and hence loss of  adiabaticity.

To address this question, we initially prepare the system in the scar eigenstate of $H^-(0)$, i.e., the AKLT state  $|\Phi_0(0)\rangle$, and then evolve the system from time zero to time $T$ with the Hamiltonian $H^-[s(t)]$, where $s(t)$ specifies the ramp function. We here consider a linear ramp with ramp speed $v$, i.e.,
\begin{equation}
s=s(t)=vt, \qquad v=\frac{1}{T}, \qquad t\in[0,T].
\end{equation}
According to the time-dependent Schr\"odinger equation, the evolved state $|\psi(t)\rangle$ at time $t$ can be expressed in terms of the unitary time-evolution operator,
\begin{align}\label{eq:TSE}
	|\psi(t)\rangle=\mathcal{T}\exp\left\{-i \int_0^t H^-[s(t')]\, dt'\right\} |\Phi_0(0)\rangle,
\end{align}
where $\mathcal{T}$ denotes time-ordering. For an ideal adiabatic evolution, the time evolved state $|\psi(T)\rangle$ coincides with the instantaneous exact scar state at the end of the ramp $|\Phi_0[s(T)]\rangle=|\Phi_0(1)\rangle$ given by the MPS \eqref{eq:MPS_manifold}. We hence introduce the adiabatic fidelity
\begin{equation}\label{eq:ad_fid}
	F = |\langle \Phi_0[s(T)] | \psi(T) \rangle|^2
\end{equation}
to quantify how close the time evolved state is to the ideal adiabatic result. The fidelity is unity in the ideal case, and deviations from unity quantify the amount of leakage to other states.

We note that the initial state $|\Phi_0(0)\rangle$ considered above can also be recast as the exact ground state of the Hamiltonian $H^+(0)$. This enables us to directly compare the dynamics obtained by starting from the scar state $|\Phi_0(0)\rangle$ and evolving with $H^-[s(t)]$ to the dynamics obtained by starting from the ground state $|\Phi_0(0)\rangle$ and evolving with $H^+[s(t)]$. This can be achieved by replacing $H^-(s)$ with $H^+(s)$ in Eq.\ \eqref{eq:TSE} and computing $F$ with respect to $|\Phi_0[s(T)]\rangle$, which is now the instantaneous ground state at the end of the ramp. We can of course also choose the initial state $|\psi(0)\rangle$ to be a thermal state and time evolve with $H^-[s(t)]$ in order to compare the performance of scar states and thermal states with respect to adiabatic time evolution. Altogether, we hence investigate and compare three different cases:
\begin{enumerate}\label{list:GST}
\item \textbf{Ground state dynamics.} The initial state is $|\Phi_0(0)\rangle$, i.e.\ the ground state of $H^+(0)$, and the system is time evolved with $H^+[s(t)]$. Plotted data are shown in blue and/or labeled `GS'.
\item \textbf{Scar state dynamics.} The initial state is $|\Phi_0(0)\rangle$, i.e.\ the scar state of $H^-(0)$, and the system is time evolved with $H^-[s(t)]$. Plotted data are shown in red and/or labeled `Scar'.
\item \textbf{Thermal state dynamics.} The initial state is the eigenstate of $H^-(0)$ that is closest in energy to the scar state, and the system is time evolved with $H^-[s(t)]$. Plotted data are shown in green and/or labeled `Thermal'.
\end{enumerate}

We numerically compute the time evolved state by using time-integration based on the Chebyshev expansion technique \cite{Ezer:1984,schaefer2017}. We consider evolution times up to $T=10^5$ and system sizes up to $N=20$ with Hilbert space dimension of order $\mathcal{D}_{\pm}\sim 10^7$ within the relevant symmetry sector.

\subsubsection{Adiabatic fidelity}

\begin{figure}
	\includegraphics[width=\linewidth]{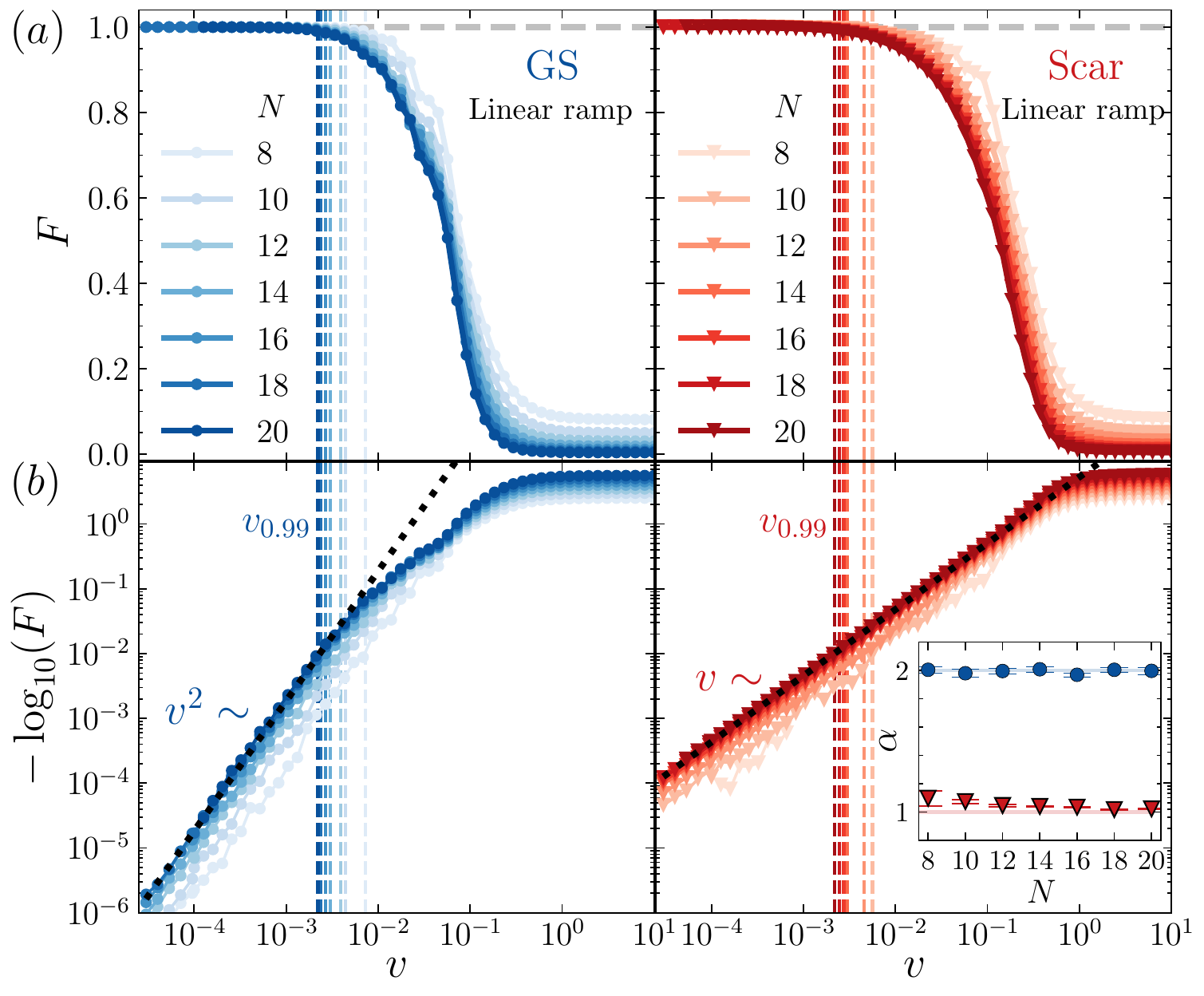}
	\caption{\textbf{Adiabatic fidelity.} (a) Adiabatic fidelity as a function of ramp speed for ground state and scar state dynamics (see the list in Sec.\ \ref{list:GST}). The vertical dashed lines indicate the adiabatic velocity $v_{0.99}$, which we define to be the ramp speed at which the adiabatic fidelity drops below $0.99$. It is remarkable, and one of the key results of this paper, that the adiabatic velocity for the scar state is about the same as that for the ground state, despite the fact that the scar state is a highly excited state with several thermal states nearby in energy. (b) The logarithmic fidelity, $-\log_{10}(F)$, shows the deviation of $F$ from unity more clearly in the high fidelity region. For both ground state and scar state dynamics, the logarithmic fidelity is seen to follow a power law scaling $\propto v^\alpha$ for small enough ramp speeds. The logarithmic fidelity for the ground state dynamics scales as $v^2$ for ramp speeds $v\lesssim v_{0.99}$. In contrast, the logarithmic fidelity for the scar state dynamics scales linearly in $v$ for a much broader range of ramp speeds. The dotted lines denote a $v^\alpha$ fit to the results obtained for $N=20$. The inset indicates the persistence of these scalings for different system sizes.}
	\label{fig:Fv_AKLT_δs}
\end{figure}

\begin{figure}
\includegraphics[width=1\linewidth]{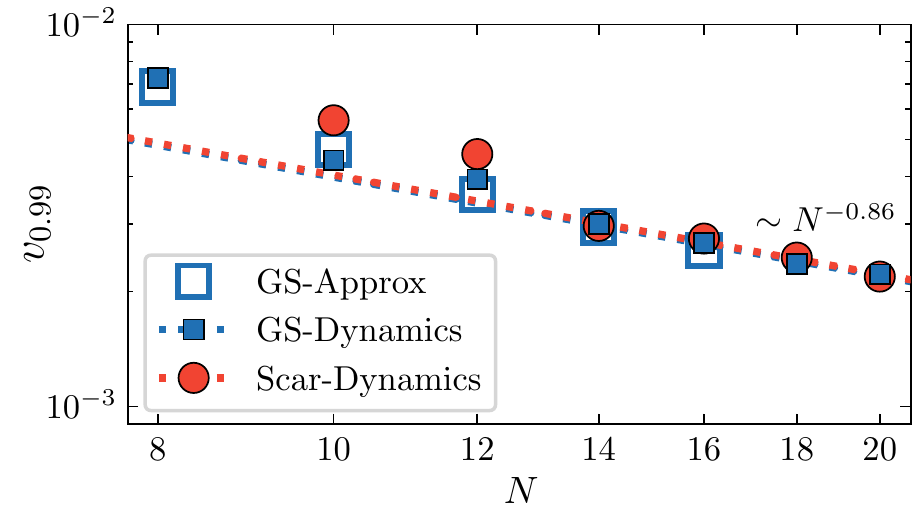}
\caption{\textbf{Scaling of adiabatic velocity with system size.} The adiabatic velocities $v_{0.99}$ for the ground state and the scar state dynamics extracted from the data in Fig.\ \ref{fig:Fv_AKLT_δs} display power law decay $v_{0.99}\approx e^{-3.5}\,N^{-0.86}$ with system size. This is in contrast to thermal state dynamics, for which the adiabatic velocity is expected to decay exponentially with system size (see Eq.\ \eqref{eq:vad_th}). The symbols labeled GS-Approx show the adiabatic velocities obtained from adiabatic perturbation theory by putting $F=0.99$ in Eqs.\ \eqref{eq:cmssq} and \eqref{eq:defF}.} \label{fig:HAC_Dyn}
\end{figure}

The adiabatic fidelity for the ground state and scar state dynamics is shown as a function of the ramp speed and for different system sizes in Fig.\ \ref{fig:Fv_AKLT_δs}(a). Let us first discuss the plot for the ground state dynamics. It is seen that the adiabatic fidelity decreases monotonically with ramp speed. As the ground state of $H^+(s)$ is isolated from the first excited state by an energy gap, we expect the adiabatic fidelity to be close to unity for low enough ramp speeds due to the adiabatic theorem, and this is indeed the case. For larger ramp speeds, the adiabatic fidelity is lower due to a larger amount of leakage to other states. As a quantitative measure for how fast the evolution can be done while still remaining close to adiabaticity, we define the adiabatic velocity $v_{0.99}$ to be the ramp speed at which the fidelity drops below $0.99$. The adiabatic velocity, which is also marked in the figure, is seen to decrease slowly with increasing system size. From Fig.\ \ref{fig:HAC_Dyn} we extract the behavior $v_{0.99}\approx e^{-3.5}\,N^{-0.86}$ on a log-log scale. It is important that the adiabatic velocity decreases as a power law rather than exponentially with system size, as this means that much larger systems can be manipulated adiabatically on reasonable time scales. In Sec.\ \ref{sec:AKLT_QSL} below, we find $v_{1/e}\sim N^{-0.5}$, where $v_{1/e}$ is defined as the ramp speed at which the fidelity drops below $e^{-1}$ and $e$ is the base of the natural logarithm.

Turning to the plot of adiabatic fidelity for the scar state dynamics, we observe that the behavior is similar to that for the ground state dynamics, and the adiabatic velocity is almost the same. This is remarkable, as the scar state is surrounded by a large number of thermal states with similar energies in the spectrum. Nevertheless, leakage to these states is very limited. We will discuss the leakage further in Secs.\ \ref{sec:dynamics_AKLT_leakage} and \ref{sec:interpretation_scar} below. It is also remarkable that the scaling of the adiabatic velocity with system size is about the same for the scar state dynamics as for the ground state dynamics, as seen in Fig.\ \ref{fig:HAC_Dyn}.

\begin{figure*}
\includegraphics[width=1\linewidth]{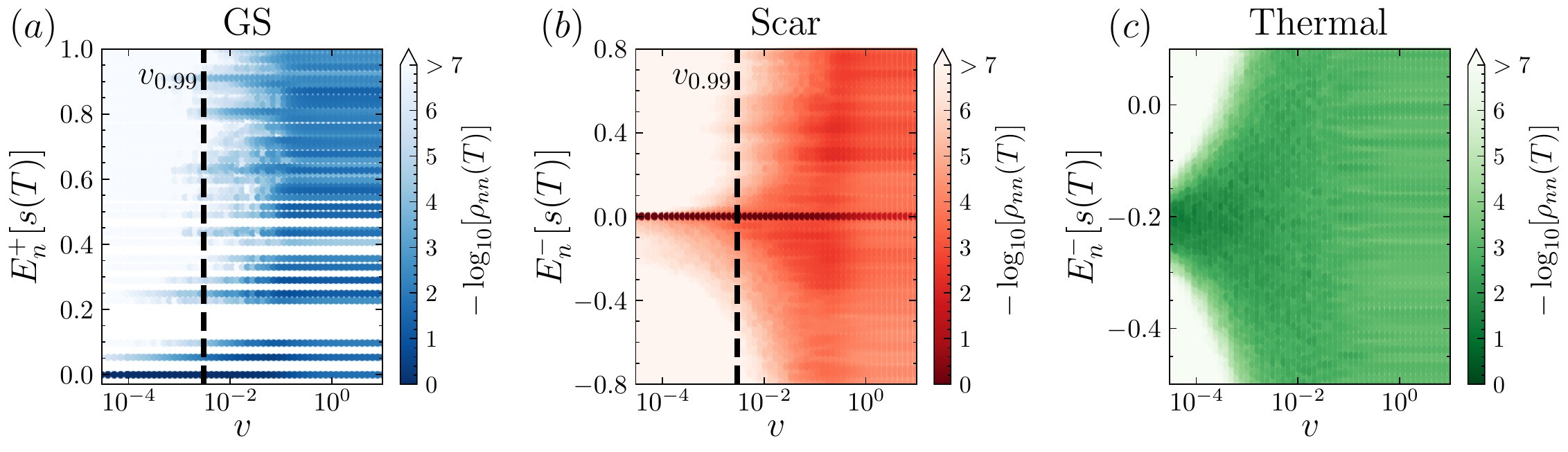}
  \caption{\textbf{Leakage to other states.} Population of the energy eigenstates at the end of the ramp for ground state, scar state, and thermal state dynamics (see the list in Sec.\ \ref{list:GST}) and $N=14$, obtained by exact diagonalization. The ground state and the scar state are seen as a dark line at zero energy in (a) and (b), respectively, because most of the population is in this state. For the thermal state dynamics, in contrast, a large number of states are populated even at low ramp speeds.}
  \label{fig:overlap_AKLT}
\end{figure*}

To further investigate the deviation of $F$ from unity in the high fidelity region, we plot the logarithmic fidelity, $-\log_{10}(F)$, in Fig.\ \ref{fig:Fv_AKLT_δs}(b). For ramp speeds that are comparable to $v_{0.99}$ or less, the logarithmic fidelity for the ground state dynamics is seen to scale quadratically with ramp speed, as expected for gapped systems \cite{grandi2010}. For the scar state, the scaling is instead linear and applies for a broader range of ramp speeds. Such deviations from quadratic scaling are typically observed in low-dimensional gapless systems \cite{grandi2010,Polkovnikov:2008} or in dynamics passing through quantum critical points \cite{Zurek:2005, Polkovnikov:2006,Grandi:2008}, where there is a crossover energy scale separating low-energy excitations with fast (diabatic) response and high-energy states having adiabatic response. The difference in scaling of the logarithmic fidelity with ramp speed for the ground and scar state dynamics means that, if one considers sufficiently low ramp speeds, higher fidelities are obtained for the ground state dynamics than for the scar state dynamics with the same ramp speed. For instance, at $v\approx 10^{-5}$ and $N=20$, the error is $-\log_{10}(F)\approx 10^{-6}$ for the ground state, but $-\log_{10}(F)\approx 10^{-4}$ for the scar state. This is relevant if very high fidelities are needed, but not for fidelities around $0.99$ as discussed above.

\subsubsection{Leakage to other states} \label{sec:dynamics_AKLT_leakage}

So far we have only quantified the total amount of leakage, and we now investigate how the leakage is distributed over the instantaneous eigenstates at the end of the ramp, $t=T$. This is shown in Fig.\ \ref{fig:overlap_AKLT}, where
\begin{equation}\label{eq:rhonn}
\rho_{nn}(t)= |\langle \Phi^{\pm}_n [s(t)]|\psi(t)\rangle|^2
\end{equation}
are the diagonal elements of the time-evolved density matrix written in the instantaneous energy basis. The observed behaviors are quite different for the ground state, scar state, and thermal state dynamics. For the ground state dynamics, the leakage is seen to happen mainly to one low-lying excited state for small ramp speeds. For the scar state dynamics, it is mainly the states that are slightly below the scar state in energy that carry population at the end of the ramp for small ramp speeds. For thermal state dynamics, there is a large amount of spreading even for quite small ramp speeds.

\begin{figure}
	\includegraphics[width=0.9\linewidth]{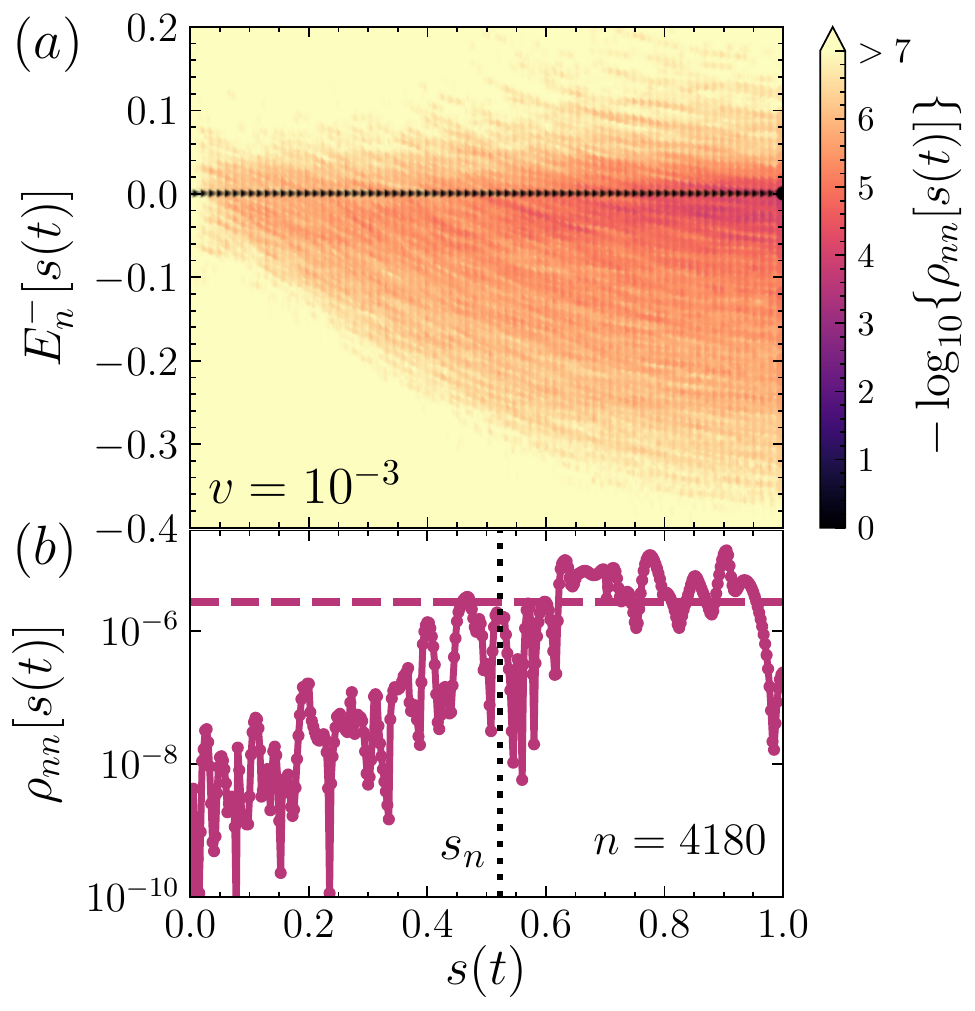}
	\caption{\textbf{Leakage dynamics.} (a) The population in each of the eigenstates as a function of the energy of the eigenstate and the ramp parameter $s(t)$ for $N=14$ spins and ramp speed $v=10^{-3}$. Leakage is seen to primarily happen when the thermal states cross the scar state, and the leaked population is then transported to lower energies because the energy of the thermal states in the middle of the spectrum decreases with $s$ as seen in Fig.\ \ref{fig:AKLT_static}(a). This explains why the population is mainly in states with energies below zero at the end of the ramp in Fig.\ \ref{fig:overlap_AKLT}(b). (b) Population in state number $4180$ as a function of the parameter $s$. The dotted line, labeled $s_n$, marks the point, where the state crosses the scar state. The population is seen to increase in a region around $s_n$. There are also several fluctuations, which are due to exchange of population with other thermal states. The dashed horizontal line is the estimate for the final population in Eq.\ \eqref{eq:cnapprox}.}
	\label{fig:leakage_AKLT}
\end{figure}

To understand how the population ends up where it does, we plot the evolution of the population $\rho_{nn}(t)$ as a function of the parameter $s(t)$ for the scar state dynamics at fixed ramp speed $v=10^{-3}$ and $N=14$ in Fig.\ \ref{fig:leakage_AKLT}. This figure indicates that leakage from the scar state to thermal states mainly happens, when the thermal states cross the scar state in the spectrum. Since the thermal states move down in energy as $s$ increases (see zoomed panel of Fig.\ \ref{fig:AKLT_static}(a)), the leaked population is mainly present in the states below zero energy at the end of the ramp (Fig.\ \ref{fig:overlap_AKLT}(b)). We also observe in the dynamics that population can transfer between thermal states, when they are close in energy.

\begin{figure}
	\includegraphics[width=0.9\linewidth]{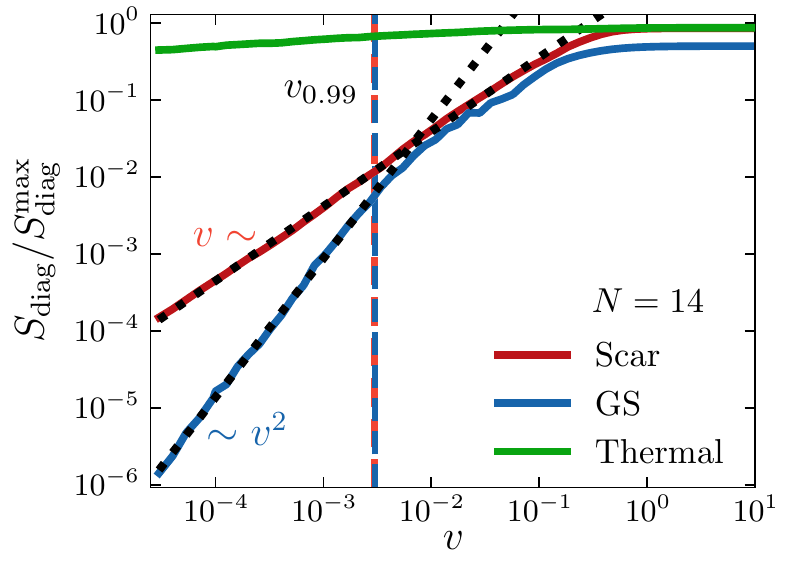}
	\caption{\textbf{Scaled diagonal entropy.} Scaled diagonal entropy obtained from exact diagonalization for ground state, scar state, and thermal state dynamics (see the list in Sec.\ \ref{list:GST}) and system size $N=14$. When the system starts in the thermal state, the scaled diagonal entropy is close to unity. This means that the population is spread over a large part of the spectrum at the end of the ramp. The scaled diagonal entropy is much lower for both the ground state and scar state dynamics at low ramp speeds because the system remains mainly in a single energy eigenstate. The vertical, dashed lines show $v_{0.99}$ determined from the dynamics, and the dotted lines are quadratic and linear fits, respectively, showing the scaling with $v$.}
	\label{fig:Sdiag_AKLT}
\end{figure}

We finally consider the diagonal (or participation) entropy \cite{Polkovnikov:2011},
\begin{equation}
	S_{\mathrm{diag}}=-\sum_n \rho_{nn}(T) \ln[\rho_{nn}(T)],
\end{equation}
as a measure for to what extent the population is spread over the energy eigenstates at the end of the ramp. If the time evolution is perfectly adiabatic, the system remains in a single instantaneous energy eigenstate and $S_{\mathrm{diag}}=0$. At the other extreme, if the population spreads equally over all the $\mathcal{D}_{\pm}$ energy eigenstates in the considered sector, the diagonal entropy takes the maximum value $S_{\mathrm{diag}}^{\mathrm{max}}=\ln(\mathcal{D}_{\pm})$. Figure \ref{fig:Sdiag_AKLT} presents the diagonal entropy as a function of the ramp speed $v$ for $N=14$. For the ground state dynamics, $S_{\mathrm{diag}}(T)$ displays a $v^2$ scaling for small ramp speeds, while for the scar state dynamics,  $S_{\mathrm{diag}}(T)$ grows linearly with ramp speed in a broader region, upon approaching $S_{\mathrm{diag}}^{\mathrm{max}}$. The thermal state dynamics on the other hand is far from adiabatic, and $|\psi(T)\rangle$ spreads substantially over the eigenstates of the final Hamiltonian even for quite low ramp speeds. The figure highlights the remarkable differences between scar states and thermal states with respect to adiabatic time evolution.

\subsection{Physical interpretation} \label{sec:interpretation}

In the previous section, we have observed different dynamical behaviors, depending on whether the initial state is the ground state, the scar state, or a thermal state. In this section, we explain the physics behind these differences and in particular show why it is much easier to manipulate scar states adiabatically than thermal states. We start with a brief summary of standard adiabatic perturbation theory and then discuss each of the three cases in turn in this context.

\subsubsection{Brief summary of adiabatic perturbation theory} \label{sec:interpretation_theory}

Adiabatic perturbation theory \cite{grandi2010} provides us with an approximate expression for the time evolved state $|\psi(t)\rangle$, which applies when the dynamics stays close to adiabaticity, i.e.\ when the adiabatic fidelity remains close to unity. The steps of the derivation are as follows (see \cite{grandi2010} for further details).

The time evolved state \eqref{eq:TSE} is written in the instantaneous eigenbasis,
\begin{equation}
|\psi(t)\rangle = \sum_{m=0}^{\mathcal{D}_{\pm}-1} c_m[s(t)]e^{-i\int_0^tE^{\pm}_m[s(t')]dt'}|\Phi^{\pm}_m[s(t)]\rangle.
\end{equation}
This expression is inserted into the time-dependent Schr{\"o}dinger equation
\begin{equation}
i|\dot{\psi}(t)\rangle=H^{\pm}[s(t)]|\psi(t)\rangle,
\end{equation}
where the dot means derivative with respect to time. After multiplying from the left by $\langle\Phi_n^{\pm}[s(t)]|$ and rewriting, one obtains
\begin{multline}
\dot{c}_n[s(t)]=-\sum_{m=0}^{\mathcal{D}_{\pm}-1}c_m[s(t)] \langle\Phi^{\pm}_n[s(t)]|\dot{\Phi}^{\pm}_m[s(t)]\rangle\\
\times e^{i\int_0^t \{E^{\pm}_n[s(t')]-E^{\pm}_m[s(t')]\}dt'}.
\end{multline}
Inserting the adiabatic approximation $c_m[s(t)]\approx\delta_{m0}$ on the right hand side of the equation and integrating both sides with respect to time leads to
\begin{multline}\label{eq:cms}
c_n(s)\approx
-\int_0^s \langle\Phi^{\pm}_n(s')|\partial_{s'}|\Phi_0(s')\rangle\\
\times e^{i\int_0^{s'} \dot{s}^{-1} [E^{\pm}_n(s'')-E_0(s'')]ds''}ds', \quad n\neq 0.
\end{multline}
For the linear ramp considered here, $\dot{s}=v=1/T$.

From this, it becomes clear that there are two main quantities determining the leakage to state $n$, namely the matrix element of the adiabatic gauge potential \cite{Kolodrubetz:2017}
\begin{equation} \label{eq:AGP}
\mathcal{A}^{\pm}_{n0}(s)=i\langle\Phi^{\pm}_n(s)|\partial_{s}|\Phi_0(s)\rangle.
\end{equation}
and the instantaneous energy difference $E_n^{\pm}(s)-E_0(s)$ divided by the ramp speed $v$. If the ratio between the energy gap and the ramp speed is large, the phase factor oscillates rapidly, which suppresses leakage. This is the source of the commonly quoted rule of thumb that one should move slowly compared to the energy gap to obtain adiabatic evolution. Another way to suppress leakage is by ensuring that the matrix element of the adiabatic gauge potential is low.

\subsubsection{Ground state} \label{sec:interpretation_GS}

We now apply these general considerations to our system. We start by considering time evolution with $H^+(s)$ starting from the ground state $|\Phi_0(0)\rangle$. We observe numerically that the ground state and the first excited state of $H^+(s)$ are separated by a gap for the whole range of $s$. For small enough ramp speed, the phase factor in \eqref{eq:cms} hence oscillates fast for all $n$. In this case, one can use methods for integrals of fast oscillating functions to obtain \cite{grandi2010}
\begin{multline}\label{eq:cmssq}
|c_n[s(T)]|^2\approx v^2
\frac{|\mathcal{A}^{+}_{n0}(0)|^2}{|E^{+}_n(0)-E_0(0)|^2}\\
+v^2 \frac{|\mathcal{A}^{+}_{n0}(1)|^2}{|E^{+}_n(1)-E_0(1)|^2}
\end{multline}
where a fast oscillating term has been averaged out. Putting this into the definition
\begin{equation}\label{eq:defF}
F=1-\sum_{n(\neq0)}|c_n[s(T)]|^2
\end{equation}
for the fidelity and observing that
\begin{equation}
-\log_{10}(F)\approx \sum_{n(\neq0)}|c_n[s(T)]|^2
\end{equation}
when the fidelity is close to unity, we arrive at the well-known result that $-\log_{10}(F)$ scales with $v^2$ for gapped systems undergoing adiabatic time evolution with a linear ramp \cite{grandi2010}. This is consistent with the numerical results in Fig.\ \ref{fig:Fv_AKLT_δs}(b).

We can also combine \eqref{eq:cmssq} with \eqref{eq:defF} and solve for $v$ to obtain an approximation for the adiabatic velocity $v_{0.99}$. The results, which are shown in Fig.\ \ref{fig:HAC_Dyn}, fit quite well with the adiabatic velocity obtained from dynamics. For $N=16$, e.g., we find $v_{0.99}\approx2.56\times 10^{-3}$ from \eqref{eq:cmssq} and $v_{0.99}\approx 2.68 \times 10^{-3}$ from the dynamics. For $N\leq 14$, we use full exact diagonalization to obtain the results from \eqref{eq:cmssq}. For $N=16$, the Hilbert space dimension $\mathcal{D}_{+}=163181$ is too large for full diagonalization, and we instead use the Lanczos algorithm to find the $2000$ lowest eigenstates of $H^+(s)$.

\subsubsection{Scar state} \label{sec:interpretation_scar}

We next consider time evolution with $H^-(s)$ starting from the scar state $|\Phi_0(0)\rangle$. For this system, we cannot proceed in the same way as for the ground state, since for some values of $s$, the scar state has the same energy as one of the thermal states, and hence the exponential in \eqref{eq:cms} is not necessarily oscillating fast for all $n$. We can hence not use \eqref{eq:cmssq}, but need to go back to \eqref{eq:cms}. This also tells us that a different mechanism is responsible for the ability to do adiabatic time evolution with the scar state, and we identify this mechanism in this section.

\begin{figure}
\includegraphics[width=1\linewidth]{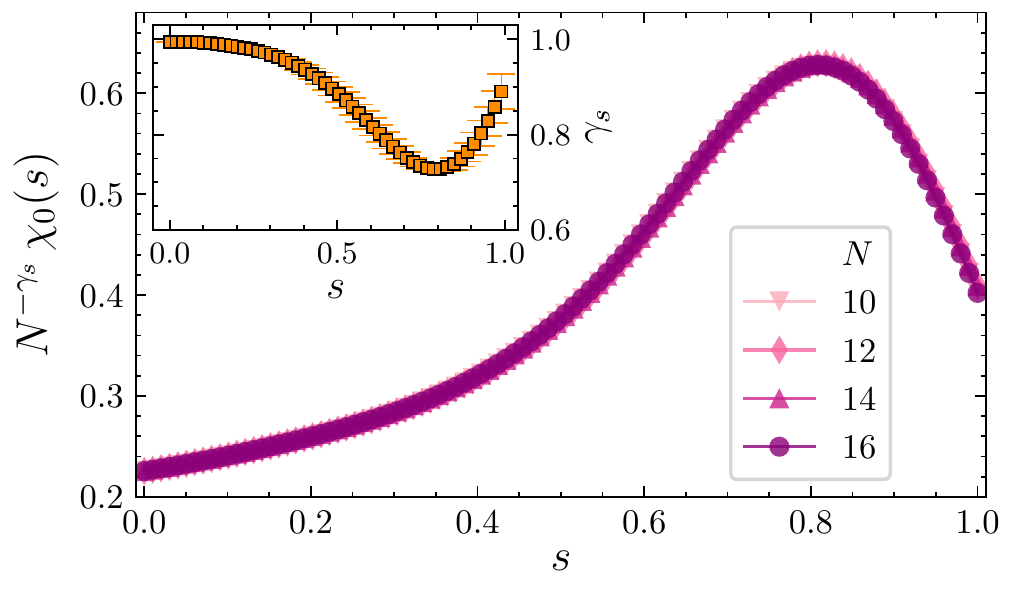}
\caption{\textbf{Smoothness of the fidelity susceptibility.} The scaled fidelity susceptibility of the state $|\Phi_0(s)\rangle$, embedded either as the ground state of $H^{+}(s)$ or the scar state of $H^{-}(s)$, varies smoothly with $s$ and exhibits a power law growth $\propto N^{\gamma_s}$ with system size similar to that of integrable systems. Note that the results shown in this figure are the same for the ground state and the scar state, as $\chi_0(s)$ is determined from $|\Phi_0(s)\rangle$ alone according to Eq.\ \eqref{eq:chi0s}. The inset displays the nonuniversal exponent $\gamma_s$ for which the fidelity susceptibilities for different system sizes collapse onto a single curve.}
\label{fig:chi_s_AKLT}
\end{figure}

We first take a closer look at $\mathcal{A}^{-}_{n0}(s)$. For each value of $s$, we are free to choose the global phase of $|\Phi^{-}_n(s)\rangle$, and it is hence possible to ensure that $\mathcal{A}^{-}_{n0}(s)$ is purely imaginary for all $s$. It is therefore sufficient to consider $|\mathcal{A}^{-}_{n0}(s)|$. Equation \eqref{eq:cms} only applies if $c_n(s)$ remains low for all $n\neq0$. As a consistency check, we therefore first show that $|\mathcal{A}^{-}_{n0}(s)|$ is finite for all $s$ despite the crossings between the scar state and thermal states seen in Fig.\ \ref{fig:AKLT_static}(a). If $|\mathcal{A}^{-}_{n0}(s)|$ diverges for some $n$ and $s$, the fidelity susceptibility \cite{you2007,Zanardi:2007}
\begin{multline}\label{eq:chi0s}
\chi_0(s)=\sum_{n(\neq 0)}|\mathcal{A}^{-}_{n0}(s)|^2=\sum_{n(\neq 0)}|\langle\Phi^{-}_n(s)|\partial_s|\Phi_0(s)\rangle|^2 \\
=\langle\Phi_0(s)|\partial_s^2|\Phi_0(s)\rangle-\langle\Phi_0(s)|\partial_s|\Phi_0(s)\rangle^2
\end{multline}
will also diverge, and it is hence sufficient to show that the latter does not diverge. The fidelity susceptibility measures the response of $| \Phi^{-}_{n}(s) \rangle$ to the variations of $H^{-}(s)$ and is a practical tool in the study of quantum chaos \cite{pandey2020,LeBlond:2021} and the stability of exact quantum many-body scars of the PXP model against static perturbations \cite{surace2021}. The absence of divergences in the fidelity susceptibility is seen in Fig.\ \ref{fig:chi_s_AKLT}. Note also that due to the rewriting in \eqref{eq:chi0s}, the fidelity susceptibility is determined from the scar state alone. This means that the fidelity susceptibility is the same for the ground state and the scar state. Figure \ref{fig:chi_s_AKLT} also shows that $\chi_0(s)$ scales as $N^{\gamma_s}$ with system size, where $\gamma_s$ is an $s$-dependent nonuniversal exponent which varies between $0.7$ and $1.0$ for the considered model. Such power law scaling of the fidelity susceptibility is characteristic of either (interacting) integrable or noninteracting (single-particle) models \cite{pandey2020,LeBlond:2021} which is now manifested in the scar state of a chaotic many-body Hamiltonian.

\begin{figure}
	\includegraphics[width=1\linewidth]{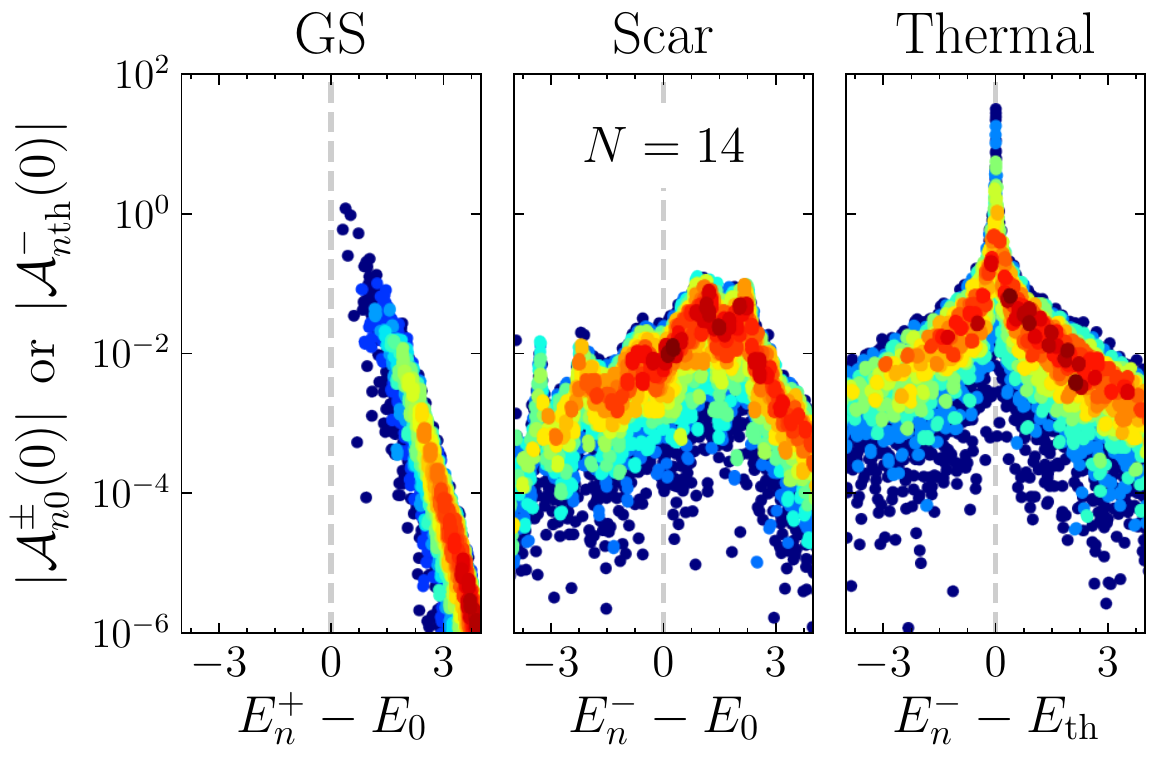}
	\caption{\textbf{Matrix elements of the adiabatic gauge potential.} This plot, and corresponding plots for other values of $s$ that look similar, provide the information needed to estimate the leakage from Eq.\ \eqref{eq:cms}. We observe that the ground state is well suited for adiabatic dynamics because only a few states that are separated from the ground state by an energy gap have a non-negligible value of $|\mathcal{A}^+_{n0}(s)|$. The scar state is well suited for adiabatic dynamics despite the absence of a protecting energy gap because $|\mathcal{A}^-_{n0}(s)|$ is generally low. The thermal state is unsuitable for adiabatic dynamics because $|\mathcal{A}^-_{n\mathrm{th}}(s)|$ has a large peak at zero energy difference. We here use $\textrm{th}$ to label the thermal state. The vertical dashed line marks the point of zero energy difference, and the color indicates the density of data points, with red being high density and blue being low density.}
	\label{fig:sus_mat}
\end{figure}

The growth of $c_n(s)$ is determined by both $|\mathcal{A}^{-}_{n0}(s)|$ and $E^{-}_n(s)-E_0(s)$, and the growth is fastest when $|\mathcal{A}^{-}_{n0}(s)|$ is large and $|E^{-}_n(s)-E_0(s)|$ is small. We hence plot the former as a function of the latter in Fig.\ \ref{fig:sus_mat}. We also show results for the ground state and thermal state dynamics for comparison. The figure is for $s=0$, but qualitatively similar results are obtained for other values of $s$. The plots are zoomed around the interesting region, as $|\mathcal{A}^{\pm}_{n0}(s)|$ generally drops exponentially fast with increasing energy difference $|E^{\pm}_n(s)-E_0(s)|$. The figure provides the key for understanding the differences in dynamics. The ground state is suitable for adiabatic time evolution, because there is a gap between the ground state and the first excited state. For the scar state, there is no such gap. Instead, adiabatic time evolution is possible because $|\mathcal{A}^{-}_{n0}(s)|$ is small for all $n$. This is the mathematical version of the statement that the nature of the scar state is so different from that of the thermal states that it is difficult to transition between the two. Finally, the thermal state is unsuitable for adiabatic dynamics, because $|\mathcal{A}^{-}_{n\textrm{th}}(s)|$ has a high peak at zero energy difference whose magnitude grows exponentially with system size, as given by Eq.\ \eqref{eq:A_th}.

As $|\mathcal{A}^{-}_{n0}(s)|$ does not have large variations in the region where $|E^{-}_n(s)-E_0(s)|$ is small, we expect the transition rate to be primarily determined by how fast the exponential in \eqref{eq:cms} oscillates. This means that transitions are expected to happen primarily for $E^{-}_n(s)-E_0(s)\approx0$. Looking at Fig.\ \ref{fig:AKLT_static}(a), we observe that the thermal states in the spectrum tend to cross the scar state in downward direction as $s$ increases. As they cross the scar state, they gain a little population, and they then continue downwards in energy (see Fig.\ \ref{fig:leakage_AKLT}(a)). We hence expect that at the end of the ramp, it is primarily states just below the scar state that are populated, and this is indeed observed in Fig.\ \ref{fig:overlap_AKLT}(b) for low $v$.

Combining this qualitative description with Eq.\ \eqref{eq:cms} enables us to obtain an approximation for the final population in the $n$th eigenstate for small $v$. Let $s_n$ be the value of $s$ for which the $n$th eigenstate crosses the scar state, i.e.\ $E^{-}_n(s_n)=E_0(s_n)=0$. We reduce the limits of the integral over $s'$ in \eqref{eq:cms} to go from $s_n-\delta$ to $s_n+\delta$ for some small $\delta$, as we assume that the contributions to $|c_n[s(T)]|$ outside this region are negligible because the energy difference between the thermal state and the scar state is too large. We then Taylor expand
\begin{align}
&\mathcal{A}^{-}_{n0}(s)\approx \mathcal{A}^{-}_{n0}(s_n)
+\left.\frac{d\mathcal{A}^{-}_{n0}(s)}{ds}\right|_{s=s_n} (s-s_n),\\
&E^{-}_n(s) - E_0(s) \approx \alpha_{n} (s - s_{n}),\quad \alpha_n\equiv \left.\frac{dE_n^{-}(s)}{ds}\right|_{s=s_n}, \nonumber
\end{align}
to first order for $s\in[s_n-\delta,s_n+\delta]$. Further observing that
\begin{multline}
e^{\frac{i}{v}\int_0^{s'} [E^{-}_n(s'') - E_0(s'')]ds''} \approx\\
e^{\frac{i}{v}\int_0^{s_n-\delta} [E^{-}_n(s'') - E_0(s'')]ds''
+\frac{i\alpha_{n}}{v}\int_{s_n-\delta}^{s'}  (s'' - s_{n})ds''}\\
=e^{i\phi}e^{\frac{i\alpha_{n}}{2v}(s' - s_{n})^2},
\end{multline}
where $e^{i\phi}$ is a phase factor that does not depend on $s'$, reduces \eqref{eq:cms} to
\begin{multline}
|c_{n}[s(T)]| \approx \left|\int_{s_n-\delta}^{s_n+\delta} \mathcal{A}^{-}_{n0}(s') e^{\frac{i\alpha_{n}}{2v}(s' - s_{n})^2} ds'\right|\\
\approx |\mathcal{A}^{-}_{n0}(s_n)| \left|\int_{s_n-\delta}^{s_n+\delta} e^{\frac{i\alpha_{n}}{2v}(s' - s_{n})^2} ds'\right|\\
=\sqrt{\frac{2v}{|\alpha_n|}}|\mathcal{A}^{-}_{n0}(s_n)| \left|\int_{-\delta\sqrt{\frac{\alpha_n}{2v}}}^{\delta\sqrt{\frac{\alpha_n}{2v}}} e^{i\tilde{s}^2} d\tilde{s}\right|.
\end{multline}
When $v$ is small compared to $\delta^2\alpha_n/2$, we can approximate the limits with minus and plus infinity and evaluate the integral using the Fresnel integrals \cite{abramowitz1988}. This leads to
\begin{equation}\label{eq:cnapprox}
|c_n[s(T)]|^2\approx \dfrac{2\pi |\mathcal{A}^{-}_{n0}(s_n)|^2}{|\alpha_n|}\,v.
\end{equation}
This explains the linear scaling of the logarithmic fidelity with $v$ for small ramp speeds observed in Fig.\ \ref{fig:Fv_AKLT_δs}(b). The above arguments apply under quite mild assumptions, suggesting that the linear scaling is universal and independent of ramp protocol for sufficiently small $v$. If a thermal state crosses the scar state several times, the different contributions to $c_n$ will have phase factors that depend on $v$, but the norm of each contribution to $c_n$ will again be proportional to $\sqrt{v}$.

We compare \eqref{eq:cnapprox} to numerical data for fixed $v=10^{-3}$, $N=14$, and state number $n=4180$ as a typical example of a thermal state crossing the scar state (at least) once during the evolution. The prediction provides the right order of magnitude for the population in the considered state, but the dynamics show several fluctuations that are not captured by the approximation. The numerics suggests that the fluctuations happen because of transfer of population between thermal states of similar energies. Such transfer does not change the total amount of leakage out of the scar state, and we hence expect that \eqref{eq:cnapprox} applies in an average sense.

\subsubsection{Thermal state}

The energy levels of thermal states tend to repel one another, which leads to Gaussian statistics rather than Poisson statistics of the energy spacings. The level repulsion also prevents levels from crossing each other as $s$ is varied. This means that Eq.\  \eqref{eq:cmssq} should also apply for thermal states for low enough $v$, with the ground state replaced by the thermal state, but since the energy difference to the neighboring states is much smaller for the thermal state than for the ground state, $v$ should be much smaller. Very small $v$ is, however, difficult to handle numerically, because it means time evolution for very long times. Here, we therefore look for scaling behaviors of the adiabatic velocity rather than obtaining it from numerical simulations.

We first determine the scaling behavior of the relevant matrix elements of the adiabatic gauge potential. Considering two instantaneous eigenstates $|\Phi^{-}_m(s)\rangle$ and $|\Phi^{-}_n(s)\rangle$ with different energies, $E^{-}_m(s)\neq E^{-}_n(s)$, one can derive the Feynman-Hellmann equation
\begin{equation}\label{eq:FeynmanHellman}
	\langle \Phi^{-}_m (s) | \partial_s | \Phi^{-}_n (s) \rangle = \frac{\langle\Phi^{-}_m (s) |\partial_{s}H^{-}(s)| \Phi^{-}_n (s) \rangle}{E^{-}_n(s)-E^{-}_m(s)}
\end{equation}
from the chain rule and the observations
\begin{eqnarray}
&&\partial_s[\langle \Phi^{-}_m (s) | H^{-}(s) | \Phi^{-}_n (s) \rangle]=0, \quad m\neq n,\\
&&\partial_s[\langle \Phi^{-}_m (s) | \Phi^{-}_n (s) \rangle]=0, \quad m\neq n,
\end{eqnarray}
which apply because the eigenstates of the Hamiltonian at a given instant are orthogonal. Here, $|E^{-}_m(s)-E^{-}_n(s)|$ is the energy difference between the two considered states, while $|\langle \Phi^{-}_m (s) | \partial_sH^{-}(s) | \Phi^{-}_n (s) \rangle|$ is a measure for how strongly the system reacts to perturbations. The latter can be seen by writing $H^-(s+ds)\approx H^-(s)+\partial_s H^-(s) ds$ and noting that $\langle \Phi^{-}_m (s) | H^{-}(s) | \Phi^{-}_n (s) \rangle=0$.

For a generic thermal state satisfying the eigenstate thermalization hypothesis, the off-diagonal matrix elements of local operators, such as $\partial_sH^{-}(s)$, scale as $\mathcal{D}_{-}^{-1/2} $ for eigenstates at similar energies. The typical separation between adjacent energy levels in a thermal spectrum of a local Hamiltonian scales as $ \sim \sqrt{N}/\mathcal{D}_{-}$ \cite{Alessio:2014_rev}. Dropping factors that are only a power of $N$, this simple argument implies
\begin{equation}\label{eq:A_th}
|\mathcal{A}^{-}_{mn}(s)|= |\langle \Phi^{-}_m (s) | \partial_s | \Phi^{-}_n (s) \rangle| \sim \mathcal{D}_{-}^{1/2} \sim e^{N\ln(3)/2}
\end{equation}
if $|\Phi^{-}_m(s)\rangle$ and $|\Phi^{-}_n(s)\rangle$ are thermal states with similar energies. In other words, the peak in the panel for thermal state dynamics in Fig.\ \ref{fig:sus_mat} diverges exponentially with system size.

We obtain an expression for the adiabatic velocity $v_{0.99}$ by combining Eqs.\ \eqref{eq:cmssq} with the ground state replaced by the thermal state and \eqref{eq:defF} with $F=0.99$. For ramp speeds that are small enough to ensure adiabatic time evolution even for thermal states, population is only transferred to a small number of states. We therefore approximate the sum in \eqref{eq:defF} by only the largest term. After dropping factors that are only a power of $N$, this leads to the scaling
\begin{equation} \label{eq:vad_th}
v_{0.99} \sim \frac{|E^{-}_m(s)-E^{-}_n(s)|}{|\mathcal{A}^-_{mn}(s)|}
\sim\mathcal{D}_{-}^{-3/2}\sim e^{-3N\ln(3)/2}.
\end{equation}
The most important conclusion from this is that the adiabatic velocity decreases exponentially with system size for thermal states, which means that thermal states of many-body systems are in practice not suitable for adiabatic dynamics.

\subsection{Comparison to the quantum speed limit} \label{sec:AKLT_QSL}

\begin{figure}
  \includegraphics[width=1\linewidth]{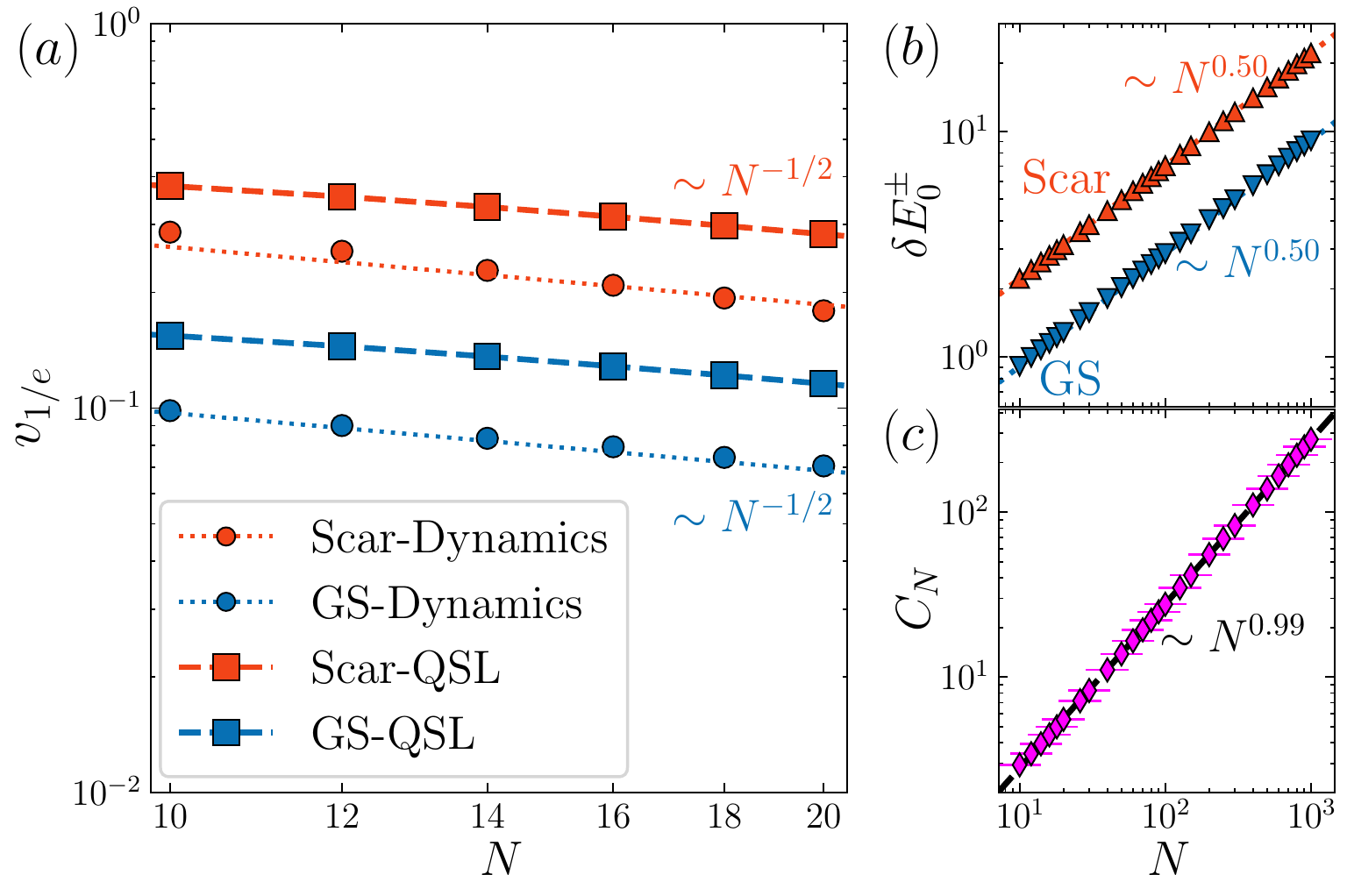}
  \caption{\textbf{Scaling of adiabatic velocity with system size.}  (a) Comparison between the adiabatic velocity $v_{1/e}$ obtained from the ground state and scar state dynamics (circles), where $v_{1/e}$ is the ramp speed at which the adiabatic fidelity \eqref{eq:ad_fid} drops below $e^{-1}$, and the upper bound $v_{1/e}^{\textrm{QSL}}$ in \eqref{eq:AC_QSL} (squares) predicted by the quantum speed limit (labeled as QSL). All four curves scale as $N^{-0.50}$ with system size. (b) $\delta E^{\pm}_0$ scales as $N^{0.50}$ for both the ground state and the scar state. (c) The catastrophe exponent $C_N$ scales approximately as $N$.}
\label{fig:QSL_Dyn}
\end{figure}

We now turn our attention to the quantum speed limit, which is a bound on the maximum speed at which a system can evolve from an initial state to an orthogonal state. The condition establishes a connection between the timescale for breakdown of adiabaticity in a many-body system and the intrinsic orthogonality catastrophe through the energy-time uncertainty relation. The necessary condition \cite{Lychkovskiy:2017} reads
\begin{equation} \label{eq:AC_QSL}
v\lesssim v_{1/e}^{\textrm{QSL}} \equiv \frac{\delta E^{\pm}_0}{2 C_N},
\end{equation}
where $C_N$ is the catastrophe exponent introduced in Sec.\ \ref{subsubsec:CE}, and
\begin{equation}\label{eq:delta_E}
\delta E^{\pm}_0 = \sqrt{\langle \psi_0| (V^{\pm})^2 | \psi_0 \rangle - \langle \psi_0| V^{\pm} | \psi_0 \rangle^2}
\end{equation}
is the initial quantum uncertainty of the generalized force operator at the beginning of the ramp, $V^{\pm}\equiv\partial_s H^{\pm}(s)|_{s=0}$. This criterion does not rely on a gap condition and is valid for a wide class of time-dependent, many-body Hamiltonians provided
\begin{equation}\label{eq:AC_QSL_cond}
   \frac{\delta E^{\pm}_0}{C_N}\to 0 \quad\textrm{for}\quad N\to\infty
\end{equation}
holds and the adiabatic fidelity \eqref{eq:ad_fid} monotonically decreases with time \cite{Lychkovskiy:2017}. The condition \eqref{eq:AC_QSL} uses a reference fidelity of $e^{-1}$, and the resulting upper bound for the adiabatic velocity $v_{1/e}^{\textrm{QSL}}$ should hence be compared to $v_{1/e}$ obtained from the dynamics.

Now consider the evolution starting from the AKLT state $|\psi_0\rangle\equiv|\Phi_0(0)\rangle$ embedded as the ground state or scar state of $H^{\pm}(0)$. As $V^{\pm}$ is a sum of $N$ local terms, one can write $(\delta E^{\pm}_0)^2$ as a sum of $N^2$ terms. Since the translationally invariant MPS $|\Phi_0(0)\rangle$ does not have long-range connected correlations, however, only of order $N$ of these terms are nonzero. Furthermore, only the part of the Hamiltonian that is proportional to $\mathcal{J}_0^{\pm}$ contributes to $V^{\pm}$. Putting these observations together, we conclude that the quantum uncertainty has the asymptotic scaling $\delta E^{\pm}_0\sim \mathcal{J}_0^{\pm} \sqrt{N}$ for large $N$. Recalling that the catastrophe exponent scales as $C_N\sim N$ (see Fig.\ \ref{fig:AKLT_static}(c)), we conclude that the criterion \eqref{eq:AC_QSL_cond} is met for both the ground and scar state dynamics. Hence, according to the necessary condition \eqref{eq:AC_QSL}, the bound on the ramp speed scales as $v_{1/e}^{\textrm{QSL}} \sim \mathcal{J}^{\pm}_0 N^{-1/2}$. The above arguments do not depend on the details of the model, and we hence expect the power law scaling with system size to apply for a wide class of time-dependent scarring Hamiltonians.

The scalings are numerically confirmed for the MPS model in Fig.\ \ref{fig:QSL_Dyn} by utilizing the MPS form of the state $|\Phi_0(s)\rangle$ for large system sizes (up to $N=10^3$). As shown by a log-log plot in Fig.\ \ref{fig:QSL_Dyn}(a), the adiabatic velocity $v_{1/e}$  is upper bounded by $v_{1/e}^{\textrm{QSL}}$ for both ground state and scar state dynamics and displays the same $N^{-1/2}$ power-law scaling as $v_{1/e}^{\textrm{QSL}}$, for numerically accessible system sizes. Besides, $v_{1/e}$ is seen to be slightly larger for the scar state dynamics than for the ground state dynamics. This difference can be traced back to the choice of the parameters $\mathcal{J}_m$ in the Hamiltonian. The important one is $\mathcal{J}_0$, as $\mathcal{J}_0$ is the coefficient multiplying the terms in the Hamiltonian that depend on $s$. Specifically, for $N=20$, we find
\begin{equation}\label{eq:vad_ratio}
\frac{v^{+}_{1/e}}{v^{-}_{1/e}}\approx 0.395 \approx \frac{\mathcal{J}^{+}_0}{\mathcal{J}^{-}_0}\approx0.408,
\end{equation}
where we use a subscript $\pm$ to explicitly distinguish between the ground and scar state dynamics. We also have $v_{1/e}^{\textrm{QSL},+}/v_{1/e}^{\textrm{QSL},-}=\mathcal{J}^{+}_0/\mathcal{J}^{-}_0$, because $\partial_sH^+(s)/\mathcal{J}_0^+=\partial_sH^-(s)/\mathcal{J}_0^-$.

We also note that the power law scaling of the adiabatic velocity with system size observed in the scar states differs from the case of nonthermal, disorder-driven, localized states (either Anderson localized or many-body localized systems) for which the adiabatic velocity decreases exponentially with system size \cite{Altshuler:2010,Khemani:2015}.

\section{Effect of perturbations}\label{sec:KAM}

As it is difficult to exactly realize a particular Hamiltonian in an experiment, we next investigate the effect of perturbations on the model discussed in the previous section. We first note that the physics observed in the previous section is generally robust for perturbations $H'$ that can be written in the form
\begin{equation}\label{eq:robustper}
H'=\sum_ih'_i(s), \quad h'_i(s)=\sum_{n=-2}^{2}\sum_{m=-2}^{2} P_{n,i} H_{nm,i} P_{m,i},
\end{equation}
where $P_{0,i}=|K_0(s)\rangle\langle K_0(s)|$, $P_{m,i}=|\mathcal{J}_{2,m}\rangle\langle \mathcal{J}_{2,m}|$ for $m=\pm1,\pm2$ [see Eqs.\ \eqref{eq:Jm} and \eqref{eq:K0}], and the operators $H_{nm,i}$ satisfy $H_{mn,i}=H_{nm,i}^\dag$ to ensure Hermiticity. This is so, because the scar state is an exact eigenstate of \eqref{eq:robustper}, and the model hence remains a scar model as long as we do not hit an integrable point or perturb the model so strongly that the scar state is no longer in the middle of the spectrum. As an example, for $s=0$, this means that perturbing the model with Hermitian operators that act only within the spin-2 subspace of two coupled, neighboring spin-1 in the chain, will still lead to a model with similar properties.

Let us next consider a perturbation that is not of the form \eqref{eq:robustper}. Specifically, we consider the Hamiltonian
\begin{equation}\label{eq:Hp_clean}
	H_\epsilon^{\pm}(s)= H^{\pm}(s) -\epsilon \sum_{i=1}^N S^z_i S^z_{i+1},
\end{equation}
where $\epsilon>0$ is the strength of the perturbation. We have chosen this perturbation, because it is local and respects all the symmetries of the model, which simplifies the numerical computations. The MPS \eqref{eq:MPS_manifold} is not an exact eigenstate of \eqref{eq:Hp_clean}.

We diagonalize the Hamiltonian numerically,
\begin{equation}
H_\epsilon^{\pm}(s)|\Phi^{\pm,\epsilon}_n(s)\rangle=E^{\pm,\epsilon}_{n}(s)|\Phi^{\pm,\epsilon}_n(s)\rangle,
\end{equation}
in the symmetry sector $(S^z_{\textrm{tot}},k,\mathcal{I},\mathcal{Z})=(0,0,+1,+1)$ to obtain the instantaneous eigenstates $|\Phi^{\pm,\epsilon}_n(s)\rangle$ and their energies $E^{\pm,\epsilon}_{n}(s)$. Here, we label the eigenstate, which has the highest fidelity with the MPS \eqref{eq:MPS_manifold}, by $n=0$. In the following, we shall refer to this state as the perturbed scar state or the perturbed ground state. The other eigenstates in the considered symmetry sector are labeled by $n\in\{1,2,\ldots,\mathcal{D}_{\pm}-1\}$ in order of increasing energy.

\begin{figure}
\includegraphics[width=1\linewidth]{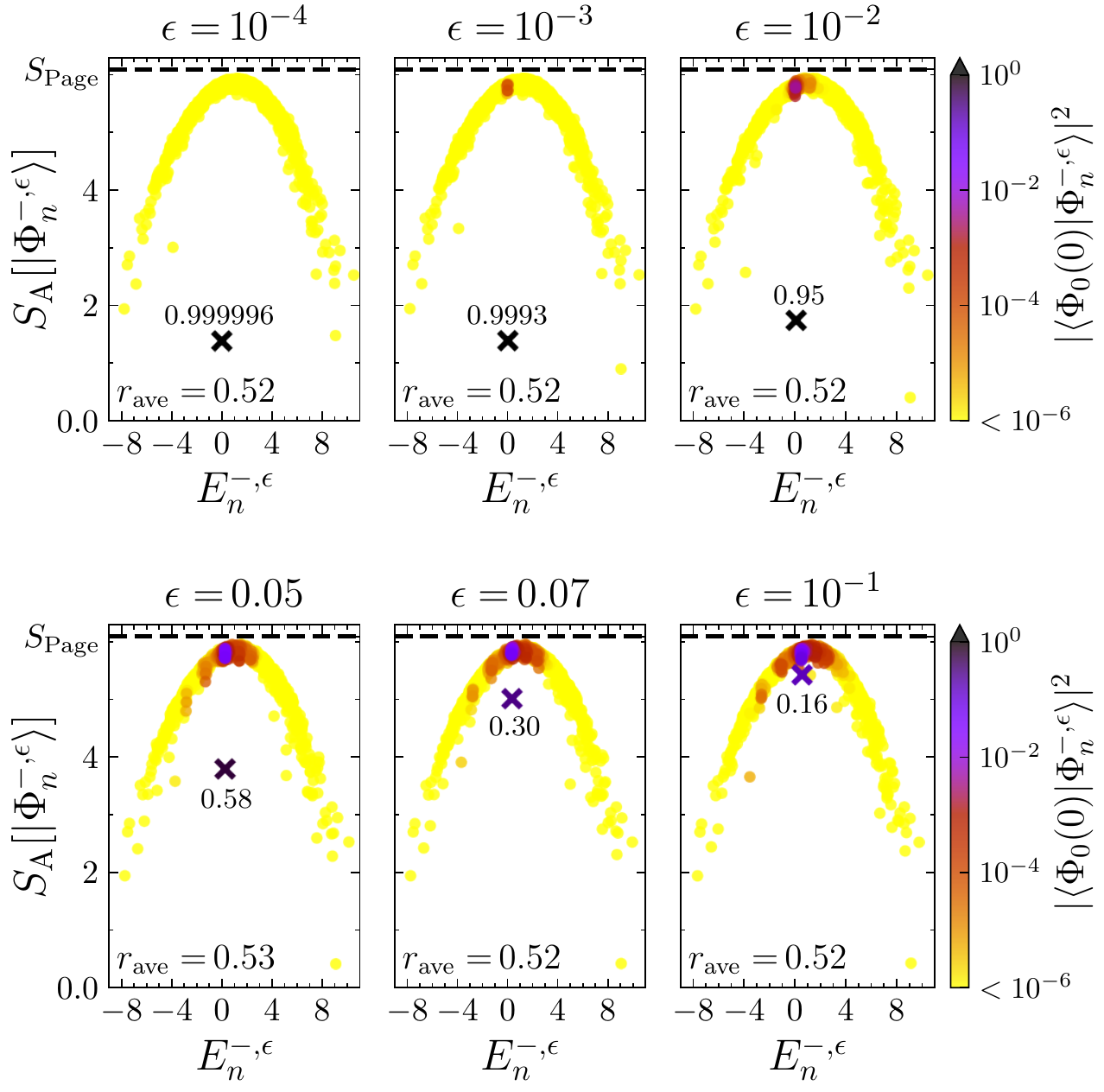}
\caption{\textbf{Scar state model under perturbation.} Half-chain von Neumann entanglement entropy as a function of energy for the eigenstates of the perturbed Hamiltonian $H_{\epsilon}^-(0)$ [see \eqref{eq:Hp_clean}] with $N=12$ in the symmetry sector $(S^z_{\textrm{tot}},k,\mathcal{I},\mathcal{Z})=(0,0,+1,+1)$. The states are colored according to their fidelities with the unperturbed scar state at $s=0$, i.e.\ the AKLT state $|\Phi_0(0)\rangle$. The state with the highest fidelity, which we call the perturbed scar state, is marked with a cross, and the fidelity is written next to the state. The averaged level spacing ratio $r_{\textrm{ave}}$, which is given in each panel, provides further evidence that the spectra are thermal also after adding the perturbation as expected.}
\label{fig:Sent_Hp}
\end{figure}

Figure \ref{fig:Sent_Hp} shows the entropy versus energy for the instantaneous eigenstates of $H_\epsilon^{-}(0)$ for $N=12$ sites. For $\epsilon\lesssim 0.01$, we observe that there is a state in the middle of the spectrum with a low entanglement entropy, which has a high fidelity with the MPS \eqref{eq:MPS_manifold}. The model hence remains a scar model for moderate $\epsilon$, while for larger $\epsilon$ the scar state disappears.

\begin{figure*}
\includegraphics[width=1\linewidth]{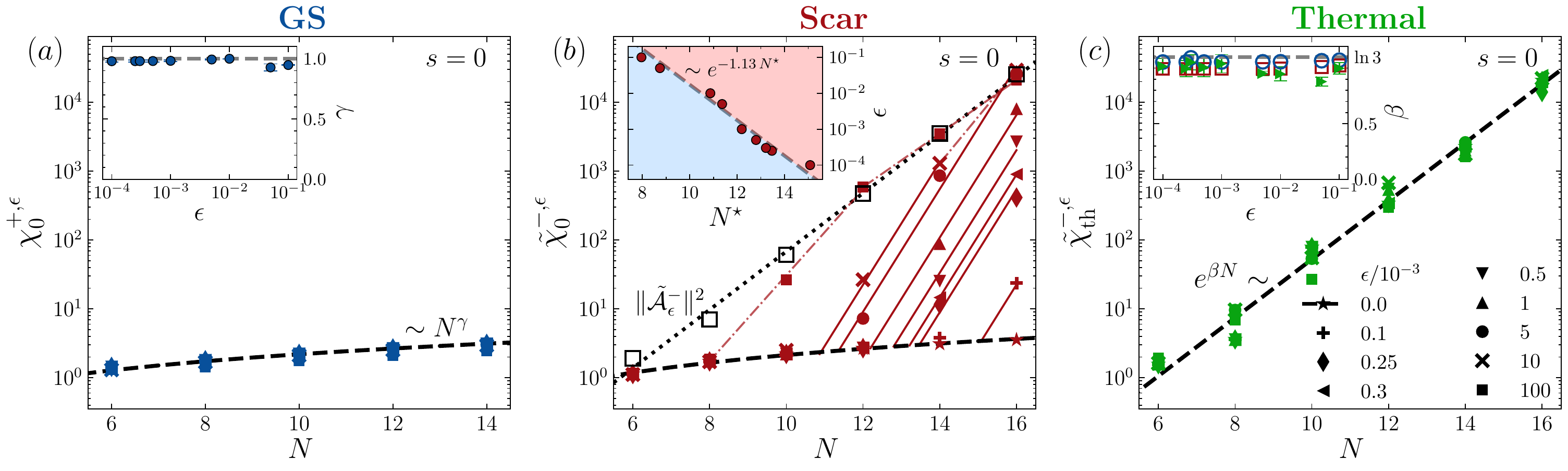}
\caption{\textbf{Scaling of the (regularized) fidelity susceptibility with system size}. (a) The fidelity susceptibility of the ground state of $H_{\epsilon}^+(0)$ scales as a power law $N^\gamma$ with system size for both the unperturbed and perturbed cases. The different marker types correspond to different perturbation strengths $\epsilon$ as listed in panel (c), and the dashed line is the fit for $\epsilon=0$. The inset in panel (a) shows that the nonuniversal exponent $\gamma$ extracted from the fits remains close to one across the considered range of $\epsilon$. (b) The regularized fidelity susceptibility for the perturbed scar state of $H_{\epsilon}^-(0)$ as a function of system size $N$ for different perturbation strengths $\epsilon$. The scaling of the regularized fidelity susceptibility with system size changes from power law to exponential at $N\approx N^\star$, and the relation between $\epsilon$ and $N^\star$ is shown in the inset. The value of $N^\star$ has been extracted as the point, where the power law fits (shown as a dashed line for $\epsilon=0$) and the exponential fits (solid lines) cross. Since there are three data points for the exponential fit for $\epsilon=0.005$, and only one or two data points for the other values of $\epsilon$, we have fixed the slope of the solid lines based on the fit at $\epsilon=0.005$ and only kept the constant term as a free parameter in the other fits. The dash-dotted lines are a guide to the eye. The open squares show the adiabatic gauge norm at $\epsilon=0.1$, and the dotted line is an exponential fit $e^{0.97N}$ to these data points. (c) The regularized fidelity susceptibility of the thermal state closest in energy to the perturbed scar state grows exponentially with system size as expected from Eq.\ \eqref{eq:chithscale}. The inset shows the coefficient $\beta$, where the green markers are for $\tilde{\chi}_{\textrm{th}}^{-,\epsilon}$, while the red squares are for $\lVert \tilde{\mathcal{A}}^{-}_\epsilon(s) \rVert^2\sim e^{\beta N}$, and the blue circles are for $\lVert \tilde{\mathcal{A}}^{+}_\epsilon(s) \rVert^2\sim e^{\beta N}$.}
\label{fig:KAM_robustness}
\end{figure*}

As we have seen above, the adiabatic gauge potential is an important quantity in determining the feasibility of adiabatic time evolution. At a same time, it can serve as a highly sensitive indicator to sharply detect transitions from nonergodic to ergodic behavior at the level of individual eigenstates \cite{pandey2020,LeBlond:2021}. Since we compute $|\Phi^{\pm,\epsilon}_0(s)\rangle$ numerically, we shall here use Eq.\ \eqref{eq:FeynmanHellman} to express the adiabatic gauge potential. To resolve a numerical instability in the computation of this quantity, the so-called ``small denominator problem'' \cite{Kolodrubetz:2017}, we further introduce a regularization and write
\begin{eqnarray}\label{eq:reg_AGP}
\tilde{\mathcal{A}}^{\pm,\epsilon}_{nm}(s)=
\frac{\langle \Phi^{\pm,\epsilon}_n(s)  |[\partial_{s} H_\epsilon^{\pm}(s)] | \Phi^{\pm,\epsilon}_m(s)  \rangle}{\mu^{\pm}+i[E^{\pm,\epsilon}_{n}(s)-E^{\pm,\epsilon}_{m}(s)]}
\end{eqnarray}
such that
\begin{equation}
\mathcal{A}^{\pm,\epsilon}_{nm}(s)=\lim_{\mu^{\pm}\to0^+}\tilde{\mathcal{A}}^{\pm,\epsilon}_{nm}(s).
\end{equation}
In the numerical computations, following Ref.\ \cite{pandey2020}, we take $\mu^{\pm}$ to be of the order of $\sqrt{N}$ times the typical spacing between energy levels. Specifically, we put
\begin{equation}
\mu^\pm=N\mathcal{D}^{-1}_{\pm}.
\end{equation}
From the regularized adiabatic gauge potential, we compute the regularized fidelity susceptibility
\begin{equation}\label{eq:reg_chi0s}
\tilde{\chi}^{\pm,\epsilon}_m(s)=\sum_{n(\neq m)}|\tilde{\mathcal{A}}^{\pm,\epsilon}_{nm}(s)|^2
\end{equation}
and also the regularized adiabatic gauge norm,
\begin{equation}\label{eq:reg_AGPnorm}
\lVert \tilde{\mathcal{A}}^{\pm}_\epsilon(s) \rVert^2=\frac{1}{\mathcal{D}_{\pm}} \sum_n \tilde{\chi}^{\pm,\epsilon}_n(s)
\end{equation}
which is the average of the regularized fidelity susceptibility over the considered symmetry sector. When the spectrum is thermal as here, the regularized adiabatic gauge norm gives a typical value for the regularized fidelity susceptibility of a thermal state. Let us also note that the regularization is not needed for computing the fidelity susceptibility of the ground state in the ground state model due to the energy gap, and for this case we hence take $\mu^+=0$ and write the fidelity susceptibility without the tilde, i.e.\ as $\chi_0^{+,\epsilon}$.

For the unperturbed case, we found in Eq.\ \eqref{eq:chi0s} that the fidelity susceptibility of the state \eqref{eq:MPS_manifold} is the same for the ground state model and the scar state model, and Fig.\ \ref{fig:chi_s_AKLT} showed that the fidelity susceptibility scales approximately linearly with system size $N$ for $s=0$. In contrast, for thermal states, the fidelity susceptibility scales exponentially with system size, which can be seen by noting that the main contributions to the sum in \eqref{eq:reg_chi0s} are the terms for which the states $m$ and $n$ are close in energy, and for such terms \eqref{eq:A_th} apply. This gives
\begin{equation}\label{eq:chithscale}
\tilde{\chi}^{-,\epsilon}_m(s)\sim e^{N\ln(3)},
\end{equation}
which is unfavorable for adiabatic dynamics. Figure \ref{fig:KAM_robustness} shows how these results are modified, when the model is perturbed. The results for the ground state model are largely unchanged for small perturbations, which is expected due to the energy gap between the ground state and the first excited state, and the scaling of the fidelity susceptibility is still approximately linear in $N$ for $s=0$. For the thermal state, we also continue to see an exponential scaling in agreement with \eqref{eq:chithscale}.

The regularized fidelity susceptibility of the perturbed scar state in the perturbed scar model, which is shown in Fig.\ \ref{fig:KAM_robustness}(b), has a more complicated behavior. Based on Fig.\ \ref{fig:Sent_Hp}, one could expect that the regularized fidelity susceptibility of the perturbed scar state follows the behavior of the ground state model for small enough perturbations and follows the behavior of a thermal state for large enough perturbations. We hence draw a few reference lines in the figure. The dashed line shows a fit of the form $N^\gamma$ to the data for $\epsilon=0$. This line follows the corresponding line for the ground state in Fig.\ \ref{fig:KAM_robustness}(a), except for a small deviation due to the regularization. The dotted line is an exponential fit of the adiabatic gauge norm and gives the typical behavior for a thermal state in the scar model. For a given perturbation strength, the regularized fidelity susceptibility of the perturbed scar state displays a quite sharp change in behavior at a system size that we shall denote by $N^\star$. For $N<N^\star$, the regularized fidelity susceptibility follows the behavior of the ground state model, i.e.\ $\tilde{\chi}_0^{-,\epsilon}\sim N$, while for $N>N^\star$, it scales exponentially with system size, i.e.\ $\tilde{\chi}^{-,\epsilon}_0\sim e^{\beta N}$. For data points that are far from the dotted line, $\beta\approx2.05$, so the growth with system size is larger than for a thermal state in this region. For most of the considered perturbation strengths, it is not clear from the numerics what happens when the data points approach the dotted line, as the system sizes that can be considered are not large enough. For the largest $\epsilon$ considered, however, it is seen that $\beta$ decreases and approaches $\ln(3)$, such that the regularized fidelity susceptibility of the perturbed scar state follows the dotted line for large enough $N$, which means that it behaves like a thermal state. The inset of Fig.\ \ref{fig:KAM_robustness}(b) shows that $N^\star$ scales as $-\ln(\epsilon)$, or alternatively that $\epsilon$ scales as $e^{-1.13 N^\star}$. In summary, the conclusion from Fig.\ \ref{fig:KAM_robustness} is that the physics is similar to the unperturbed case, as long as $\epsilon$ is small enough that $N<N^\star$. For $N=12$, e.g., significant deviations start to happen for $\epsilon=0.005$. Referring to Fig.\ \ref{fig:Sent_Hp}, this is also the point, where the fidelity between the perturbed and the unperturbed scar state starts to deviate significantly from unity.

\begin{figure}
\includegraphics[width=1\linewidth]{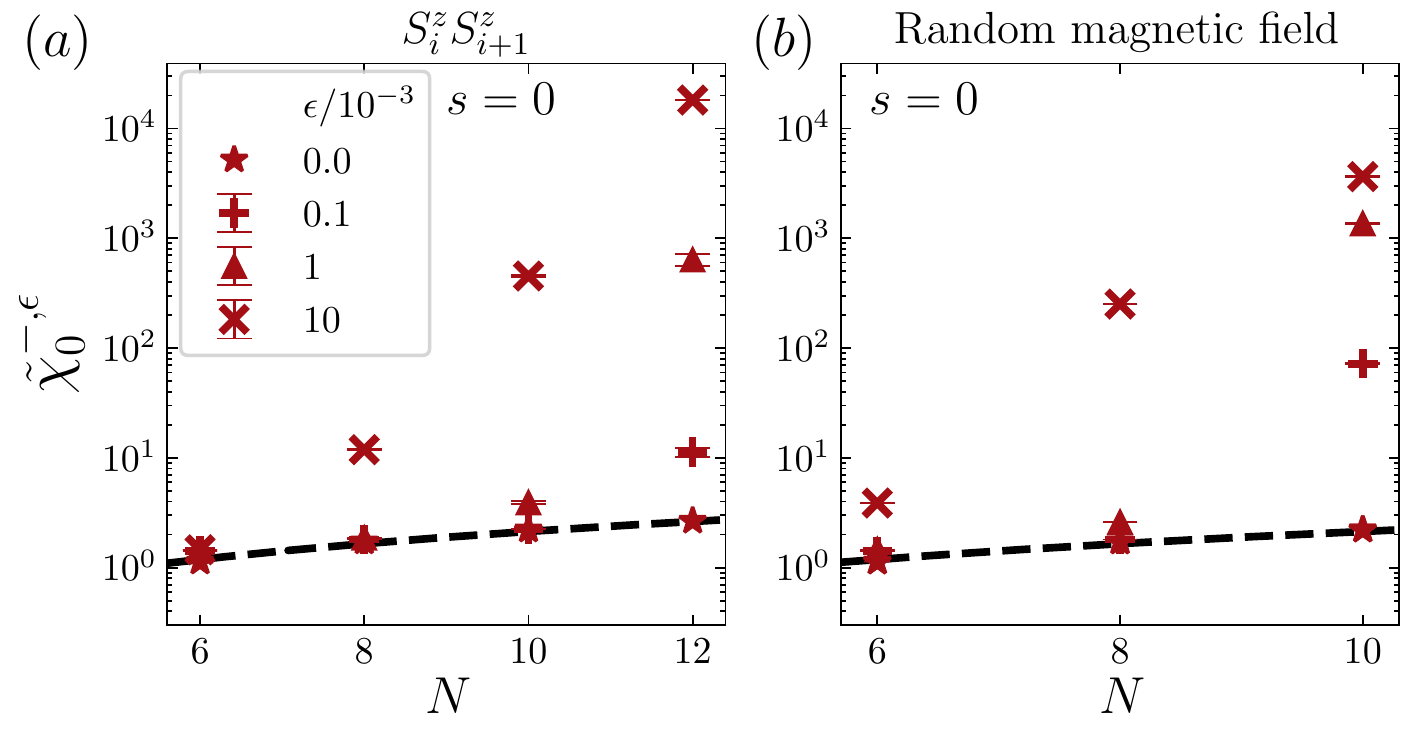}
\caption{\textbf{Regularized fidelity susceptibility for the models \eqref{eq:Hzzeps} and \eqref{eq:Hzeps}.} The regularized fidelity susceptibility for the perturbed scar state of the disordered Hamiltonians (a) $H_{zz,\epsilon}^{-}(0)$ and (b) $H_{zz,\epsilon}^{-}(0)$ as a function of system size at different disorder strengths $\epsilon$. The dashed lines represent power law fits to the unperturbed data at $\epsilon=0$. For both models, the change from power law to exponential scaling occurs at smaller system sizes than in Fig.\ \ref{fig:KAM_robustness}(b). The results for $\epsilon\neq0$ and $N=6,8,10,12$ are averaged over $10^5,10^4,10^3,50$ disorder realizations, respectively. The error bars represent the standard error of the mean.}
\label{fig:x_dis}
\end{figure}

Finally, we also consider perturbations that do not respect the symmetries of the scar model. Such perturbations can introduce an additional leakage source for the scar dynamics, through which thermal states that belong to different symmetry sectors can potentially mix with the scar state and have nonzero contribution on its fidelity susceptibility, and hence, its adiabatic dynamics. To see the impact of such a mixing, we consider the Hamiltonians
\begin{equation}\label{eq:Hzzeps}
H_{zz,\epsilon}^{-} (s)=H^{-}(s)+\sum_j\epsilon_jS_j^zS_{j+1}^z
\end{equation}
and
\begin{equation}\label{eq:Hzeps}
H_{z,\epsilon}^{-} (s)=H^{-}(s)+\sum_j\epsilon_jS_j^z
\end{equation}
where $\epsilon_j$ is a random variable drawn from the Gaussian distribution with zero mean and standard deviation $\epsilon$. $H_{zz,\epsilon}^{-}(s)$ is thus a disordered version of Eq.\ \eqref{eq:Hp_clean}, and the random magnetic field in $H_{z,\epsilon}^{-}(s)$ breaks all symmetries except the total magnetization. Figure \ \ref{fig:x_dis} shows $\tilde{\chi}_0^{-,\epsilon}$ as a function of system size for the models $H_{zz,\epsilon}^{-}(0)$ and $H_{zz,\epsilon}^{-}(0)$ at different disorder strengths in the range $0\leq\epsilon\leq 0.01$. We have checked that over this region, the energy spectrum remains thermal with $r_{\mathrm{ave}}$ around $0.52$ despite the presence of disorder. The change from power law scaling to exponential scaling is visible for both models. However, for a fixed $\epsilon$, $N^\star$ is smaller than observed for the clean, symmetry-preserving perturbation \eqref{eq:Hp_clean} in Fig.\ \ref{fig:KAM_robustness}(b) as expected due to the larger dimension of the relevant sector of the Hilbert space.

\section{Tower of scars}\label{sec:tower}

The model we have analyzed so far has a single scar state, and we now briefly discuss how the results generalize to the case of a tower of scar states. This in turn opens up the possibility for \textit{parallel} adiabatic manipulation of quantum states. Models with a tower of scar states can be systematically constructed using the MPS-based approach of Ref.\ \cite{Moudgalya:2020}. This is done by first forming a tower of exact eigenstates with bounded entanglement entropy based on the so-called MPS tangent space method \cite{Haegeman:2013}, and the tower is then embedded into the thermal spectrum of a parent Hamiltonian.

Starting from an exact uniform MPS as a root state, the tower of exact eigenstates is constructed by successive application of a quasiparticle creation operator $Q^\dagger_k=\sum_j e^{ik j} \hat{q}_j$. The quasiparticle creation operator produces a superposition with momentum $k$ of perturbations generated by the action of a local operator $\hat{q}_j$ centered around site $j$. Here, we take the exact MPS $|\Phi_0(s)\rangle$ as the root state and consider the quasiparticle creation operator
\begin{equation}
Q^\dagger_\pi=\sum_j (-1)^j (S^{+}_j)^2.
\end{equation}
This gives the tower
\begin{equation}\label{eq:gen_AKLT_tower}
	|\Phi_\ell(s)\rangle= \mu_{N,\ell} (Q_\pi^\dagger)^\ell|\Phi_0(s)\rangle,
\end{equation}
where $\ell=0,1,\cdots,N/2$ and $\mu_{N,\ell}$ is a normalization factor. The state $|\Phi_\ell(s)\rangle$ describes $\ell$ noninteracting quasiparticles with momentum $\pi$ on top of the exact MPS background $|\Phi_0(s)\rangle$, parameterized by the ramp parameter $s$. At $s=0$, the set of $\{|\Phi_\ell(0)\rangle \}$ reproduces the well-known AKLT tower of states \cite{moudgalya2018a}, which starts from the AKLT state $|\Phi_0(0)\rangle$ at $\ell=0$ and ends on the (ferromagnetic) product state $|+\cdots+\rangle$ at $\ell=N/2$ when $N/2$ is even (note $|\Phi_{N/2}(s)\rangle=0$ for $N/2$ odd). Given the fact that the operator $(Q_\pi^\dagger)^\ell$ has an exact matrix product operator representation (with virtual bond dimension $\chi_Q = \ell+1$ \cite{moudgalya2018b}), the states $|\Phi_\ell(s)\rangle$ have the structure of a matrix product operator acting on an MPS. They can hence be expressed as an MPS with subvolume law entanglement scaling  $S_A \leq \ln (\chi \chi_Q)$, where $\chi$ is the virtual bond dimension of the MPS and $\ln (\chi \chi_Q)$ is of order $\ln(N)$ when $\ell$ is of order $N$ and $N\gg 1$ \cite{moudgalya2018b}.

Following the MPS construction of Ref.\ \cite{Moudgalya:2020}, the parent Hamiltonian with the tower of states $\{|\Phi_\ell(s)\rangle\}$ is composed of local terms with the form
\begin{equation}\label{eq:hi_AKLTtower}
	h'_{j}(s) = \frac{\omega_0}{2} \sum_{m =+1}^{+2} |K_{m}\rangle\langle K_{m}|
	+ \sum_{m,m'=-2}^0 \mathcal{J}_{mm'} |K_{m}\rangle\langle K_{m'}|,
\end{equation}
where $|K_{m(\neq 0)}\rangle\equiv|J_{2,m}\rangle$ denotes the angular momentum eigenstates defined in Eq.\ \eqref{eq:Jm}, and $|K_{0}\rangle\equiv|K_{0}(s)\rangle$ represents the basis state given by Eq.\ \eqref{eq:K0}, whose dependence on time is through the ramp parameter $s(t)$. Here $\mathcal{J}_{mm'}={\mathcal{J}}^\star_{m'm}\in \mathbb{C}$ are the elements of a Hermitian coefficient matrix and $\omega_0$ determines the energy spacing between the consecutive scar states in the tower.

\begin{figure}
	\includegraphics[width=1\linewidth]{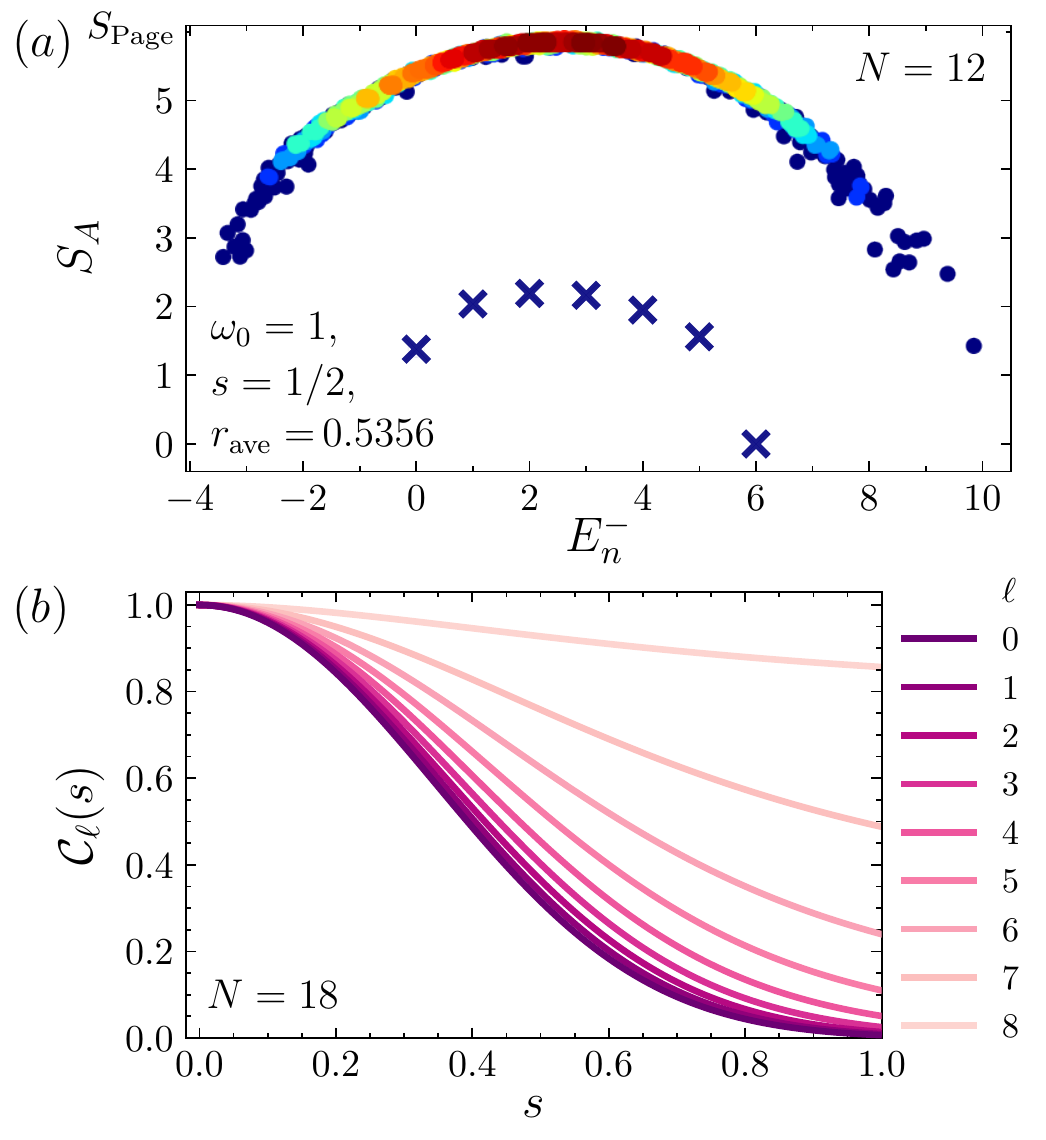}
	\caption{\textbf{Tower of scars.} (a) Half-chain entanglement entropy for the eigenstates of the Hamiltonian \eqref{eq:Htower_AKLT} with $N=12$, $\omega_0=1$, and $s=1/2$ in the sector $(S^z_{\textrm{tot}},k,\mathcal{I})=(0,0,+1)$. The states are colored according to the density of data points with red (blue) being high (low) density. The corresponding mean level spacing ratio $r_{\mathrm{ave}}$ follows the GOE value. The crosses denote an equally-spaced tower of exact scars which appear at energies $E^{-}_\ell = \ell \omega_0 $ with $\ell=0,1,\ldots, N/2$ and symmetry sectors $(S^z_{\textrm{tot}},e^{ik},\mathcal{I})=(2\ell,(-1)^\ell,(-1)^\ell)$. (b) Instantaneous fidelity  $\mathcal{C}_\ell(s)=|\langle \Phi_\ell(0)|\Phi_\ell(s)\rangle|^2$ for the $\ell$th scar state as a function of the ramp parameter $s$ at fixed system size $N=18$.}
	\label{fig:ScarTower}
\end{figure}

In the following, we will specifically consider the Hamiltonian
\begin{equation}\label{eq:Htower_AKLT}
	H_{\textrm{ST}}^{-}(s)= \sum_j h'_j(s),\quad \mathcal{J}^-_{mm'}=(-1)^m \delta_{m,m'},
\end{equation}
where we set the coefficients $\mathcal{J}^-_{mm'}$ to be diagonal and site-independent, following the same reasoning that leads to Eq.\ \eqref{eq:Hm}. The scar tower in the spectrum of the Hamiltonian \eqref{eq:Htower_AKLT} is seen in Fig.\ \ref{fig:ScarTower}(a), where we plot the entanglement entropy of the eigenstates within the sector $(S^z_{\textrm{tot}},k,\mathcal{I})=(0,0,+1)$ at typical values of $\omega_0=1$, $s=1/2$, and $N=12$ as well as the entanglement entropy of the scar states. The states of the tower $\{|\Phi_\ell(s)\rangle \}$ are exact eigenstates of $H_{\textrm{ST}}^{-}(s)$, which appear as outliers at energies $E^{-}_\ell = \ell \omega_0 $. They belong to \textit{different} symmetry sectors $(S^z_{\textrm{tot}},e^{ik},\mathcal{I})=(2\ell,(-1)^\ell,(-1)^\ell)$ due to the action of the creation operator $(Q_\pi^\dagger)^\ell$. In Fig.\ \ref{fig:ScarTower}(b), we evaluate the instantaneous fidelity of the $\ell$th scar state $\mathcal{C}_\ell(s)=|\langle \Phi_\ell(0)|\Phi_\ell(s)\rangle|^2$, measured with respect to the corresponding state in the AKLT tower $\{|\Phi_\ell(0)\rangle \}$. The result reveals that the structure of the scar states changes continuously as the ramp parameter $s$ is varied. These results imply that the considered model \eqref{eq:Htower_AKLT} fulfils the essential ingredients needed for investigating adiabatic dynamics of scar states given in Sec.\ \ref{sec:scar_MPS}.

Let us first consider the time evolution starting from one of the scar states in the AKLT tower, i.e.\ $|\psi_0\rangle \in \{|\Phi_\ell(0)\rangle\}$. Owing to the block diagonal structure of $H_{\textrm{ST}}^{-}(s)$, the evolution of $|\psi_0\rangle$ is restricted to a single symmetry sector of the Hamiltonian. In other words, the initially prepared scar state cannot evolve into scar or thermal states in different symmetry sectors. Note also that since all the scar states belong to different symmetry sectors, we have $\mathcal{A}^-_{\ell' \ell}(s)= i\langle \Phi_{\ell'}(s)|\partial_{s}|\Phi_\ell(s)\rangle = 0$ for all $s$ and $\ell\neq\ell'$. The problem hence reduces to the type of problem studied in Sec.\ \ref{sec:scar_MPS}, and the primary mechanism for population transfer is hence the leakage from the initial scar state to nearby thermal states within the same symmetry sector (followed by transition between nearby thermal states).

\begin{figure}
	\includegraphics[width=1\linewidth]{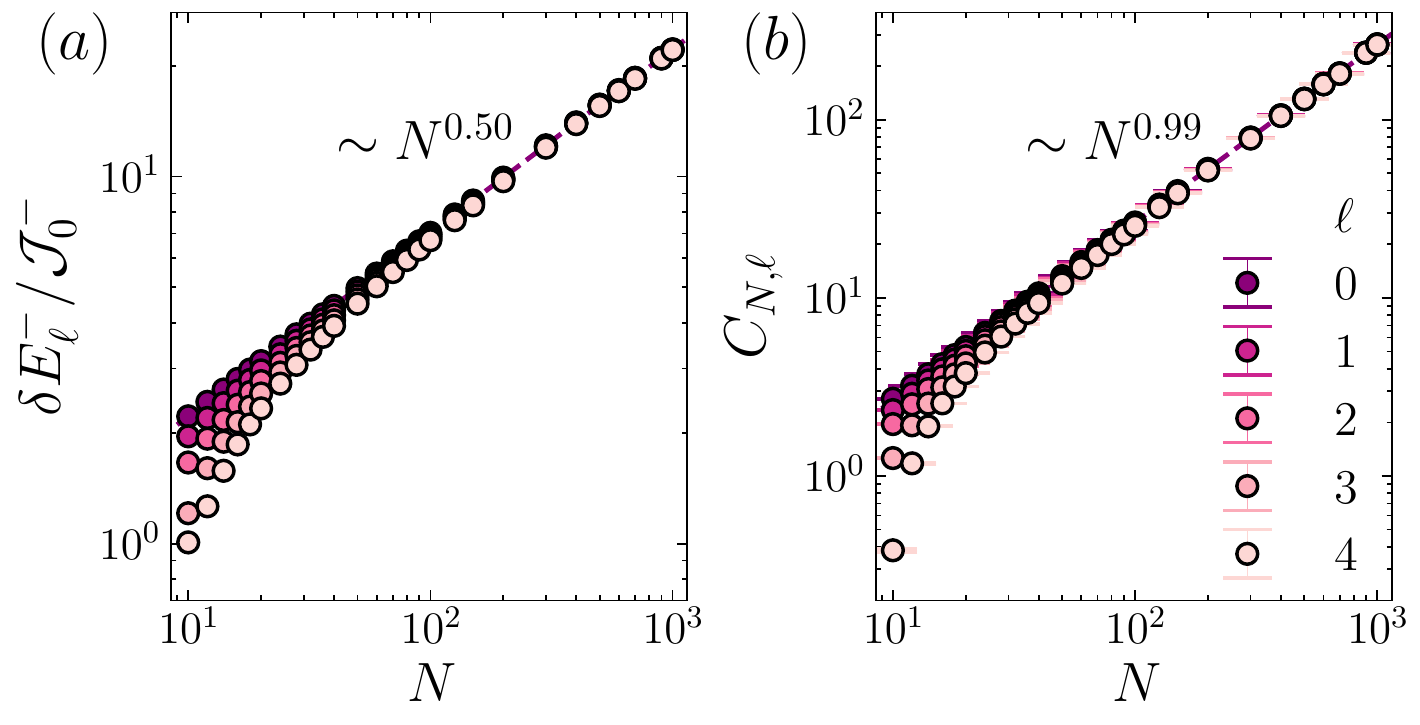}
    \caption{\textbf{Power law scaling of the quantum speed limit.} (a) The system-size dependence of the quantum uncertainty of the generalized force operator $\delta E_\ell^{-}$ and (b) the catastrophe exponent $C_{N,\ell}$ associated to the $\ell$th scar state in the tower display the same asymptotic behaviors observed for the single embedded scar state in Fig.\ \ref{fig:QSL_Dyn}(b,c). This signifies the validity of the condition \eqref{eq:AC_QSL_cond}, which in turn puts the upper bound $v_{1/e}^{\textrm{QSL}} \sim \mathcal{J}^{-}_0 N^{-1/2}$ on the adiabatic velocity for the evolution starting from each individual scar state of the AKLT tower.}
	\label{fig:VQSL_tower}
\end{figure}

When the initial state is $|\psi_0\rangle\in \{|\Phi_\ell(s)\rangle\}$ and the Hamiltonian is $H_{\textrm{ST}}^{-}(s)$, we find numerically in Fig.\ \ref{fig:VQSL_tower} that the quantum uncertainty $\delta E^{-}_\ell$ given by Eq.\ \eqref{eq:delta_E} scales as $\delta E^{-}_\ell\sim \mathcal{J}_0^{-} \sqrt{N}$ for large $N$ and that the catastrophe exponent $C_{N,\ell}$ extracted by fitting the instantaneous fidelity of the first five scar states in the tower to the function $-\ln [\mathcal{C}_\ell(s)] \approx C_{N,\ell} s^2$ scales as $C_{N,\ell} \sim N$ for large $N$. These asymptotic scaling behaviors are similar to those previously observed in Fig.\ \ref{fig:QSL_Dyn}(b,c) for the model with a single embedded scar state and are rooted in the locality of the Hamiltonian $H_{\textrm{ST}}^{-}(s)$ and the fact that all the scar eigenstates $\{|\Phi_\ell(s)\rangle\}$ have a natural MPS representation with finite correlation length (see the discussion below Eq.\ \eqref{eq:AC_QSL_cond}). The scaling results signify the validity of the condition \eqref{eq:AC_QSL_cond} for the evolution starting from an individual scar state of the AKLT tower, i.e., $\delta E^{-}_\ell/C_{N,\ell} \to 0$ as $N\to \infty$, irrespective of $\ell$. Consequently, the criterion \eqref{eq:AC_QSL} remains applicable, which in turn leads to the power law scaling of the adiabatic velocity with the upper bound $v_{1/e}^{\textrm{QSL}} \sim \mathcal{J}^{-}_0 N^{-1/2}$.

One can also do the adiabatic dynamics starting from an initial state that is a superposition of the scar states in the tower, i.e.\
\begin{equation}
|\psi_0\rangle=\sum_\ell \alpha_{N,\ell} |\Phi_\ell(0) \rangle \nonumber
\end{equation}
for some choice of the coefficients $\alpha_{N,\ell}$. Each scar state $|\Phi_\ell(0) \rangle$, appearing in this superposition, evolves independently under the relevant symmetry sector of the Hamiltonian $H_{\textrm{ST}}^{-}[s(t)]$ without leaking into other scar or thermal states of different sectors. As argued before, the dynamics of each part remains adiabatic provided $v\lesssim v_{1/e}^{\textrm{QSL}}$. In this vein, the evolution of $|\psi_0 (\xi)\rangle$ splits into a set of dynamically disconnected, adiabatic processes, each of which starts from  $|\Phi_\ell(0) \rangle$ and in the adiabatic limit reaches $|\Phi_\ell(s(T)) \rangle$ at the end of the ramp. During this \textit{parallel} adiabatic evolution, the state of the system evolves from $|\psi_0\rangle$ to
\begin{equation}
|\psi_T\rangle=\sum_\ell \alpha_{N,\ell} e^{-i\int_0^T E^-_\ell dt} |\Phi_\ell[s(T)] \rangle \nonumber
\end{equation}
and cannot be moved out of the scarred subspace spanned by the instantaneous tower of eigenstates, $\{|\Phi_\ell(s) \rangle\}$.

A superposition of MPSs is again an MPS with a virtual bond dimension that is (at most) the sum of the virtual bond dimensions \cite{Schollwock:2011}, and as mentioned above, MPSs can be generated efficiently with a quantum circuit \cite{Schon2005,Malz2023}. A particularly interesting superposition is the family of states (parameterized by a static variable $\xi\in \mathbb{C}$) given by \cite{Mark:2020},
\begin{equation}
	|\psi_0 (\xi)\rangle = e^{\xi Q_\pi^\dagger}\, |\Phi_0(0) \rangle = \prod_{i=1}^N \left[1+ \xi (-1)^i (S^{+}_i)^2 \right] |\Phi_0(0)\rangle,
\end{equation}
that lies entirely within the scarred subspace of $H_{\textrm{ST}}^{-}(0)$. This family has an MPS structure with $\chi = 4$ and $\chi=8$ for open and periodic boundary conditions, respectively, and thus exhibit area law entanglement \cite{Mark:2020}. It has been shown recently that such superpositions of scar eigenstates can be prepared with high fidelity in quantum processing units \cite{Gustafson:2023}. In time-independent scarring models, a global quench from this kind of initial states gives rise to anomalously long-lived, periodic revivals that reveal their particular importance.

It is worth noting that a more complex situation can arise in the presence of scar-preserving perturbations that break the block diagonal structure of the parent Hamiltonian (namely, by considering the terms $|K_{m}\rangle\langle K_{m'}|$ with $m\neq m'$ in Eq.\ \eqref{eq:hi_AKLTtower}), or when all scar states in the tower belong to the same symmetry sector. In such cases, population can transfer between different scar states of the same tower, which introduces an additional timescale for the leakage, $\tau_{\ell\ell'}\sim {|\mathcal{A}^-_{\ell\ell'}(s)|}/{|(\ell-\ell')\omega_0|}$.

\section{Scar model with fractional quantum Hall anyons}
\label{sec:FQHscar}

Another important application of adiabatic time evolution is transformations that demonstrate topological properties of quantum states, e.g.\ braiding of anyons. We therefore next construct a quantum scar model, in which the scar state is a lattice Laughlin state with two quasiholes and show that the quasiholes can be moved adiabatically.

\begin{figure}
\includegraphics[width=0.8\linewidth]{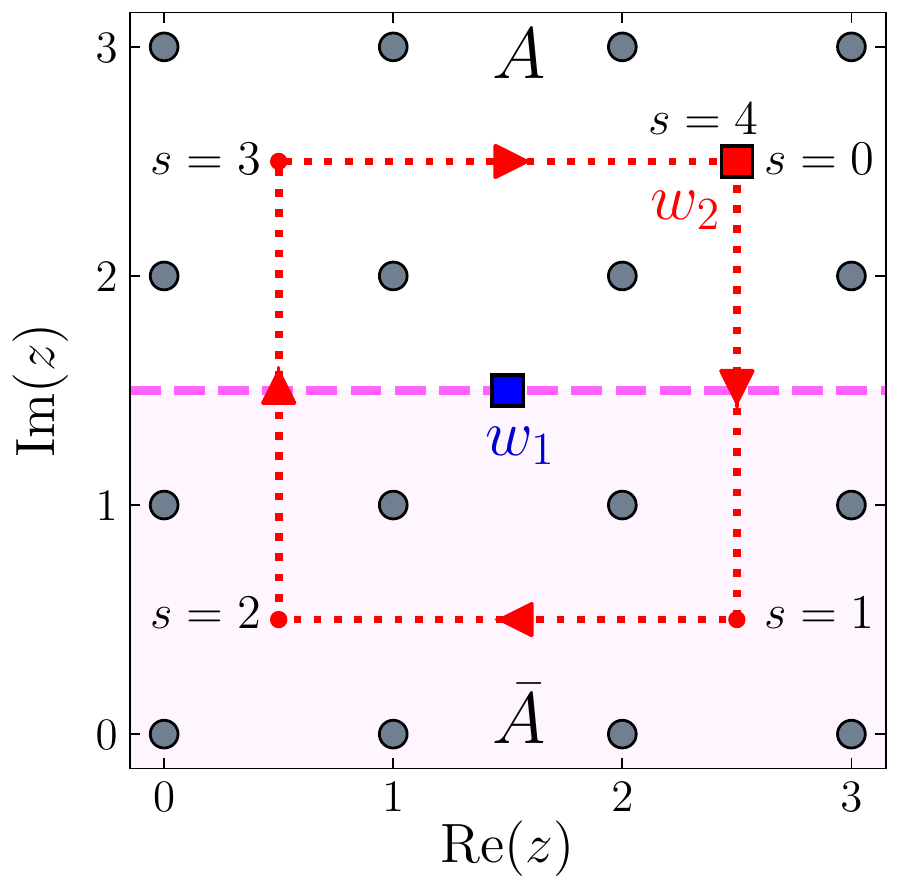}
\caption{Geometry of the system for the case of a $4 \times 4$ square lattice. We fix the coordinate $w_{1}$ to be at the center of the lattice, whereas the coordinate $w_{2}$ moves along the square path shown with red, dotted lines. The latter is parametrized by $s \in [0,4]$, in such a way that the real and imaginary parts of $w_2$ change linearly with $s$. The dashed line divides the system into two parts, $A$ and $\bar{A}$, with respect to which the entanglement entropy is computed. For the lattices with size $3\times4$, $5\times4$, and $3\times6$, $w_1$ is again at the center of the lattice, and $w_2$ moves around a rectangle of width one lattice spacing and height two lattice spacings and center at $w_1$.} \label{fig:lattice-geometry}
\end{figure}

\subsection{Model}

The starting point for constructing the scar model is to define the scar state. Here, we take the scar state to be a bosonic lattice Laughlin state at half filling with two quasiholes. Specifically, we consider $N$ sites at the fixed positions $z_j$ in the complex plane, where $N$ is even, and $N/2-1$ hardcore bosons distributed on these sites. The basis states in the Hilbert space are hence $|n_1,n_2,\ldots,n_N\rangle$, where $n_j\in\{0,1\}$ is the number of particle on the site at $z_j$ and $\sum_j n_j=N/2-1$. The scar state also depends on two additional coordinates, $w_1$ and $w_2$, that can be anywhere in the complex plane. The anyons in this state are extended objects that form on the lattice sites that are in the vicinity of $w_1$ and $w_2$. The coordinates $w_1$ and $w_2$ constitute the parameters that we vary during the adiabatic time evolution, i.e.\ $w_1=w_1(s)$ and $w_2=w_2(s)$.

The Laughlin state with anyons can be formulated in terms of conformal blocks in rational conformal field theories \cite{moore1991}, and starting from the conformal blocks there is a natural way to convert the Laughlin state with anyons to a state defined on a lattice \cite{nielsen2012,nielsen2015}. In short, the difference is that both the particles and the background charge are restricted to the sites at $z_j$ rather than to a disk-shaped region in the complex plane. This leads to the following expression \cite{nielsen2012,nielsen2015}
\begin{align}\label{toppsi}
    |\Phi_0(s) \rangle &= C_s\sum_{n_1, \ldots,n_N}
    (-1)^{\sum_{j=1}^{N}(j-1)n_{j}}
    \delta_{n} \\ \nonumber
    &\times \prod_{j = 1}^{N} \left\{ [z_{j} - w_{1}(s)] [z_{j} - w_{2}(s)] \right\}^{n_{j}} \\ \nonumber
    &\times \prod_{j<k}^{N} (z_{j} - z_{k})^{2n_{j}n_{k} - n_{j} - n_{k}} | n_{1}, \ldots, n_{N} \rangle
\end{align}
for the Laughlin state at half filling with two quasiholes. Here $C_s$ is a normalization constant and
\begin{equation}
    \delta_{n} = \left\{
    \begin{array}{cl}
    1, & \text{for }\sum_{j=1}^{N} n_{j} = N/2-1\\
    0, & \text{otherwise}
    \end{array}\right.
\end{equation}
fixes the number of particles.

Having defined the scar state, the next step is to identify operators that annihilate the scar state. Such operators have been derived in \cite{tu2014a,nielsen2015}, and they take the form
\begin{align}
    \Lambda_{j} &= \sum_{k(\neq j)}^{N} \frac{1}{z_{j} - z_{k}} \left[ d_{k} - d_{j}(2n_{k} - 1)\right] - \sum_{l=1}^{2} \frac{1}{z_{j} - w_{l}(s)}d_{j},\nonumber\\
    \Gamma_{j} &= \sum_{k(\neq j)}^{N} \frac{1}{z_{j} - z_{k}} d_{j}d_{k},\label{eq:LambdaGamma}
\end{align}
where $d_{j}$ is the annihilation operator of a hardcore boson at site $j$ and $n_{j} = d_{j}^{\dagger}d_{j}$, i.e.\
\begin{equation}\label{eq:CFTannihilate}
\Lambda_j|\Phi_0(s)\rangle=\Gamma_j|\Phi_0(s)\rangle=0, \quad \forall j.
\end{equation}
From these operators one can construct the Hamiltonian
\begin{equation}
\label{topHam}
    H_{\beta}(s) = \sum_{j=1}^{N} \Lambda_{j}^{\dagger} \Lambda_{j} + \beta \sum_{j=1}^{N} \Gamma_{j}^{\dagger} \Gamma_{j}, \quad \beta\in\mathbb{R}.
\end{equation}
It follows from \eqref{eq:CFTannihilate} that the state $|\Phi_0(s)\rangle$ is an exact eigenstate with zero energy. We utilize the parameter $\beta$ to adjust the position of this eigenstate in the spectrum, i.e.,\ how many states have lower energy than that of $|\Phi_0(s)\rangle$. If we choose $\beta\geq 0$, the state \eqref{toppsi} is the ground state of the model for all $s$, while for appropriately chosen negative $\beta$ it is a scar state in the middle of the spectrum. A similar construction was used in \cite{Srivatsa2020} to construct a topological scar model, but for the case of a Laughlin state without anyons.

Inserting \eqref{eq:LambdaGamma} into \eqref{topHam}, we obtain
\begin{widetext}
\begin{equation}\label{topHamexp}
    H_{\beta}(s) = \sum_{i \neq j}^{N} F^{A}_{ij}(s) d_{i}^{\dagger}d_{j} + \sum_{i \neq j}^{N} F^{B}_{ij}(s) n_{i}n_{j}
    + \sum_{i \neq j \neq k}^{N} F^{C}_{ijk} d_{i}^{\dagger}d_{j}n_{k}
    + \sum_{i \neq j \neq k}^{N} F^{D}_{ijk} n_{i}n_{j}n_{k}
    + \sum_{i = 1}^{N} F^{E}_{i} (s) n_{i}
\end{equation}
after some algebra, where
\begin{align}
    &F^{A}_{ij}(s) = 2 | c_{ij} |^{2} + \sum_{k(\neq i, \neq j)}^{N} \left( c_{ik}^{*}c_{ij} + c_{ji}^{*}c_{jk} + c_{ki}^{*}c_{kj} \right) - \sum_{l=1}^{2} \left( c_{ij}\Tilde{c}_{il}^{*} + c_{ji}^{*}\Tilde{c}_{jl} \right), \quad c_{jk} \equiv \frac{1}{z_{j} - z_{k}}, \quad \Tilde{c}_{jl} \equiv \frac{1}{z_{j} - w_{l}(s)}, \nonumber\\
    &F^{B}_{ij}(s) = \beta | c_{ij} |^{2} - 2\sum_{k(\neq i, \neq j)}^{N} \left( c_{ij}^{*}c_{ik} + \textrm{c.c.} \right) + 2\sum_{l=1}^{2} \left( c_{ij}\Tilde{c}_{il}^{*} + \textrm{c.c.} \right), \qquad
    F^{C}_{ijk} = -2 \left( c_{ik}^{*}c_{ij} + c_{ji}^{*}c_{jk} \right) + \beta c_{ki}^{*}c_{kj},\nonumber\\
    &F^{D}_{ijk} = 4 c_{ij}^{*}c_{ik},\qquad
    F^{E}_{i}(s) = \sum_{j(\neq i)}^{N} | c_{ij} |^{2} + \sum_{j(\neq i), k(\neq i)}^{N} c_{ij}^{*}c_{ik} - \sum_{j(\neq i)}^{N} \sum_{l=1}^{2} \left( c_{ij}\Tilde{c}_{il}^{*} + \textrm{c.c.} \right) + \sum_{l,m=1}^{2} \Tilde{c}_{il}^{*} \Tilde{c}_{im},
\end{align}
\end{widetext}
and c.c.\ denotes the complex conjugate. It is seen that the Hamiltonian conserves the number of particles, is nonlocal, and includes interaction terms of up to three particles.

The state \eqref{toppsi} is an exact eigenstate of the Hamiltonian \eqref{topHamexp} for general choices of the positions $z_i$ of the lattice sites in the complex plane. In the following, we shall consider a square lattice with $L_x\times L_y$ sites. We fix $w_1$ at the center of the lattice and move $w_2$ along the edges of a rectangle. The center of the rectangle coincides with $w_1$. For $L_x$ even (odd) the width of the rectangle is two (one) lattice spacings, and for $L_y$ even (odd) the height of the rectangle is two (one) lattice spacings. The movement is parametrized by $s \equiv s(t)$, which changes linearly from $0$ to $1$ when $w_2$ moves along the first edge, from $1$ to $2$ when $w_2$ moves along the second edge, from $2$ to $3$ when $w_2$ moves along the third edge, and from $3$ to $4$ when $w_2$ moves along the fourth edge. The geometry is illustrated for the case of a $4\times4$ lattice in Fig.\ \ref{fig:lattice-geometry}. Note that $w_1$ and $w_2$ are not that far from each other. This means that the process does not represent a braiding of well-separated anyons, as the anyons overlap substantially. This is, however, not of importance for our purposes, as our aim is not to compute the result of anyon braiding, but rather to investigate how slowly one needs to move the anyons to ensure close-to-perfect adiabaticity.

In the following, we consider two Hamiltonians on the $L_x\times L_y$ lattice, namely
\begin{equation}
H^+(s)\equiv H_{+5}(s)
\end{equation}
with $\beta=+5$, for which the state \eqref{toppsi} is the ground state and
\begin{equation}
H^-(s)\equiv H_{-5}(s)
\end{equation}
with $\beta=-5$, for which \eqref{toppsi} is a scar state in the middle of the spectrum. We first establish that $H^-(s)$ is indeed a scar model for the considered parameters, and then study the dynamics when $w_2$ moves along the first edge.

\subsubsection{Tower of scar states}

The above model can be extended to involve a tower of scar states. The Hamiltonian \eqref{topHamexp} only describes the sector with $(N-2)/2$ particles. One can, however, use the considered approach to also derive Hamiltonians for sectors with $(N-2k)/2$ particles and $2k$ quasiholes, where $k$ is a nonnegative integer. For $k\geq2$, the coordinates $w_1$ and $w_2$ can be chosen to be the coordinates of two of the quasiholes, and for $k=0$ the Hamiltonian can be chosen to be independent of $w_1$ and $w_2$. If we further add the term $\Delta \sum_{i}n_i$ to the Hamiltonian, the scar states with $2k$ and $2k+2$ quasiholes will differ in energy by $\Delta$, and we hence have a tower of equidistant scar states.

\subsection{Scarring properties of $H^-(s)$}

\begin{figure*}
\includegraphics[width=\linewidth]{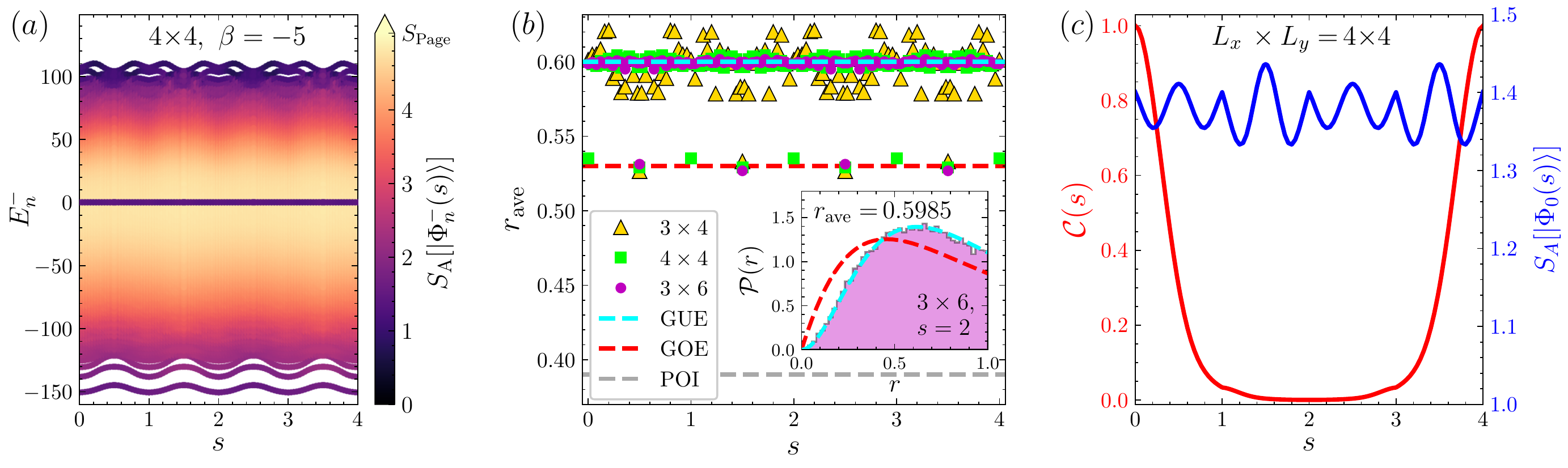}
\caption{\textbf{Static properties of the scar model.} (a) Von Neumann entanglement entropy of the eigenstates of the scar Hamiltonian $H^{-}(s)$ on a $4\times4$ lattice with region $A$ chosen as shown in Fig.\ \ref{fig:lattice-geometry}. The dark line at zero energy is produced by the low-entangled scar state. The entanglement entropy of the other eigenstates near the middle of the spectrum is close to the Page value, which indicates that they are thermal. (b) The mean level spacing ratio $r_{\textrm{ave}}$ of the eigenstates also indicates thermal behavior. Except for certain anyon positions with enhanced symmetry, at which the result is close to $r_{\textrm{GOE}} \approx 0.536$, the mean level spacing ratio shows agreement with $r_{\textrm{GUE}} \approx 0.603$. The inset shows GUE level statistics for the $3\times6$ lattice for $s=2$. The value for the Poisson distribution $r_{\textrm{POI}} \approx 0.386$ is also shown for comparison. (c) The entanglement entropy $\mathcal{S}_A[|\Phi_0(s)\rangle]$ of the scar state $|\Phi_0(s)\rangle$ as well as the instantaneous fidelity $\mathcal{C}(s) = |\langle \Phi_0(0) | \Phi_0(s) \rangle|^2$. While the instantaneous fidelity shows that the structure of the scar state changes as the second quasihole moves around, the entanglement entropy remains low.}
\label{fig:static_CFT}
\end{figure*}

\subsubsection{Entanglement entropy}

To compute the entanglement entropy, we divide the system into two parts consisting of, respectively, the upper and lower half of the lattice as illustrated in Fig.\ \ref{fig:lattice-geometry}. We perform an exact numerical diagonalization of $H^-(s)$ to obtain the eigenstates in the sector with $(N-2)/2$ particles and then compute the von Neumann entropy from Eq.\ \eqref{eq:vNS}. Figure \ref{fig:static_CFT}(a) shows that the scar state has a much lower entanglement entropy than the other states with similar energies (see also Fig.\ \ref{fig:static_CFT}(c) for a more detailed view). The other states in the middle of the spectrum have entanglement entropies that are close to the Page value $[N\ln(2) - 1]/2$ \cite{page1993} as expected for thermal states.

\subsubsection{Level spacing statistics}

To further confirm that the spectrum of $H^{-}(s)$ is thermal, we plot the mean level spacing ratio \eqref{eq:mean_level_spacing_ratios} as well as the distribution of level spacing ratios \eqref{eq:level_spacings} in Fig.\ \ref{fig:static_CFT}(b) for the sector with $(N-2)/2$ particles. For ergodic Hamiltonians with broken time-reversal invariance, the statistics of level spacing ratios is generally expected to obey the Gaussian unitary ensemble (GUE) with average $r_{\textrm{GUE}} \approx 0.603$. Ergodic Hamiltonians that do not break time reversal invariance are instead expected to follow the Gaussian orthogonal ensemble (GOE) with $r_{\textrm{GOE}} \approx 0.536$, in contrast to the Poisson distribution with $r_{\textrm{POI}} \approx 0.386$ for integrable and many-body localized models \cite{Poilblanc:1993,Atas:2013}. The Hamiltonian \eqref{topHamexp} breaks time reversal symmetry, and the level spacing ratios in Fig.\ \ref{fig:static_CFT}(b) indeed follow GUE statistics for most values of $s$.

Even systems with broken time reversal symmetry can exhibit GOE level spacing statistics if the Hamiltonian remains invariant under a combined symmetry consisting of time reversal followed by a spatial unitary operation, such as reflection or rotation. Such anomalous GOE statistics was pointed out in Ref.\ \cite{Robnik:1986} for single-particle Hamiltonians and has also been observed in interacting fermionic models describing the lowest Landau level of fractional quantum Hall systems where time reversal symmetry is broken by an external magnetic field \cite{Fremling:2018}. When $w_1$ and $w_2$ are at special positions, the model considered here has a combined symmetry consisting of time reversal, represented by complex conjugation, followed by mirroring in an axis going through both $w_1$ and $w_2$. For the path on the $4\times4$ lattice, the special positions are when $w_2$ is at one of the four corners or halfway between two adjacent corners. For rectangular lattices, the special positions are only when $w_2$ is midway between two of the corners. These are precisely the values of $s$ for which GOE statistics is observed in Fig.\ \ref{fig:static_CFT}(b).

\subsubsection{Change of the scar state with $s$}

Finally, Fig.\ \ref{fig:static_CFT}(c) shows that the scar state changes as a function of the parameter $s$. The state at $s=4$ is the same as the state at $s=0$, and hence the instantaneous fidelity at $s=4$ is unity. If one takes $w_2$ all the way around the loop, one can end up in the desired state either because the dynamics is adiabatic or because the dynamics is so fast that the system does not have time to react and therefore stays in the initial state. To eliminate the latter possibility, we shall consider time evolution from $s=0$ to $s=1$ in the following, i.e.\ moving along the first edge only. The instantaneous fidelity for $s=1$ is seen to be close to zero, and the ideal final state is hence almost orthogonal to the initial state.

\subsection{Adiabatic time evolution}
\label{sec:dynamics_CFT}

We now turn to a study of the dynamics. We shall consider two different ramp protocols in the following, which we refer to as linear ramp and sinusoidal ramp, respectively. For the linear ramp, given by
\begin{equation}\label{eq:toplinramp}
s \equiv s(t)=vt, \qquad v=1/T, \qquad t\in[0,T],
\end{equation}
$s$ changes linearly in time from zero to one. For the sinusoidal ramp, we instead take
\begin{equation}
\label{eq:parametrisation}
    s \equiv s(t) = \mathrm{sin}^{2}\left(\frac{\pi}{2}vt\right), \quad v=1/T, \quad t\in[0,T].
\end{equation}
For this ramp, the time derivative of $s$ goes to zero at the corners of the rectangle. This eliminates abrupt changes happening because the direction of movement changes if $w_2$ goes all the way around the loop. We also consider this second type of ramp to investigate to what extent the choice of ramp affects the results. We again consider ground state, scar state, and thermal state dynamics as defined in Sec.\ \ref{list:GST}, and we simulate the dynamics using the Chebyshev expansion \cite{schaefer2017}.

\subsubsection{Adiabatic fidelity}
\label{sec:dynamics_CFT_fid}

\begin{figure}
\includegraphics[width=\linewidth]{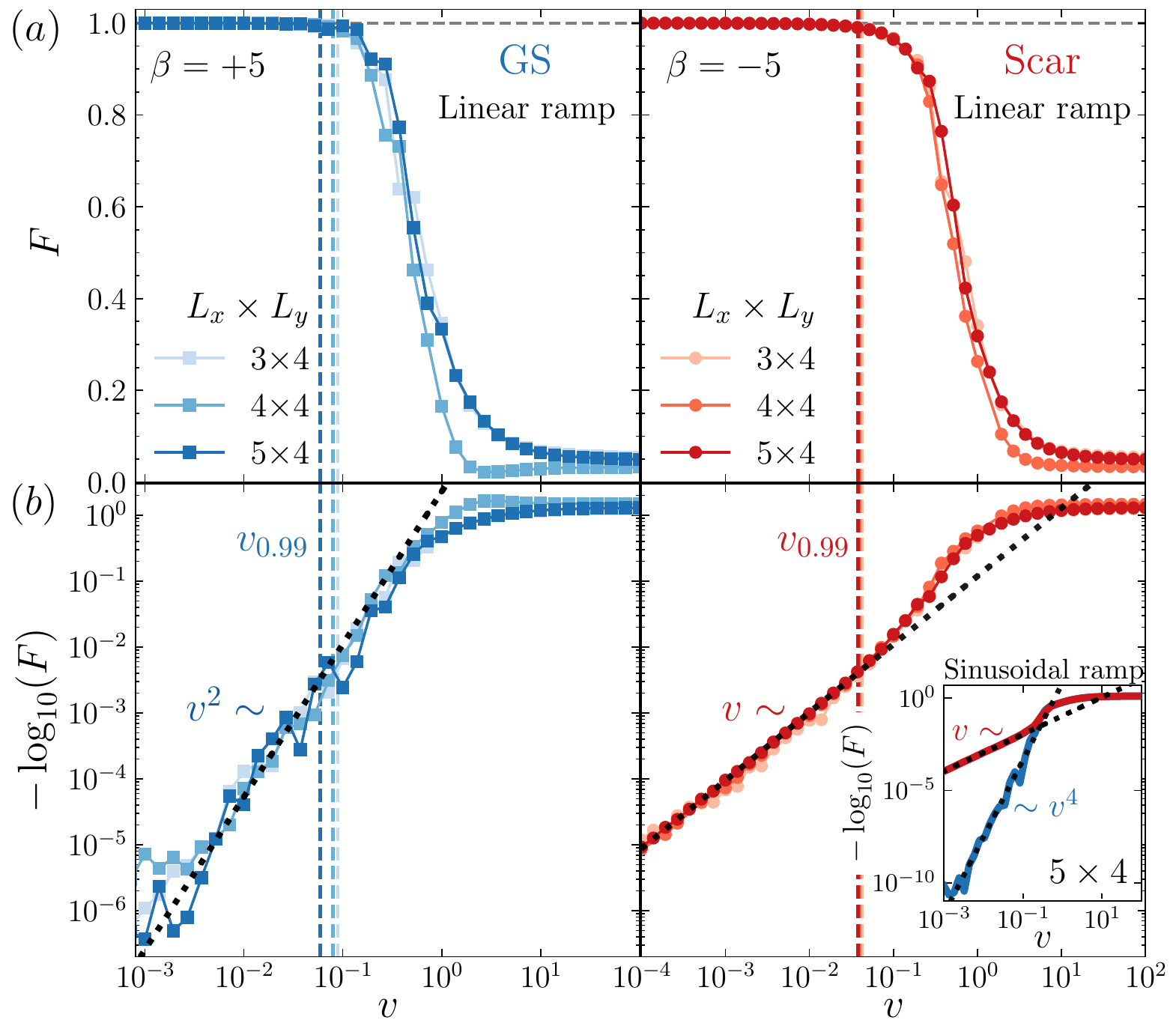}
\caption{\textbf{Adiabatic fidelity.} (a) Adiabatic fidelity \eqref{eq:ad_fid} as a function of ramp speed $v$ for ground state and scar state dynamics (see the list in Sec.\ \ref{list:GST}) and different system sizes. The vertical dashed lines show the adiabatic velocity $v_{0.99}$ defined as the ramp speed for which the fidelity drops below $0.99$. It is remarkable that $v_{0.99}$ is about the same for the scar state as for the ground state, despite the fact that the scar state sits in the middle of a thermal spectrum with no protecting energy gap.
(b) In the adiabatic regime, the logarithmic fidelity, $-\log_{10}(F)$, scales as $v^2$ for the ground state dynamics and $v$ for the scar state dynamics. The inset shows that the logarithmic fidelity for the sinusoidal ramp scales differently with $v$ for the ground state dynamics, but not for the scar state dynamics.} \label{fig:fidelity_CFT}
\end{figure}

The adiabatic fidelity \eqref{eq:ad_fid} of the time evolved state is shown in Fig.\ \ref{fig:fidelity_CFT}. The plot to the left shows the adiabatic fidelity, when the initial state is the ground state, and the system is evolved with $H^+(s)$. The plot to the right instead shows the same quantity when the initial state is the scar state and the system is evolved with $H^-(s)$. We observe that the adiabatic fidelity in the latter case is close to unity for ramp speeds as high as $10^{-1}$, despite the fact that the scar state is in the middle of the spectrum and is surrounded by thermal states having similar energies. In fact, the two plots for $F$ look rather similar, indicating that the ground state and the scar state behave similarly with respect to adiabaticity.

To get a more detailed view and to determine the scaling behavior with $v$ in the adiabatic regime, we plot the logarithmic fidelity, $-\mathrm{log}_{10}(F)$, in Fig.\ \ref{fig:fidelity_CFT}(b). For both cases and for both of the considered ramp functions, the logarithmic fidelity follows a power law scaling with $v$ in the regime of small ramp speed. For the ground state dynamics, $-\mathrm{log}_{10}(F)$ scales as $v^{2}$ for the linear ramp and as $v^{4}$ for the sinusoidal ramp. For the scar state dynamics, we instead find that $-\mathrm{log}_{10}(F)$ scales as $v$ for both of the considered ramp functions. We will discuss this scaling further in Sec.\ \ref{sec:CFT_theory}. The differences in scaling mean that the ground state performs better than the scar state if one is interested in very high fidelities. If one, e.g., wants $1-F\approx 10^{-10}$, one should choose ground state dynamics and optimize the ramp function.

For both the linear ramp and the sinusoidal ramp, there is a change in behavior of the logarithmic fidelity for the scar state dynamics from linear scaling in $v$ at low ramp speeds to a behavior almost coinciding with the ground state dynamics for intermediate ramp speeds. This suggests that there are two leakage mechanisms in play: one that only applies to the scar state dynamics and is dominant at low ramp speeds, and one that applies to both ground and scar state dynamics and is dominant at intermediate ramp speeds. We will identify these mechanisms in Sec.\ \ref{sec:CFT_theory} below.

For the model considered here, it is not practicable to investigate in detail how the adiabatic velocity scales with system size. First, the complexity of quantum many-body systems means that only small systems can be simulated numerically. If, e.g., we want to consider quadratic lattices, $2\times2$ is too small, $4\times 4$ is possible, and $6\times6$ is already too big. Second, the Hamiltonian \eqref{topHamexp} is long-range and the value of some of the coefficients depends on the positions of all the lattice sites. What happens deep in the interior of the system hence continues to change as the system size grows. We can, however, make a few observations from the results obtained for lattices with sizes $3\times4$, $4\times4$, and $5\times4$. For the ground state dynamics, the adiabatic velocity changes only slightly with system size for these three lattices, and for the scar state dynamics the change is even less. This suggests that adiabatic dynamics is possible even for relatively large system sizes.

\subsubsection{Leakage to other states}

\begin{figure}
\includegraphics[width=1\linewidth]{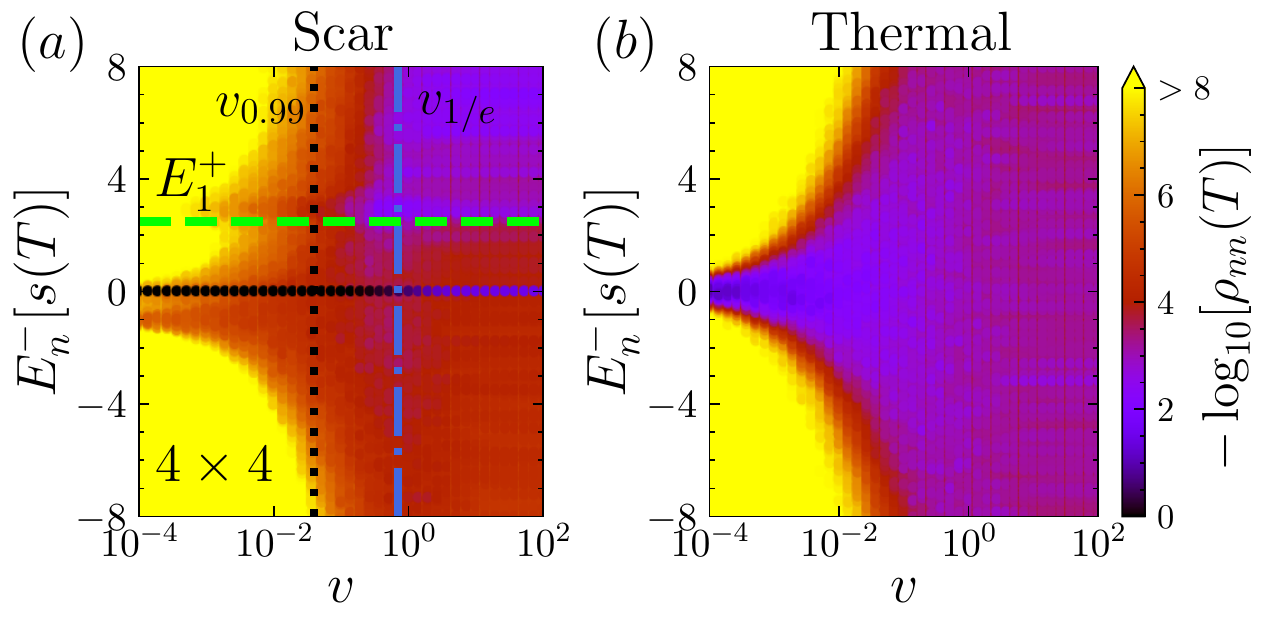}
\caption{\textbf{Leakage to other states.} Population in each of the instantaneous eigenstates at the end of the ramp as a function of the ramp speed $v$ and the final instantaneous energies $E^{-}_n[s(T)]$. The computations are for (a) scar state and (b) thermal state dynamics on the $4 \times 4$ lattice with the linear ramp. While the transfer to other states is small for the scar state dynamics at small $v$, it is much larger for the thermal state dynamics. This shows that the scar state is suitable for adiabatic dynamics, while the thermal states are not. The dashed green line shows the energy of the first excited state of $H^+[s(T)]$, and the vertical lines show the adiabatic velocities $v_{0.99}$ and $v_{1/e}$ obtained from the dynamics.}
\label{fig:leakage}
\end{figure}

\begin{figure}
\includegraphics[width=1\linewidth]{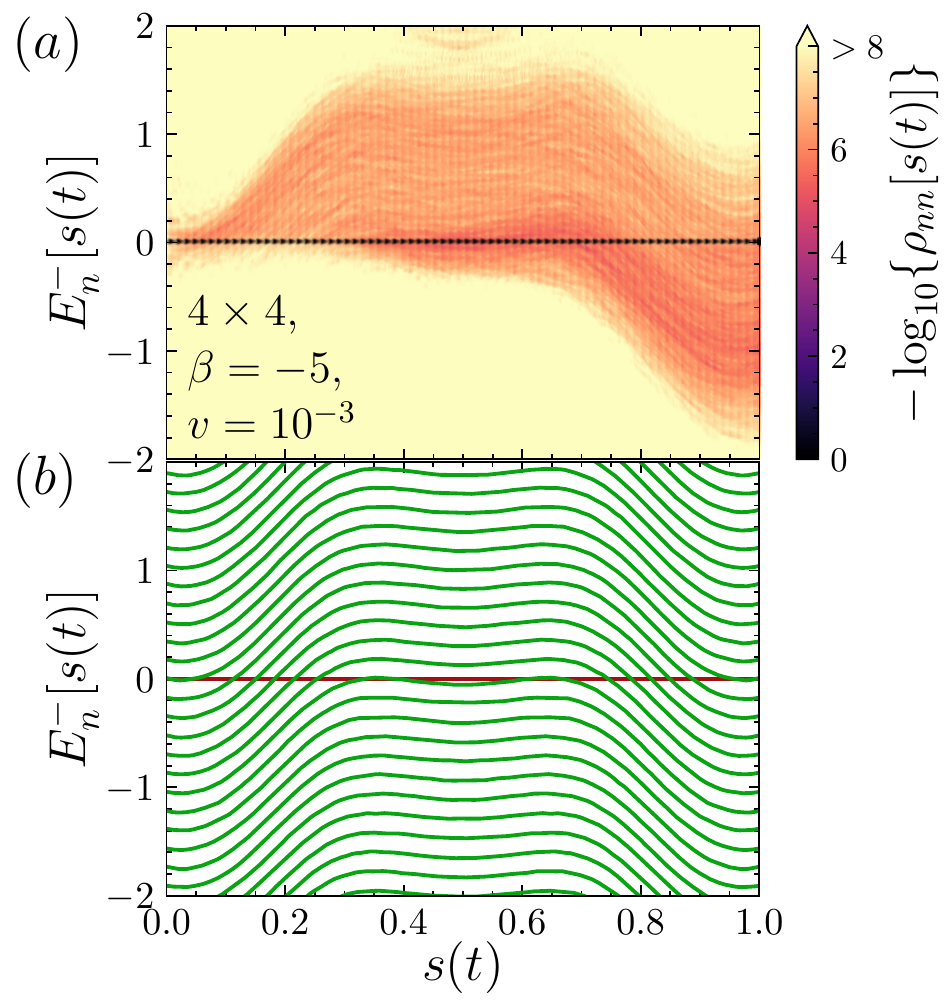}
\caption{\textbf{Leakage dynamics.} (a) The evolution of the population in each eigenstate as a function of the eigenstate energy and the parameter $s$ for the scar state dynamics on the $4\times4$ lattice with a linear ramp with ramp speed $v=10^{-3}$. (b) The evolution of the energy spectrum as a function of $s$ for the $4\times 4$ lattice. The scar state is shown in red and the thermal states in green. For the thermal states, we plot only every 30 state for clarity. Leakage from the scar state to thermal states happens primarily for the thermal states that are close in energy to the scar state. During the first about one third of the ramp, the population already leaked to the thermal states is transported upwards in energy, because the energies of the thermal eigenstates increase with $s$. During the last about one third of the ramp, this transport is downwards. In addition, population can transfer between thermal states when they are close in energy.}
\label{fig:leakage_CFT}
\end{figure}

\begin{figure}
\includegraphics[width=1\linewidth]{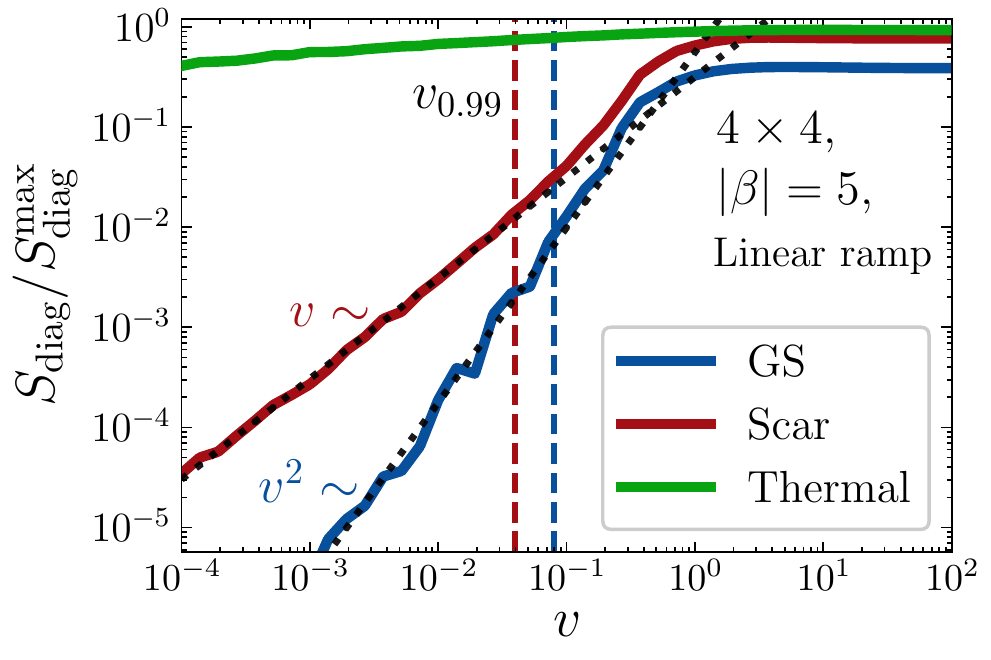}
\caption{\textbf{Scaled diagonal entropy.} The scaled diagonal entropy as a function of ramp speed for the ground state, scar state, and thermal state dynamics on a $4 \times 4$ lattice. The vertical dashed lines show the adiabatic velocity $v_{0.99}$. The plot shows that the time-evolved state at the end of the ramp spreads over several energy eigenstates for the thermal state dynamics, while the time-evolved state at the end of the ramp is almost equal to a single energy eigenstate for the ground state and scar state dynamics at small ramp speeds.}
\label{fig:S-diag}
\end{figure}

The amount of leakage to other states as quantified by $\rho_{nn}(T)$ in Eq.\ \eqref{eq:rhonn} is shown in Fig.\ \ref{fig:leakage} for scar state and thermal state dynamics. The leakage to other states is seen to be much smaller for the scar state dynamics than for the thermal state dynamics. For the scar state dynamics, we observe that there is an enhanced population transfer to states with energies around or slightly larger than the first excited state of $H^{+}[s(T)]$, denoted by $E^+_1[s(T)]$. We discuss this observation further in Sec.\ \ref{sec:dynamics} below.

For the scar state dynamics at low ramp speeds, the leaked population is seen to mainly end up in instantaneous eigenstates with energies slightly below zero. This can be understood from Fig.\ \ref{fig:leakage_CFT}, which shows both how the energies of the eigenstates change with $s(t)$ and how the population of the eigenstates depends on the energies of the eigenstates and $s(t)$. Note that the spectra at $s=0$ and $s=1$ are identical. It is seen that leakage from the scar state to the thermal states happens mainly for the thermal states that are close in energy to the scar state. The population that has already leaked to the thermal states is moved up and down in energy following the energy of the thermal states. In addition, population can also be transferred between thermal states that are close in energy. The leaked population ends up mainly in states that have energies slightly below zero at the end of the ramp, because these states are the states that have crossed the scar state during the ramp.

The scaled diagonal entropy is shown in Fig.\ \ref{fig:S-diag}. The main conclusion from this plot is that the population quickly spreads over several states for the thermal state dynamics, while the population spreading is much smaller for the ground state and scar state dynamics at low ramp speeds. The thermal states are hence not suitable for adiabatic dynamics.

\subsubsection{Emergence of the spectral gap}
\label{sec:dynamics}

\begin{figure}
	\includegraphics[width=1\linewidth]{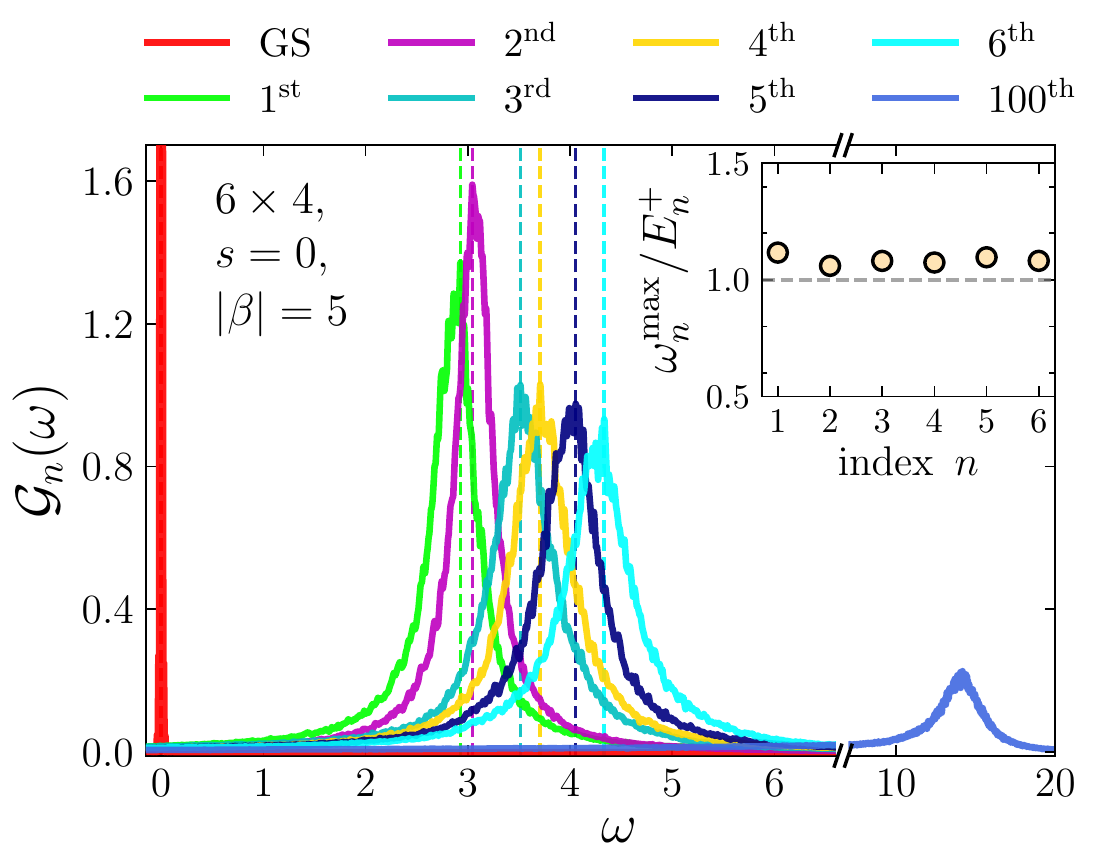}
	\caption{\textbf{Spectral gap.} Dynamical spectral function \eqref{eq:G} measured with respect to the ground state (GS), $1$st, $2$nd, $3$rd, $4$th, $5$th, and $6$th excited state as well as the $100$th eigenstate of $H^+(0)$. Each state produces one peak. The vertical, dashed lines mark the maxima of the peaks, and the values of $\omega$ at the maxima are denoted $\omega_n^{\textrm{max}}$. The inset shows that $\omega_n^{\textrm{max}}\approx E_n^{+}$. In other words, if a transfer from the scar state $|\Phi_0(s)\rangle$ to the state $|\Phi_n^{+}(s)\rangle$ happens in the scar model, it is seen as a leakage to a group of states with energies around $E_n^{+}$.}
	\label{fig:G_CFT_64}
\end{figure}

In Fig.\ \ref{fig:leakage}(a), we have seen that there is a slightly enhanced leakage to states at certain energies starting from $E^{+}_1[s(T)]$. We here provide an explanation for this effect. For the case of ground state dynamics, leakage happens primarily to low-lying excited states. This is, in part, because the energy difference is lower for these states, but also because the low-lying excited states have structural similarities to the ground state, which leads to larger matrix elements of the adiabatic gauge potential. When we convert the ground state model into the scar model by changing the sign of $\beta$, the state $|\Phi_0(s)\rangle$ becomes a highly excited state. One may speculate what happens to the low-lying excited states from the ground state model, when the sign of $\beta$ is switched. To find out, we use the kernel polynomial method \cite{Weise:2006} to compute the dynamical spectral function
\begin{align}\label{eq:G}
	\mathcal{G}_{n}(\omega)=\sum_{m} |\langle \Phi^+_n(s)|\Phi_m^-(s)\rangle|^2 \delta[\omega-E^-_{m}(s)]
\end{align}
for the $6\times 4$ lattice with Hilbert space dimension $\mathcal {D}^{\pm}=2496144$. The spectral resolution of the kernel polynomial method is inversely controlled by the number of Chebyshev moments which we set to $2^{17}$ to achieve high resolution.

The result of the computation for $s=0$ as an example is shown in Fig.\ \ref{fig:G_CFT_64}. The ground state $|\Phi_0^{+}(0)\rangle$ of course maps perfectly to the scar state $|\Phi_0^{-}(0)\rangle$ by construction, and as a result the spectral function for $n=0$ is a sharp peak at $\omega=0$. The spectral function for the $n$th low-lying excited state $|\Phi_n^{+}(0)\rangle$ is a peak centered around $E_n^{+}(0)$, which means that the state $|\Phi_n^{+}(0)\rangle$ only has significant weight on eigenstates of $H^{-}(0)$ with energies around $E_n^{+}(0)$. It is also seen that the peak broadens as $n$ increases.

The picture that emerges from this computation is the following. Because the eigenstates $|\Phi_n^{+}(s)\rangle$ of $H^{+}(s)$ with low $n$ are close in energy to the state $|\Phi_0(s)\rangle$, we expect $|\Phi_n^{+}(s)\rangle$ to have some similarities with $|\Phi_0(s)\rangle$. All these states, e.g., have low entanglement entropy compared to thermal states in the middle of the spectrum. As a result, we expect leakage to these states to happen more easily. In the scar model, some eigenstates of $H^{-}(s)$ may still couple to the states $|\Phi_n^{+}(s)\rangle$ with low $n$ because of their similarity to $|\Phi_0(s)\rangle$. But now these states are superpositions of a group of instantaneous eigenstates with energies around $E_n^{+}(s)$ with low $n$. The result is an enhanced leakage to groups of states with energies around $E_n^{+}(s)$, e.g., $E_1^{+}$, as seen in Fig.\ \ref{fig:leakage}(a). We note also that population transferred from $|\Phi_0(s)\rangle$ to $|\Phi_n^{+}(s)\rangle$ will evolve further since leakage easily happens between thermal states that are nearby in energy and since the different instantaneous eigenstates evolve with different dynamical phases.

\subsection{Comparison to adiabatic perturbation theory}
\label{sec:CFT_theory}

\begin{figure}
\includegraphics[width=1\linewidth]{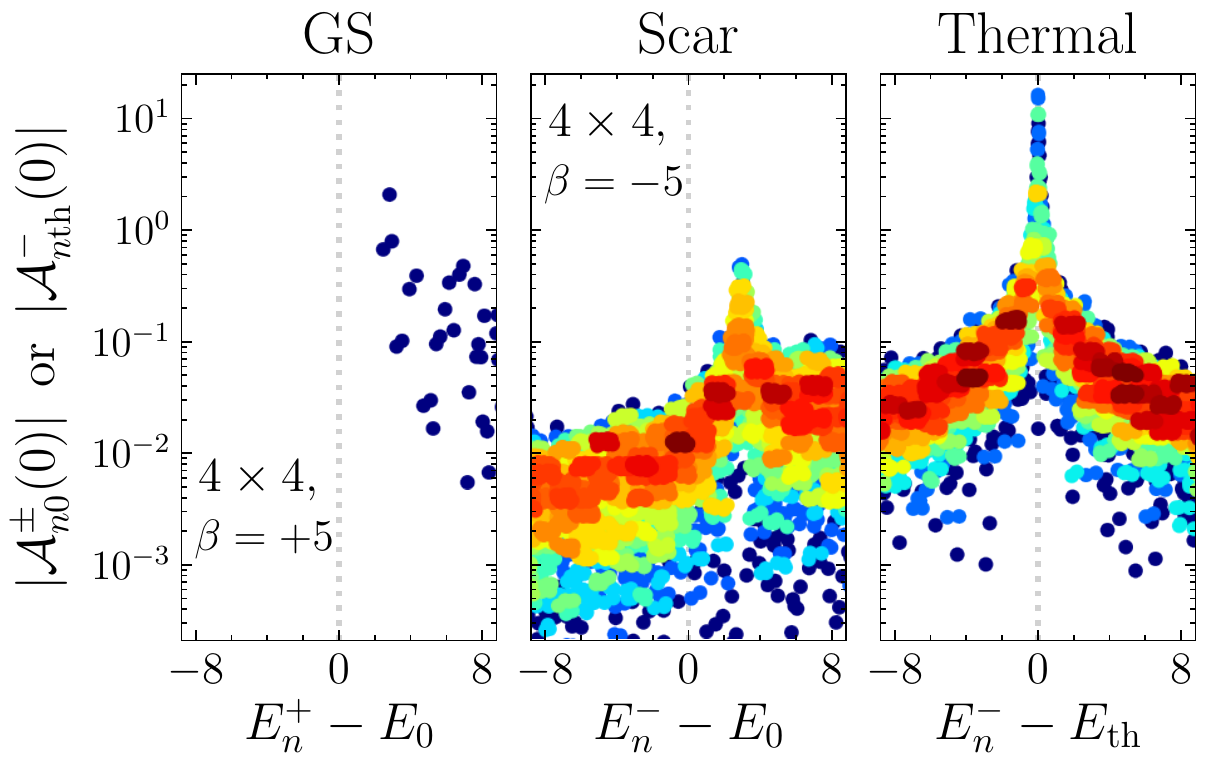}
\caption{\textbf{Matrix elements of the adiabatic gauge potential.} This plot, and corresponding plots for other values of $s$ that look similar, provides the information needed to estimate the leakage from Eq.\ \eqref{eq:cms}. For the ground state dynamics, leakage happens primarily to low-lying excited states, but since these are separated from the ground state by a gap, the leakage is low. For the scar state leakage is low, because it happens through two inefficient mechanisms, namely leakage to states that are close to zero energy $E_0=0$, but have quite low matrix elements of the adiabatic gauge potential, and leakage to states that have different energies, but particularly high matrix elements of the adiabatic gauge potential. For the thermal state dynamics, the matrix elements of the adiabatic gauge potential show a large peak at zero energy difference, and this gives rise to rapid leakage. The color indicates the density of data points, with red being high density and blue being low density. The vertical dashed lines mark the points of zero energy difference.}
\label{fig:fidelity_CFT-susceptibility}
\end{figure}

Returning to the discussion of adiabatic perturbation theory in Sec.\ \ref{sec:interpretation_theory}, the rate of leakage to other states is determined approximately by Eq.\ \eqref{eq:cms}. The central quantities are the matrix element of the adiabatic gauge potential \eqref{eq:AGP}, and the energy difference. Both of these are shown for $s=0$ in Fig.\ \ref{fig:fidelity_CFT-susceptibility}. Qualitatively similar plots are obtained for other values of $s$. These plots illustrate how well adiabatic time evolution works for the different states and the main reasons for the observed leakage.

The ground state is suitable for adiabatic time evolution, because all the other states are separated from the ground state by an energy gap. In this case, Eq.\ \eqref{eq:cmssq} applies for the linear ramp for low enough ramp speeds, which leads to a quadratic scaling of the logarithmic fidelity with ramp speed.

The scar state is suitable for adiabatic time evolution because all states that are not separated from the scar state by a significant energy difference have low matrix elements of the adiabatic gauge potential. In this case, there are two mechanisms for leakage. Leakage can happen to thermal states that are close in energy to the scar state, but have low matrix elements of the adiabatic gauge potential. This gives rise to a logarithmic fidelity that scales linearly in $v$ as argued in Sec.\ \ref{sec:interpretation_scar}, and this mechanism is dominant at small ramp speeds. The other mechanism is leakage to states that are at different energies than the scar state, but have particularly high values of the matrix elements of the adiabatic gauge potential, typically of the order $|\mathcal{A}^{-}_{n0}|\sim O(1)$. This type of leakage is similar to the leakage from a gapped ground state, and hence scales quadratically in ramp speed when the ramp is linear and the ramp speed is low. As seen in Fig.\ \ref{fig:fidelity_CFT}(b), this mechanism can be dominant at intermediate ramp speeds, or it may not become dominant within the high fidelity regime, as seen for the MPS model. Whether the second mechanism is important or not for intermediate ramp speeds depends on the heights and locations of the peaks of the matrix elements of the adiabatic gauge potential in a given model. For the model considered here, the energy at which the peak appears (e.g., $E^{-}_n\approx2.97$ at $s=0$) is close to the gap between the ground state and the low-lying excited state of $H^+$ with the largest absolute value of the adiabatic gauge potential (e.g., $E^{+}_2\approx2.82$ at $s=0$). This coincidence happens for all $s$ and is consistent with the observation of the spectral gap in Sec.\ \ref{sec:dynamics}. For even higher ramp speeds, the exponential in \eqref{eq:cms} oscillates at a slower rate for a given energy difference and leakage happens to a larger range of energy eigenstates.

Finally, we observe that the thermal state is not suitable for adiabatic dynamics because the matrix elements of the adiabatic gauge potential has a large peak at zero energy difference.

\subsection{Comments on the quantum speed limit}

For reasons discussed already in Sec.\ \ref{sec:dynamics_CFT_fid}, we are not able to judge how $\delta E^{\pm}_0/C_N$ scales with system size for the model considered in this section, but if $\delta E^{\pm}_0/C_N\to 0$ for $N\to\infty$, we note that the upper bound $v_{1/e}^{\textrm{QSL}}$ of the adiabatic velocity computed from \eqref{eq:AC_QSL} is the same for the ground and scar state dynamics. This is so, because the ground state of $H^+(s)$ is the same state as the scar state of $H^{-}(s)$, and because $\partial_s H_{\beta}(s)$ is independent of $\beta$, such that $\partial_s H^+(s)=\partial_sH^-(s)$.

\section{Conclusion}

We have shown that scar states embedded in the middle of a thermal spectrum without a protecting energy gap are well suited for adiabatic time evolution. The reason is that leakage to the nearby thermal states is impeded because of the structural differences between the scar state and the thermal states.

By investigating two rather different scar models, one based on an MPS construction in one dimension and the other being a fractional quantum Hall scar model in two dimensions, we have identified two main mechanisms that lead to leakage out of the scar state. One is leakage to thermal states that are close to the scar state in energy. This leakage happens only slowly, because of the structural differences between the scar state and the thermal state, which mathematically is seen as a low value of the matrix elements of the adiabatic gauge potential between the states. Since the scar states are exact eigenstates for all values of the considered parameters, the thermal states can cross the scar state in the spectrum without hybridizing. Such crossings give rise to a leakage $1-F$ that scales linearly with ramp speed in the limit of small ramp speeds. The linear scaling is practically universal and independent of ramp protocol for small enough $v$ as it relies on that one can Taylor expand the matrix element of the adiabatic gauge potential and the energy difference to first order in the changing parameter over an interval whose width scales as the square root of the ramp speed. Different behaviors could be observed, if the first derivative of the energy difference is zero when a thermal state is close to the scar state.

The other mechanism for leakage is leakage to states that have energies different from the scar state, but matrix elements of the adiabatic gauge potential that are particularly high. Such leakage is reminiscent of the leakage out of a gapped ground state undergoing adiabatic evolution, and for small and constant ramp speeds, the leakage $1-F$ scales quadratically with the ramp speed. Scar models are often produced by lifting a ground state higher up in the spectrum, and peaks in the matrix elements of the adiabatic gauge potential can appear as remnants in the scar model of the low-lying excitations of the original model. How significant this mechanism is, depends on the model and can be judged from the highest peaks in the matrix elements of the adiabatic gauge potential. In addition to these two mechanisms, population that has already leaked to thermal states can redistribute through leakage between thermal states in the spectrum that happens relatively easily if the thermal states are close in energy.

At very low ramp speeds, the first mechanism is dominant. As a result, ground states of gapped systems perform better than scar states in this regime, as only the second mechanism is at play for gapped ground states. This is, however, only relevant if one wants very high adiabatic fidelities, such as $0.9999$ for the considered models. At intermediate ramp speeds, there may be a regime, where the second mechanism becomes dominant, and for sufficiently high ramp speeds leakage happens to a broader range of states. In these regimes, the adiabatic fidelities obtained for a given ramp speed are roughly the same for the scar and ground state dynamics. As a result, the adiabatic velocity $v_{0.99}$, defined as the ramp speed at which the adiabatic fidelity drops below $0.99$, is about the same for scar states and gapped ground states in the considered models. For the model in one dimension, we find that $v_{0.99}$ scales as a power law with system size $N$ for both the scar state and the ground state (roughly as $N^{-0.8}$). On the contrary, if the system starts in a thermal state, $v_{0.99}$ drops exponentially with system size, and a large leakage is observed even for small ramp speeds. Similar behaviors are seen for the two-dimensional model, although in that case we cannot determine the precise scaling behaviors with system size due to the complexity.

The identified mechanisms provide knowledge that one can utilize to judge and improve the expected performance of scar models with respect to adiabatic time evolution. The first mechanism, e.g., tells us that one should avoid that the same thermal state remains close to the scar state during a significant fraction of the evolution, as this would lead to a larger leakage to that state. One could also think of engineering echoes, where leakage happening to a thermal state in a first crossing is returned in a second crossing. It would also be interesting to investigate to what extent shortcuts to adiabaticity \cite{Torrontegui2013,Odelin2019,Ljubotina:2020,Benseny2021} can be applied for scar states.

We also studied the effect of perturbations for the one-dimensional model. We showed that the scar state remains an exact eigenstate for a family of perturbations, and for this family of perturbations the physics is robust. For perturbations outside this family, we found that the physics is unaltered as long as the perturbations are small enough that the fidelity between the perturbed scar state of the perturbed model and the exact scar state of the unperturbed model remains high. This is easier to achieve for small systems. In some scar models, the most natural perturbations belong to the family of perturbations for which the scar state remains an exact eigenstate \cite{Kolb2023}, and it would be interesting to construct suitable paths for adiabatic dynamics in such models.

The ability to manipulate scar states adiabatically opens up interesting possibilities. In particular, instead of having a model with a single ground state manifold isolated from the rest of the spectrum, one can have models with a tower of states that are all isolated from other states in the spectrum, behave like ground states, and can be manipulated like ground states. This, e.g., opens a door for a different type of parallelism in adiabatic optimization processes as well as interference among different ground states and revival dynamics that can be adiabatically modified.

\begin{acknowledgments}
This work has been supported by Carlsbergfondet under Grant No.\ CF20-0658 and by Danmarks Frie Forskningsfond under Grant No.\ 8049-00074B.
\end{acknowledgments}	

\bibliography{references}

\begin{thebibliography}{89}%
\makeatletter
\providecommand \@ifxundefined [1]{%
 \@ifx{#1\undefined}
}%
\providecommand \@ifnum [1]{%
 \ifnum #1\expandafter \@firstoftwo
 \else \expandafter \@secondoftwo
 \fi
}%
\providecommand \@ifx [1]{%
 \ifx #1\expandafter \@firstoftwo
 \else \expandafter \@secondoftwo
 \fi
}%
\providecommand \natexlab [1]{#1}%
\providecommand \enquote  [1]{``#1''}%
\providecommand \bibnamefont  [1]{#1}%
\providecommand \bibfnamefont [1]{#1}%
\providecommand \citenamefont [1]{#1}%
\providecommand \href@noop [0]{\@secondoftwo}%
\providecommand \href [0]{\begingroup \@sanitize@url \@href}%
\providecommand \@href[1]{\@@startlink{#1}\@@href}%
\providecommand \@@href[1]{\endgroup#1\@@endlink}%
\providecommand \@sanitize@url [0]{\catcode `\\12\catcode `\$12\catcode
  `\&12\catcode `\#12\catcode `\^12\catcode `\_12\catcode `\%12\relax}%
\providecommand \@@startlink[1]{}%
\providecommand \@@endlink[0]{}%
\providecommand \url  [0]{\begingroup\@sanitize@url \@url }%
\providecommand \@url [1]{\endgroup\@href {#1}{\urlprefix }}%
\providecommand \urlprefix  [0]{URL }%
\providecommand \Eprint [0]{\href }%
\providecommand \doibase [0]{https://doi.org/}%
\providecommand \selectlanguage [0]{\@gobble}%
\providecommand \bibinfo  [0]{\@secondoftwo}%
\providecommand \bibfield  [0]{\@secondoftwo}%
\providecommand \translation [1]{[#1]}%
\providecommand \BibitemOpen [0]{}%
\providecommand \bibitemStop [0]{}%
\providecommand \bibitemNoStop [0]{.\EOS\space}%
\providecommand \EOS [0]{\spacefactor3000\relax}%
\providecommand \BibitemShut  [1]{\csname bibitem#1\endcsname}%
\let\auto@bib@innerbib\@empty
\bibitem [{\citenamefont {Born}\ and\ \citenamefont {Fock}(1928)}]{born1928}%
  \BibitemOpen
  \bibfield  {author} {\bibinfo {author} {\bibfnamefont {M.}~\bibnamefont
  {Born}}\ and\ \bibinfo {author} {\bibfnamefont {V.}~\bibnamefont {Fock}},\
  }\bibfield  {title} {\bibinfo {title} {{Beweis des Adiabatensatzes}},\ }\href
  {https://doi.org/10.1007/BF01343193} {\bibfield  {journal} {\bibinfo
  {journal} {Z. Phys.}\ }\textbf {\bibinfo {volume} {51}},\ \bibinfo {pages}
  {165} (\bibinfo {year} {1928})}\BibitemShut {NoStop}%
\bibitem [{\citenamefont {Schwinger}(1937)}]{schwinger1937}%
  \BibitemOpen
  \bibfield  {author} {\bibinfo {author} {\bibfnamefont {J.}~\bibnamefont
  {Schwinger}},\ }\bibfield  {title} {\bibinfo {title} {On nonadiabatic
  processes in inhomogeneous fields},\ }\href
  {https://doi.org/10.1103/PhysRev.51.648} {\bibfield  {journal} {\bibinfo
  {journal} {Phys. Rev.}\ }\textbf {\bibinfo {volume} {51}},\ \bibinfo {pages}
  {648} (\bibinfo {year} {1937})}\BibitemShut {NoStop}%
\bibitem [{\citenamefont {Kato}(1950)}]{kato1950}%
  \BibitemOpen
  \bibfield  {author} {\bibinfo {author} {\bibfnamefont {T.}~\bibnamefont
  {Kato}},\ }\bibfield  {title} {\bibinfo {title} {On the adiabatic theorem of
  quantum mechanics},\ }\href {https://doi.org/10.1143/JPSJ.5.435} {\bibfield
  {journal} {\bibinfo  {journal} {Journal of the Physical Society of Japan}\
  }\textbf {\bibinfo {volume} {5}},\ \bibinfo {pages} {435} (\bibinfo {year}
  {1950})}\BibitemShut {NoStop}%
\bibitem [{\citenamefont {Aspuru-Guzik}\ \emph {et~al.}(2005)\citenamefont
  {Aspuru-Guzik}, \citenamefont {Dutoi}, \citenamefont {Love},\ and\
  \citenamefont {Head-Gordon}}]{aspuru-guzik2005}%
  \BibitemOpen
  \bibfield  {author} {\bibinfo {author} {\bibfnamefont {A.}~\bibnamefont
  {Aspuru-Guzik}}, \bibinfo {author} {\bibfnamefont {A.~D.}\ \bibnamefont
  {Dutoi}}, \bibinfo {author} {\bibfnamefont {P.~J.}\ \bibnamefont {Love}},\
  and\ \bibinfo {author} {\bibfnamefont {M.}~\bibnamefont {Head-Gordon}},\
  }\bibfield  {title} {\bibinfo {title} {Simulated quantum computation of
  molecular energies},\ }\href {https://doi.org/10.1126/science.1113479}
  {\bibfield  {journal} {\bibinfo  {journal} {Science}\ }\textbf {\bibinfo
  {volume} {309}},\ \bibinfo {pages} {1704} (\bibinfo {year}
  {2005})}\BibitemShut {NoStop}%
\bibitem [{\citenamefont {Farhi}\ \emph {et~al.}(2000)\citenamefont {Farhi},
  \citenamefont {Goldstone}, \citenamefont {Gutmann},\ and\ \citenamefont
  {Sipser}}]{farhi2000}%
  \BibitemOpen
  \bibfield  {author} {\bibinfo {author} {\bibfnamefont {E.}~\bibnamefont
  {Farhi}}, \bibinfo {author} {\bibfnamefont {J.}~\bibnamefont {Goldstone}},
  \bibinfo {author} {\bibfnamefont {S.}~\bibnamefont {Gutmann}},\ and\ \bibinfo
  {author} {\bibfnamefont {M.}~\bibnamefont {Sipser}},\ }\bibfield  {title}
  {\bibinfo {title} {Quantum computation by adiabatic evolution},\ }\href
  {https://arxiv.org/abs/quant-ph/0001106} {\bibfield  {journal} {\bibinfo
  {journal} {arXiv preprint quant-ph/0001106}\ } (\bibinfo {year}
  {2000})}\BibitemShut {NoStop}%
\bibitem [{\citenamefont {Albash}\ and\ \citenamefont
  {Lidar}(2018)}]{albash2018}%
  \BibitemOpen
  \bibfield  {author} {\bibinfo {author} {\bibfnamefont {T.}~\bibnamefont
  {Albash}}\ and\ \bibinfo {author} {\bibfnamefont {D.~A.}\ \bibnamefont
  {Lidar}},\ }\bibfield  {title} {\bibinfo {title} {Adiabatic quantum
  computation},\ }\href {https://doi.org/10.1103/RevModPhys.90.015002}
  {\bibfield  {journal} {\bibinfo  {journal} {Rev. Mod. Phys.}\ }\textbf
  {\bibinfo {volume} {90}},\ \bibinfo {pages} {015002} (\bibinfo {year}
  {2018})}\BibitemShut {NoStop}%
\bibitem [{\citenamefont {Kadowaki}\ and\ \citenamefont
  {Nishimori}(1998)}]{Annealing}%
  \BibitemOpen
  \bibfield  {author} {\bibinfo {author} {\bibfnamefont {T.}~\bibnamefont
  {Kadowaki}}\ and\ \bibinfo {author} {\bibfnamefont {H.}~\bibnamefont
  {Nishimori}},\ }\bibfield  {title} {\bibinfo {title} {Quantum annealing in
  the transverse {I}sing model},\ }\href
  {https://doi.org/10.1103/PhysRevE.58.5355} {\bibfield  {journal} {\bibinfo
  {journal} {Phys. Rev. E}\ }\textbf {\bibinfo {volume} {58}},\ \bibinfo
  {pages} {5355} (\bibinfo {year} {1998})}\BibitemShut {NoStop}%
\bibitem [{\citenamefont {Thouless}(1983)}]{thouless1983}%
  \BibitemOpen
  \bibfield  {author} {\bibinfo {author} {\bibfnamefont {D.~J.}\ \bibnamefont
  {Thouless}},\ }\bibfield  {title} {\bibinfo {title} {Quantization of particle
  transport},\ }\href {https://doi.org/10.1103/PhysRevB.27.6083} {\bibfield
  {journal} {\bibinfo  {journal} {Phys. Rev. B}\ }\textbf {\bibinfo {volume}
  {27}},\ \bibinfo {pages} {6083} (\bibinfo {year} {1983})}\BibitemShut
  {NoStop}%
\bibitem [{\citenamefont {Marzlin}\ and\ \citenamefont
  {Sanders}(2004)}]{Marzlin:2004}%
  \BibitemOpen
  \bibfield  {author} {\bibinfo {author} {\bibfnamefont {K.-P.}\ \bibnamefont
  {Marzlin}}\ and\ \bibinfo {author} {\bibfnamefont {B.~C.}\ \bibnamefont
  {Sanders}},\ }\bibfield  {title} {\bibinfo {title} {{Inconsistency in the
  Application of the Adiabatic Theorem}},\ }\href
  {https://doi.org/10.1103/PhysRevLett.93.160408} {\bibfield  {journal}
  {\bibinfo  {journal} {Phys. Rev. Lett.}\ }\textbf {\bibinfo {volume} {93}},\
  \bibinfo {pages} {160408} (\bibinfo {year} {2004})}\BibitemShut {NoStop}%
\bibitem [{\citenamefont {Tong}\ \emph {et~al.}(2005)\citenamefont {Tong},
  \citenamefont {Singh}, \citenamefont {Kwek},\ and\ \citenamefont
  {Oh}}]{Tong:2005}%
  \BibitemOpen
  \bibfield  {author} {\bibinfo {author} {\bibfnamefont {D.~M.}\ \bibnamefont
  {Tong}}, \bibinfo {author} {\bibfnamefont {K.}~\bibnamefont {Singh}},
  \bibinfo {author} {\bibfnamefont {L.~C.}\ \bibnamefont {Kwek}},\ and\
  \bibinfo {author} {\bibfnamefont {C.~H.}\ \bibnamefont {Oh}},\ }\bibfield
  {title} {\bibinfo {title} {{Quantitative Conditions Do Not Guarantee the
  Validity of the Adiabatic Approximation}},\ }\href
  {https://doi.org/10.1103/PhysRevLett.95.110407} {\bibfield  {journal}
  {\bibinfo  {journal} {Phys. Rev. Lett.}\ }\textbf {\bibinfo {volume} {95}},\
  \bibinfo {pages} {110407} (\bibinfo {year} {2005})}\BibitemShut {NoStop}%
\bibitem [{\citenamefont {Tong}\ \emph {et~al.}(2007)\citenamefont {Tong},
  \citenamefont {Singh}, \citenamefont {Kwek},\ and\ \citenamefont
  {Oh}}]{Tong:2007}%
  \BibitemOpen
  \bibfield  {author} {\bibinfo {author} {\bibfnamefont {D.~M.}\ \bibnamefont
  {Tong}}, \bibinfo {author} {\bibfnamefont {K.}~\bibnamefont {Singh}},
  \bibinfo {author} {\bibfnamefont {L.~C.}\ \bibnamefont {Kwek}},\ and\
  \bibinfo {author} {\bibfnamefont {C.~H.}\ \bibnamefont {Oh}},\ }\bibfield
  {title} {\bibinfo {title} {{Sufficiency Criterion for the Validity of the
  Adiabatic Approximation}},\ }\href
  {https://doi.org/10.1103/PhysRevLett.98.150402} {\bibfield  {journal}
  {\bibinfo  {journal} {Phys. Rev. Lett.}\ }\textbf {\bibinfo {volume} {98}},\
  \bibinfo {pages} {150402} (\bibinfo {year} {2007})}\BibitemShut {NoStop}%
\bibitem [{\citenamefont {Altland}\ and\ \citenamefont
  {Gurarie}(2008)}]{Altland:2008}%
  \BibitemOpen
  \bibfield  {author} {\bibinfo {author} {\bibfnamefont {A.}~\bibnamefont
  {Altland}}\ and\ \bibinfo {author} {\bibfnamefont {V.}~\bibnamefont
  {Gurarie}},\ }\bibfield  {title} {\bibinfo {title} {{Many Body Generalization
  of the Landau-Zener Problem}},\ }\href
  {https://doi.org/10.1103/PhysRevLett.100.063602} {\bibfield  {journal}
  {\bibinfo  {journal} {Phys. Rev. Lett.}\ }\textbf {\bibinfo {volume} {100}},\
  \bibinfo {pages} {063602} (\bibinfo {year} {2008})}\BibitemShut {NoStop}%
\bibitem [{\citenamefont {Du}\ \emph {et~al.}(2008)\citenamefont {Du},
  \citenamefont {Hu}, \citenamefont {Wang}, \citenamefont {Wu}, \citenamefont
  {Zhao},\ and\ \citenamefont {Suter}}]{Du:2008}%
  \BibitemOpen
  \bibfield  {author} {\bibinfo {author} {\bibfnamefont {J.}~\bibnamefont
  {Du}}, \bibinfo {author} {\bibfnamefont {L.}~\bibnamefont {Hu}}, \bibinfo
  {author} {\bibfnamefont {Y.}~\bibnamefont {Wang}}, \bibinfo {author}
  {\bibfnamefont {J.}~\bibnamefont {Wu}}, \bibinfo {author} {\bibfnamefont
  {M.}~\bibnamefont {Zhao}},\ and\ \bibinfo {author} {\bibfnamefont
  {D.}~\bibnamefont {Suter}},\ }\bibfield  {title} {\bibinfo {title}
  {{Experimental Study of the Validity of Quantitative Conditions in the
  Quantum Adiabatic Theorem}},\ }\href
  {https://doi.org/10.1103/PhysRevLett.101.060403} {\bibfield  {journal}
  {\bibinfo  {journal} {Phys. Rev. Lett.}\ }\textbf {\bibinfo {volume} {101}},\
  \bibinfo {pages} {060403} (\bibinfo {year} {2008})}\BibitemShut {NoStop}%
\bibitem [{\citenamefont {Amin}(2009)}]{Amin:2009}%
  \BibitemOpen
  \bibfield  {author} {\bibinfo {author} {\bibfnamefont {M.~H.~S.}\
  \bibnamefont {Amin}},\ }\bibfield  {title} {\bibinfo {title} {{Consistency of
  the Adiabatic Theorem}},\ }\href
  {https://doi.org/10.1103/PhysRevLett.102.220401} {\bibfield  {journal}
  {\bibinfo  {journal} {Phys. Rev. Lett.}\ }\textbf {\bibinfo {volume} {102}},\
  \bibinfo {pages} {220401} (\bibinfo {year} {2009})}\BibitemShut {NoStop}%
\bibitem [{\citenamefont {Comparat}(2009)}]{comparat2009}%
  \BibitemOpen
  \bibfield  {author} {\bibinfo {author} {\bibfnamefont {D.}~\bibnamefont
  {Comparat}},\ }\bibfield  {title} {\bibinfo {title} {General conditions for
  quantum adiabatic evolution},\ }\href
  {https://doi.org/10.1103/PhysRevA.80.012106} {\bibfield  {journal} {\bibinfo
  {journal} {Phys. Rev. A}\ }\textbf {\bibinfo {volume} {80}},\ \bibinfo
  {pages} {012106} (\bibinfo {year} {2009})}\BibitemShut {NoStop}%
\bibitem [{\citenamefont {Tong}(2010)}]{Tong:2010}%
  \BibitemOpen
  \bibfield  {author} {\bibinfo {author} {\bibfnamefont {D.~M.}\ \bibnamefont
  {Tong}},\ }\bibfield  {title} {\bibinfo {title} {{Quantitative Condition is
  Necessary in Guaranteeing the Validity of the Adiabatic Approximation}},\
  }\href {https://doi.org/10.1103/PhysRevLett.104.120401} {\bibfield  {journal}
  {\bibinfo  {journal} {Phys. Rev. Lett.}\ }\textbf {\bibinfo {volume} {104}},\
  \bibinfo {pages} {120401} (\bibinfo {year} {2010})}\BibitemShut {NoStop}%
\bibitem [{\citenamefont {Bachmann}\ \emph {et~al.}(2017)\citenamefont
  {Bachmann}, \citenamefont {De~Roeck},\ and\ \citenamefont
  {Fraas}}]{Bachmann:2017}%
  \BibitemOpen
  \bibfield  {author} {\bibinfo {author} {\bibfnamefont {S.}~\bibnamefont
  {Bachmann}}, \bibinfo {author} {\bibfnamefont {W.}~\bibnamefont {De~Roeck}},\
  and\ \bibinfo {author} {\bibfnamefont {M.}~\bibnamefont {Fraas}},\ }\bibfield
   {title} {\bibinfo {title} {{Adiabatic Theorem for Quantum Spin Systems}},\
  }\href {https://doi.org/10.1103/PhysRevLett.119.060201} {\bibfield  {journal}
  {\bibinfo  {journal} {Phys. Rev. Lett.}\ }\textbf {\bibinfo {volume} {119}},\
  \bibinfo {pages} {060201} (\bibinfo {year} {2017})}\BibitemShut {NoStop}%
\bibitem [{\citenamefont {Lychkovskiy}\ \emph {et~al.}(2017)\citenamefont
  {Lychkovskiy}, \citenamefont {Gamayun},\ and\ \citenamefont
  {Cheianov}}]{Lychkovskiy:2017}%
  \BibitemOpen
  \bibfield  {author} {\bibinfo {author} {\bibfnamefont {O.}~\bibnamefont
  {Lychkovskiy}}, \bibinfo {author} {\bibfnamefont {O.}~\bibnamefont
  {Gamayun}},\ and\ \bibinfo {author} {\bibfnamefont {V.}~\bibnamefont
  {Cheianov}},\ }\bibfield  {title} {\bibinfo {title} {{Time Scale for
  Adiabaticity Breakdown in Driven Many-Body Systems and Orthogonality
  Catastrophe}},\ }\href {https://doi.org/10.1103/PhysRevLett.119.200401}
  {\bibfield  {journal} {\bibinfo  {journal} {Phys. Rev. Lett.}\ }\textbf
  {\bibinfo {volume} {119}},\ \bibinfo {pages} {200401} (\bibinfo {year}
  {2017})}\BibitemShut {NoStop}%
\bibitem [{\citenamefont {{Bachmann}}\ \emph {et~al.}(2018)\citenamefont
  {{Bachmann}}, \citenamefont {{De Roeck}},\ and\ \citenamefont
  {{Fraas}}}]{bachmann:2018}%
  \BibitemOpen
  \bibfield  {author} {\bibinfo {author} {\bibfnamefont {S.}~\bibnamefont
  {{Bachmann}}}, \bibinfo {author} {\bibfnamefont {W.}~\bibnamefont {{De
  Roeck}}},\ and\ \bibinfo {author} {\bibfnamefont {M.}~\bibnamefont
  {{Fraas}}},\ }\bibfield  {title} {\bibinfo {title} {{The adiabatic theorem in
  a quantum many-body setting}},\ }\href
  {https://ui.adsabs.harvard.edu/abs/2018arXiv180809985B} {\bibfield  {journal}
  {\bibinfo  {journal} {arXiv preprint arXiv:1808.09985}\ } (\bibinfo {year}
  {2018})}\BibitemShut {NoStop}%
\bibitem [{\citenamefont {Gogolin}\ and\ \citenamefont
  {Eisert}(2016)}]{gogolin2016}%
  \BibitemOpen
  \bibfield  {author} {\bibinfo {author} {\bibfnamefont {C.}~\bibnamefont
  {Gogolin}}\ and\ \bibinfo {author} {\bibfnamefont {J.}~\bibnamefont
  {Eisert}},\ }\bibfield  {title} {\bibinfo {title} {Equilibration,
  thermalisation, and the emergence of statistical mechanics in closed quantum
  systems},\ }\href {https://doi.org/10.1088/0034-4885/79/5/056001} {\bibfield
  {journal} {\bibinfo  {journal} {Rep. Prog. Phys.}\ }\textbf {\bibinfo
  {volume} {79}},\ \bibinfo {pages} {056001} (\bibinfo {year}
  {2016})}\BibitemShut {NoStop}%
\bibitem [{\citenamefont {Mori}\ \emph {et~al.}(2018)\citenamefont {Mori},
  \citenamefont {Ikeda}, \citenamefont {Kaminishi},\ and\ \citenamefont
  {Ueda}}]{mori2018}%
  \BibitemOpen
  \bibfield  {author} {\bibinfo {author} {\bibfnamefont {T.}~\bibnamefont
  {Mori}}, \bibinfo {author} {\bibfnamefont {T.~N.}\ \bibnamefont {Ikeda}},
  \bibinfo {author} {\bibfnamefont {E.}~\bibnamefont {Kaminishi}},\ and\
  \bibinfo {author} {\bibfnamefont {M.}~\bibnamefont {Ueda}},\ }\bibfield
  {title} {\bibinfo {title} {Thermalization and prethermalization in isolated
  quantum systems: a theoretical overview},\ }\href
  {https://doi.org/10.1088/1361-6455/aabcdf} {\bibfield  {journal} {\bibinfo
  {journal} {J. Phys. B}\ }\textbf {\bibinfo {volume} {51}},\ \bibinfo {pages}
  {112001} (\bibinfo {year} {2018})}\BibitemShut {NoStop}%
\bibitem [{\citenamefont {Shiraishi}\ and\ \citenamefont
  {Mori}(2017)}]{Shiraishi2017}%
  \BibitemOpen
  \bibfield  {author} {\bibinfo {author} {\bibfnamefont {N.}~\bibnamefont
  {Shiraishi}}\ and\ \bibinfo {author} {\bibfnamefont {T.}~\bibnamefont
  {Mori}},\ }\bibfield  {title} {\bibinfo {title} {{Systematic Construction of
  Counterexamples to the Eigenstate Thermalization Hypothesis}},\ }\href
  {https://doi.org/10.1103/PhysRevLett.119.030601} {\bibfield  {journal}
  {\bibinfo  {journal} {Phys. Rev. Lett.}\ }\textbf {\bibinfo {volume} {119}},\
  \bibinfo {pages} {030601} (\bibinfo {year} {2017})}\BibitemShut {NoStop}%
\bibitem [{\citenamefont {Bernien}\ \emph {et~al.}(2017)\citenamefont
  {Bernien}, \citenamefont {Schwartz}, \citenamefont {Keesling}, \citenamefont
  {Levine}, \citenamefont {Omran}, \citenamefont {Pichler}, \citenamefont
  {Choi}, \citenamefont {Zibrov}, \citenamefont {Endres}, \citenamefont
  {Greiner}, \citenamefont {Vuleti{\'{c}}},\ and\ \citenamefont
  {Lukin}}]{bernien2017}%
  \BibitemOpen
  \bibfield  {author} {\bibinfo {author} {\bibfnamefont {H.}~\bibnamefont
  {Bernien}}, \bibinfo {author} {\bibfnamefont {S.}~\bibnamefont {Schwartz}},
  \bibinfo {author} {\bibfnamefont {A.}~\bibnamefont {Keesling}}, \bibinfo
  {author} {\bibfnamefont {H.}~\bibnamefont {Levine}}, \bibinfo {author}
  {\bibfnamefont {A.}~\bibnamefont {Omran}}, \bibinfo {author} {\bibfnamefont
  {H.}~\bibnamefont {Pichler}}, \bibinfo {author} {\bibfnamefont
  {S.}~\bibnamefont {Choi}}, \bibinfo {author} {\bibfnamefont {A.~S.}\
  \bibnamefont {Zibrov}}, \bibinfo {author} {\bibfnamefont {M.}~\bibnamefont
  {Endres}}, \bibinfo {author} {\bibfnamefont {M.}~\bibnamefont {Greiner}},
  \bibinfo {author} {\bibfnamefont {V.}~\bibnamefont {Vuleti{\'{c}}}},\ and\
  \bibinfo {author} {\bibfnamefont {M.~D.}\ \bibnamefont {Lukin}},\ }\bibfield
  {title} {\bibinfo {title} {Probing many-body dynamics on a 51-atom quantum
  simulator},\ }\href {https://doi.org/10.1038/nature24622} {\bibfield
  {journal} {\bibinfo  {journal} {Nature}\ }\textbf {\bibinfo {volume} {551}},\
  \bibinfo {pages} {579} (\bibinfo {year} {2017})}\BibitemShut {NoStop}%
\bibitem [{\citenamefont {Turner}\ \emph
  {et~al.}(2018{\natexlab{a}})\citenamefont {Turner}, \citenamefont
  {Michailidis}, \citenamefont {Abanin}, \citenamefont {Serbyn},\ and\
  \citenamefont {Papic}}]{Turner:2018}%
  \BibitemOpen
  \bibfield  {author} {\bibinfo {author} {\bibfnamefont {C.~J.}\ \bibnamefont
  {Turner}}, \bibinfo {author} {\bibfnamefont {A.~A.}\ \bibnamefont
  {Michailidis}}, \bibinfo {author} {\bibfnamefont {D.~A.}\ \bibnamefont
  {Abanin}}, \bibinfo {author} {\bibfnamefont {M.}~\bibnamefont {Serbyn}},\
  and\ \bibinfo {author} {\bibfnamefont {Z.}~\bibnamefont {Papic}},\ }\bibfield
   {title} {\bibinfo {title} {Weak ergodicity breaking from quantum many-body
  scars},\ }\href {https://doi.org/10.1038/s41567-018-0137-5} {\bibfield
  {journal} {\bibinfo  {journal} {Nature Physics}\ }\textbf {\bibinfo {volume}
  {14}},\ \bibinfo {pages} {745} (\bibinfo {year}
  {2018}{\natexlab{a}})}\BibitemShut {NoStop}%
\bibitem [{\citenamefont {Turner}\ \emph
  {et~al.}(2018{\natexlab{b}})\citenamefont {Turner}, \citenamefont
  {Michailidis}, \citenamefont {Abanin}, \citenamefont {Serbyn},\ and\
  \citenamefont {Papi\ifmmode~\acute{c}\else \'{c}\fi{}}}]{Turner:2018_PRB}%
  \BibitemOpen
  \bibfield  {author} {\bibinfo {author} {\bibfnamefont {C.~J.}\ \bibnamefont
  {Turner}}, \bibinfo {author} {\bibfnamefont {A.~A.}\ \bibnamefont
  {Michailidis}}, \bibinfo {author} {\bibfnamefont {D.~A.}\ \bibnamefont
  {Abanin}}, \bibinfo {author} {\bibfnamefont {M.}~\bibnamefont {Serbyn}},\
  and\ \bibinfo {author} {\bibfnamefont {Z.}~\bibnamefont
  {Papi\ifmmode~\acute{c}\else \'{c}\fi{}}},\ }\bibfield  {title} {\bibinfo
  {title} {Quantum scarred eigenstates in a {R}ydberg atom chain: Entanglement,
  breakdown of thermalization, and stability to perturbations},\ }\href
  {https://doi.org/10.1103/PhysRevB.98.155134} {\bibfield  {journal} {\bibinfo
  {journal} {Phys. Rev. B}\ }\textbf {\bibinfo {volume} {98}},\ \bibinfo
  {pages} {155134} (\bibinfo {year} {2018}{\natexlab{b}})}\BibitemShut
  {NoStop}%
\bibitem [{\citenamefont {Moudgalya}\ \emph
  {et~al.}(2018{\natexlab{a}})\citenamefont {Moudgalya}, \citenamefont
  {Rachel}, \citenamefont {Bernevig},\ and\ \citenamefont
  {Regnault}}]{moudgalya2018a}%
  \BibitemOpen
  \bibfield  {author} {\bibinfo {author} {\bibfnamefont {S.}~\bibnamefont
  {Moudgalya}}, \bibinfo {author} {\bibfnamefont {S.}~\bibnamefont {Rachel}},
  \bibinfo {author} {\bibfnamefont {B.~A.}\ \bibnamefont {Bernevig}},\ and\
  \bibinfo {author} {\bibfnamefont {N.}~\bibnamefont {Regnault}},\ }\bibfield
  {title} {\bibinfo {title} {Exact excited states of nonintegrable models},\
  }\href {https://doi.org/10.1103/PhysRevB.98.235155} {\bibfield  {journal}
  {\bibinfo  {journal} {Phys. Rev. B}\ }\textbf {\bibinfo {volume} {98}},\
  \bibinfo {pages} {235155} (\bibinfo {year} {2018}{\natexlab{a}})}\BibitemShut
  {NoStop}%
\bibitem [{\citenamefont {Moudgalya}\ \emph
  {et~al.}(2018{\natexlab{b}})\citenamefont {Moudgalya}, \citenamefont
  {Regnault},\ and\ \citenamefont {Bernevig}}]{moudgalya2018b}%
  \BibitemOpen
  \bibfield  {author} {\bibinfo {author} {\bibfnamefont {S.}~\bibnamefont
  {Moudgalya}}, \bibinfo {author} {\bibfnamefont {N.}~\bibnamefont
  {Regnault}},\ and\ \bibinfo {author} {\bibfnamefont {B.~A.}\ \bibnamefont
  {Bernevig}},\ }\bibfield  {title} {\bibinfo {title} {Entanglement of exact
  excited states of {A}ffleck-{K}ennedy-{L}ieb-{T}asaki models: {E}xact
  results, many-body scars, and violation of the strong eigenstate
  thermalization hypothesis},\ }\href
  {https://doi.org/10.1103/PhysRevB.98.235156} {\bibfield  {journal} {\bibinfo
  {journal} {Phys. Rev. B}\ }\textbf {\bibinfo {volume} {98}},\ \bibinfo
  {pages} {235156} (\bibinfo {year} {2018}{\natexlab{b}})}\BibitemShut
  {NoStop}%
\bibitem [{\citenamefont {Schecter}\ and\ \citenamefont
  {Iadecola}(2019)}]{Schecter2019}%
  \BibitemOpen
  \bibfield  {author} {\bibinfo {author} {\bibfnamefont {M.}~\bibnamefont
  {Schecter}}\ and\ \bibinfo {author} {\bibfnamefont {T.}~\bibnamefont
  {Iadecola}},\ }\bibfield  {title} {\bibinfo {title} {{Weak Ergodicity
  Breaking and Quantum Many-Body Scars in Spin-1 $XY$ Magnets}},\ }\href
  {https://doi.org/10.1103/PhysRevLett.123.147201} {\bibfield  {journal}
  {\bibinfo  {journal} {Phys. Rev. Lett.}\ }\textbf {\bibinfo {volume} {123}},\
  \bibinfo {pages} {147201} (\bibinfo {year} {2019})}\BibitemShut {NoStop}%
\bibitem [{\citenamefont {Serbyn}\ \emph {et~al.}(2021)\citenamefont {Serbyn},
  \citenamefont {Abanin},\ and\ \citenamefont {Papi{\'c}}}]{serbyn2021}%
  \BibitemOpen
  \bibfield  {author} {\bibinfo {author} {\bibfnamefont {M.}~\bibnamefont
  {Serbyn}}, \bibinfo {author} {\bibfnamefont {D.~A.}\ \bibnamefont {Abanin}},\
  and\ \bibinfo {author} {\bibfnamefont {Z.}~\bibnamefont {Papi{\'c}}},\
  }\bibfield  {title} {\bibinfo {title} {Quantum many-body scars and weak
  breaking of ergodicity},\ }\href {https://doi.org/10.1038/s41567-021-01230-2}
  {\bibfield  {journal} {\bibinfo  {journal} {Nature Physics}\ }\textbf
  {\bibinfo {volume} {17}},\ \bibinfo {pages} {675} (\bibinfo {year}
  {2021})}\BibitemShut {NoStop}%
\bibitem [{\citenamefont {Moudgalya}\ \emph {et~al.}(2022)\citenamefont
  {Moudgalya}, \citenamefont {Bernevig},\ and\ \citenamefont
  {Regnault}}]{moudgalya2022}%
  \BibitemOpen
  \bibfield  {author} {\bibinfo {author} {\bibfnamefont {S.}~\bibnamefont
  {Moudgalya}}, \bibinfo {author} {\bibfnamefont {B.~A.}\ \bibnamefont
  {Bernevig}},\ and\ \bibinfo {author} {\bibfnamefont {N.}~\bibnamefont
  {Regnault}},\ }\bibfield  {title} {\bibinfo {title} {Quantum many-body scars
  and {H}ilbert space fragmentation: a review of exact results},\ }\href
  {https://doi.org/10.1088/1361-6633/ac73a0} {\bibfield  {journal} {\bibinfo
  {journal} {Rep. Prog. Phys.}\ }\textbf {\bibinfo {volume} {85}},\ \bibinfo
  {pages} {086501} (\bibinfo {year} {2022})}\BibitemShut {NoStop}%
\bibitem [{\citenamefont {Aramthottil}\ \emph {et~al.}(2022)\citenamefont
  {Aramthottil}, \citenamefont {Bhattacharya}, \citenamefont
  {Gonz\'alez-Cuadra}, \citenamefont {Lewenstein}, \citenamefont {Barbiero},\
  and\ \citenamefont {Zakrzewski}}]{Aramthottil2022}%
  \BibitemOpen
  \bibfield  {author} {\bibinfo {author} {\bibfnamefont {A.~S.}\ \bibnamefont
  {Aramthottil}}, \bibinfo {author} {\bibfnamefont {U.}~\bibnamefont
  {Bhattacharya}}, \bibinfo {author} {\bibfnamefont {D.}~\bibnamefont
  {Gonz\'alez-Cuadra}}, \bibinfo {author} {\bibfnamefont {M.}~\bibnamefont
  {Lewenstein}}, \bibinfo {author} {\bibfnamefont {L.}~\bibnamefont
  {Barbiero}},\ and\ \bibinfo {author} {\bibfnamefont {J.}~\bibnamefont
  {Zakrzewski}},\ }\bibfield  {title} {\bibinfo {title} {Scar states in
  deconfined {${\mathbb{Z}}_{2}$} lattice gauge theories},\ }\href
  {https://doi.org/10.1103/PhysRevB.106.L041101} {\bibfield  {journal}
  {\bibinfo  {journal} {Phys. Rev. B}\ }\textbf {\bibinfo {volume} {106}},\
  \bibinfo {pages} {L041101} (\bibinfo {year} {2022})}\BibitemShut {NoStop}%
\bibitem [{\citenamefont {Chandran}\ \emph {et~al.}(2023)\citenamefont
  {Chandran}, \citenamefont {Iadecola}, \citenamefont {Khemani},\ and\
  \citenamefont {Moessner}}]{chandran2023}%
  \BibitemOpen
  \bibfield  {author} {\bibinfo {author} {\bibfnamefont {A.}~\bibnamefont
  {Chandran}}, \bibinfo {author} {\bibfnamefont {T.}~\bibnamefont {Iadecola}},
  \bibinfo {author} {\bibfnamefont {V.}~\bibnamefont {Khemani}},\ and\ \bibinfo
  {author} {\bibfnamefont {R.}~\bibnamefont {Moessner}},\ }\bibfield  {title}
  {\bibinfo {title} {Quantum many-body scars: A quasiparticle perspective},\
  }\href {https://doi.org/10.1146/annurev-conmatphys-031620-101617} {\bibfield
  {journal} {\bibinfo  {journal} {Annual Review of Condensed Matter Physics}\
  }\textbf {\bibinfo {volume} {14}},\ \bibinfo {pages} {443} (\bibinfo {year}
  {2023})}\BibitemShut {NoStop}%
\bibitem [{\citenamefont {Dooley}(2021)}]{dooley2021}%
  \BibitemOpen
  \bibfield  {author} {\bibinfo {author} {\bibfnamefont {S.}~\bibnamefont
  {Dooley}},\ }\bibfield  {title} {\bibinfo {title} {Robust quantum sensing in
  strongly interacting systems with many-body scars},\ }\href
  {https://doi.org/10.1103/PRXQuantum.2.020330} {\bibfield  {journal} {\bibinfo
   {journal} {PRX Quantum}\ }\textbf {\bibinfo {volume} {2}},\ \bibinfo {pages}
  {020330} (\bibinfo {year} {2021})}\BibitemShut {NoStop}%
\bibitem [{\citenamefont {Dooley}\ \emph {et~al.}(2023)\citenamefont {Dooley},
  \citenamefont {Pappalardi},\ and\ \citenamefont {Goold}}]{dooley2023}%
  \BibitemOpen
  \bibfield  {author} {\bibinfo {author} {\bibfnamefont {S.}~\bibnamefont
  {Dooley}}, \bibinfo {author} {\bibfnamefont {S.}~\bibnamefont {Pappalardi}},\
  and\ \bibinfo {author} {\bibfnamefont {J.}~\bibnamefont {Goold}},\ }\bibfield
   {title} {\bibinfo {title} {Entanglement enhanced metrology with quantum
  many-body scars},\ }\href {https://doi.org/10.1103/PhysRevB.107.035123}
  {\bibfield  {journal} {\bibinfo  {journal} {Phys. Rev. B}\ }\textbf {\bibinfo
  {volume} {107}},\ \bibinfo {pages} {035123} (\bibinfo {year}
  {2023})}\BibitemShut {NoStop}%
\bibitem [{\citenamefont {Zhang}\ \emph {et~al.}(2023)\citenamefont {Zhang},
  \citenamefont {Dong}, \citenamefont {Gao}, \citenamefont {Zhao},
  \citenamefont {Hao}, \citenamefont {Desaules}, \citenamefont {Guo},
  \citenamefont {Chen}, \citenamefont {Deng}, \citenamefont {Liu} \emph
  {et~al.}}]{zhang2023many}%
  \BibitemOpen
  \bibfield  {author} {\bibinfo {author} {\bibfnamefont {P.}~\bibnamefont
  {Zhang}}, \bibinfo {author} {\bibfnamefont {H.}~\bibnamefont {Dong}},
  \bibinfo {author} {\bibfnamefont {Y.}~\bibnamefont {Gao}}, \bibinfo {author}
  {\bibfnamefont {L.}~\bibnamefont {Zhao}}, \bibinfo {author} {\bibfnamefont
  {J.}~\bibnamefont {Hao}}, \bibinfo {author} {\bibfnamefont {J.-Y.}\
  \bibnamefont {Desaules}}, \bibinfo {author} {\bibfnamefont {Q.}~\bibnamefont
  {Guo}}, \bibinfo {author} {\bibfnamefont {J.}~\bibnamefont {Chen}}, \bibinfo
  {author} {\bibfnamefont {J.}~\bibnamefont {Deng}}, \bibinfo {author}
  {\bibfnamefont {B.}~\bibnamefont {Liu}}, \emph {et~al.},\ }\bibfield  {title}
  {\bibinfo {title} {Many-body hilbert space scarring on a superconducting
  processor},\ }\href {https://doi.org/10.1038/s41567-022-01784-9} {\bibfield
  {journal} {\bibinfo  {journal} {Nature Physics}\ }\textbf {\bibinfo {volume}
  {19}},\ \bibinfo {pages} {120} (\bibinfo {year} {2023})}\BibitemShut
  {NoStop}%
\bibitem [{\citenamefont {Diedrich}\ \emph {et~al.}(1989)\citenamefont
  {Diedrich}, \citenamefont {Bergquist}, \citenamefont {Itano},\ and\
  \citenamefont {Wineland}}]{diedrich1989}%
  \BibitemOpen
  \bibfield  {author} {\bibinfo {author} {\bibfnamefont {F.}~\bibnamefont
  {Diedrich}}, \bibinfo {author} {\bibfnamefont {J.~C.}\ \bibnamefont
  {Bergquist}}, \bibinfo {author} {\bibfnamefont {W.~M.}\ \bibnamefont
  {Itano}},\ and\ \bibinfo {author} {\bibfnamefont {D.~J.}\ \bibnamefont
  {Wineland}},\ }\bibfield  {title} {\bibinfo {title} {{Laser Cooling to the
  Zero-Point Energy of Motion}},\ }\href
  {https://doi.org/10.1103/PhysRevLett.62.403} {\bibfield  {journal} {\bibinfo
  {journal} {Phys. Rev. Lett.}\ }\textbf {\bibinfo {volume} {62}},\ \bibinfo
  {pages} {403} (\bibinfo {year} {1989})}\BibitemShut {NoStop}%
\bibitem [{\citenamefont {Bergenfeldt}\ and\ \citenamefont
  {M\o{}lmer}(2009)}]{bergenfeldt2009}%
  \BibitemOpen
  \bibfield  {author} {\bibinfo {author} {\bibfnamefont {C.}~\bibnamefont
  {Bergenfeldt}}\ and\ \bibinfo {author} {\bibfnamefont {K.}~\bibnamefont
  {M\o{}lmer}},\ }\bibfield  {title} {\bibinfo {title} {Cooling a
  micromechanical resonator to its ground state by measurement and feedback},\
  }\href {https://doi.org/10.1103/PhysRevA.80.043838} {\bibfield  {journal}
  {\bibinfo  {journal} {Phys. Rev. A}\ }\textbf {\bibinfo {volume} {80}},\
  \bibinfo {pages} {043838} (\bibinfo {year} {2009})}\BibitemShut {NoStop}%
\bibitem [{\citenamefont {Srednicki}(1993)}]{srednicki1993}%
  \BibitemOpen
  \bibfield  {author} {\bibinfo {author} {\bibfnamefont {M.}~\bibnamefont
  {Srednicki}},\ }\bibfield  {title} {\bibinfo {title} {{Entropy and area}},\
  }\href {https://doi.org/10.1103/PhysRevLett.71.666} {\bibfield  {journal}
  {\bibinfo  {journal} {Phys. Rev. Lett.}\ }\textbf {\bibinfo {volume} {71}},\
  \bibinfo {pages} {666} (\bibinfo {year} {1993})}\BibitemShut {NoStop}%
\bibitem [{\citenamefont {Eisert}\ \emph {et~al.}(2010)\citenamefont {Eisert},
  \citenamefont {Cramer},\ and\ \citenamefont {Plenio}}]{eisert2010}%
  \BibitemOpen
  \bibfield  {author} {\bibinfo {author} {\bibfnamefont {J.}~\bibnamefont
  {Eisert}}, \bibinfo {author} {\bibfnamefont {M.}~\bibnamefont {Cramer}},\
  and\ \bibinfo {author} {\bibfnamefont {M.~B.}\ \bibnamefont {Plenio}},\
  }\bibfield  {title} {\bibinfo {title} {Colloquium: Area laws for the
  entanglement entropy},\ }\href {https://doi.org/10.1103/RevModPhys.82.277}
  {\bibfield  {journal} {\bibinfo  {journal} {Rev. Mod. Phys.}\ }\textbf
  {\bibinfo {volume} {82}},\ \bibinfo {pages} {277} (\bibinfo {year}
  {2010})}\BibitemShut {NoStop}%
\bibitem [{\citenamefont {Perez-Garcia}\ \emph {et~al.}(2007)\citenamefont
  {Perez-Garcia}, \citenamefont {Verstraete}, \citenamefont {Wolf},\ and\
  \citenamefont {Cirac}}]{perez-garcia2007}%
  \BibitemOpen
  \bibfield  {author} {\bibinfo {author} {\bibfnamefont {D.}~\bibnamefont
  {Perez-Garcia}}, \bibinfo {author} {\bibfnamefont {F.}~\bibnamefont
  {Verstraete}}, \bibinfo {author} {\bibfnamefont {M.~M.}\ \bibnamefont
  {Wolf}},\ and\ \bibinfo {author} {\bibfnamefont {J.~I.}\ \bibnamefont
  {Cirac}},\ }\bibfield  {title} {\bibinfo {title} {Matrix product state
  representations},\ }\href {https://dl.acm.org/doi/10.5555/2011832.2011833}
  {\bibfield  {journal} {\bibinfo  {journal} {Quantum Info. Comput.}\ }\textbf
  {\bibinfo {volume} {7}},\ \bibinfo {pages} {401–430} (\bibinfo {year}
  {2007})}\BibitemShut {NoStop}%
\bibitem [{\citenamefont {Moudgalya}\ \emph {et~al.}(2020)\citenamefont
  {Moudgalya}, \citenamefont {O'Brien}, \citenamefont {Bernevig}, \citenamefont
  {Fendley},\ and\ \citenamefont {Regnault}}]{Moudgalya:2020}%
  \BibitemOpen
  \bibfield  {author} {\bibinfo {author} {\bibfnamefont {S.}~\bibnamefont
  {Moudgalya}}, \bibinfo {author} {\bibfnamefont {E.}~\bibnamefont {O'Brien}},
  \bibinfo {author} {\bibfnamefont {B.~A.}\ \bibnamefont {Bernevig}}, \bibinfo
  {author} {\bibfnamefont {P.}~\bibnamefont {Fendley}},\ and\ \bibinfo {author}
  {\bibfnamefont {N.}~\bibnamefont {Regnault}},\ }\bibfield  {title} {\bibinfo
  {title} {Large classes of quantum scarred {H}amiltonians from matrix product
  states},\ }\href {https://doi.org/10.1103/PhysRevB.102.085120} {\bibfield
  {journal} {\bibinfo  {journal} {Phys. Rev. B}\ }\textbf {\bibinfo {volume}
  {102}},\ \bibinfo {pages} {085120} (\bibinfo {year} {2020})}\BibitemShut
  {NoStop}%
\bibitem [{\citenamefont {Sch\"on}\ \emph {et~al.}(2005)\citenamefont
  {Sch\"on}, \citenamefont {Solano}, \citenamefont {Verstraete}, \citenamefont
  {Cirac},\ and\ \citenamefont {Wolf}}]{Schon2005}%
  \BibitemOpen
  \bibfield  {author} {\bibinfo {author} {\bibfnamefont {C.}~\bibnamefont
  {Sch\"on}}, \bibinfo {author} {\bibfnamefont {E.}~\bibnamefont {Solano}},
  \bibinfo {author} {\bibfnamefont {F.}~\bibnamefont {Verstraete}}, \bibinfo
  {author} {\bibfnamefont {J.~I.}\ \bibnamefont {Cirac}},\ and\ \bibinfo
  {author} {\bibfnamefont {M.~M.}\ \bibnamefont {Wolf}},\ }\bibfield  {title}
  {\bibinfo {title} {{Sequential Generation of Entangled Multiqubit States}},\
  }\href {https://doi.org/10.1103/PhysRevLett.95.110503} {\bibfield  {journal}
  {\bibinfo  {journal} {Phys. Rev. Lett.}\ }\textbf {\bibinfo {volume} {95}},\
  \bibinfo {pages} {110503} (\bibinfo {year} {2005})}\BibitemShut {NoStop}%
\bibitem [{\citenamefont {Malz}\ \emph {et~al.}(2023)\citenamefont {Malz},
  \citenamefont {Styliaris}, \citenamefont {Wei},\ and\ \citenamefont
  {Cirac}}]{Malz2023}%
  \BibitemOpen
  \bibfield  {author} {\bibinfo {author} {\bibfnamefont {D.}~\bibnamefont
  {Malz}}, \bibinfo {author} {\bibfnamefont {G.}~\bibnamefont {Styliaris}},
  \bibinfo {author} {\bibfnamefont {Z.-Y.}\ \bibnamefont {Wei}},\ and\ \bibinfo
  {author} {\bibfnamefont {J.~I.}\ \bibnamefont {Cirac}},\ }\bibfield  {title}
  {\bibinfo {title} {Preparation of matrix product states with log-depth
  quantum circuits},\ }\href {https://doi.org/10.48550/arXiv.2307.01696}
  {\bibfield  {journal} {\bibinfo  {journal} {arXiv preprint arXiv:2307.01696}\
  } (\bibinfo {year} {2023})}\BibitemShut {NoStop}%
\bibitem [{\citenamefont {Affleck}\ \emph {et~al.}(1987)\citenamefont
  {Affleck}, \citenamefont {Kennedy}, \citenamefont {Lieb},\ and\ \citenamefont
  {Tasaki}}]{Affleck:1987}%
  \BibitemOpen
  \bibfield  {author} {\bibinfo {author} {\bibfnamefont {I.}~\bibnamefont
  {Affleck}}, \bibinfo {author} {\bibfnamefont {T.}~\bibnamefont {Kennedy}},
  \bibinfo {author} {\bibfnamefont {E.~H.}\ \bibnamefont {Lieb}},\ and\
  \bibinfo {author} {\bibfnamefont {H.}~\bibnamefont {Tasaki}},\ }\bibfield
  {title} {\bibinfo {title} {{Rigorous results on valence-bond ground states in
  antiferromagnets}},\ }\href {https://doi.org/10.1103/PhysRevLett.59.799}
  {\bibfield  {journal} {\bibinfo  {journal} {Phys. Rev. Lett.}\ }\textbf
  {\bibinfo {volume} {59}},\ \bibinfo {pages} {799} (\bibinfo {year}
  {1987})}\BibitemShut {NoStop}%
\bibitem [{\citenamefont {Schollwöck}(2011)}]{Schollwock:2011}%
  \BibitemOpen
  \bibfield  {author} {\bibinfo {author} {\bibfnamefont {U.}~\bibnamefont
  {Schollwöck}},\ }\bibfield  {title} {\bibinfo {title} {The density-matrix
  renormalization group in the age of matrix product states},\ }\href
  {https://doi.org/https://doi.org/10.1016/j.aop.2010.09.012} {\bibfield
  {journal} {\bibinfo  {journal} {Annals of Physics}\ }\textbf {\bibinfo
  {volume} {326}},\ \bibinfo {pages} {96} (\bibinfo {year} {2011})}\BibitemShut
  {NoStop}%
\bibitem [{\citenamefont {Orús}(2014)}]{Orus:2014}%
  \BibitemOpen
  \bibfield  {author} {\bibinfo {author} {\bibfnamefont {R.}~\bibnamefont
  {Orús}},\ }\bibfield  {title} {\bibinfo {title} {A practical introduction to
  tensor networks: {M}atrix product states and projected entangled pair
  states},\ }\href {https://doi.org/https://doi.org/10.1016/j.aop.2014.06.013}
  {\bibfield  {journal} {\bibinfo  {journal} {Annals of Physics}\ }\textbf
  {\bibinfo {volume} {349}},\ \bibinfo {pages} {117} (\bibinfo {year}
  {2014})}\BibitemShut {NoStop}%
\bibitem [{\citenamefont {Fannes}\ \emph {et~al.}(1992)\citenamefont {Fannes},
  \citenamefont {Nachtergaele},\ and\ \citenamefont
  {Werner}}]{fannes1992finitely}%
  \BibitemOpen
  \bibfield  {author} {\bibinfo {author} {\bibfnamefont {M.}~\bibnamefont
  {Fannes}}, \bibinfo {author} {\bibfnamefont {B.}~\bibnamefont
  {Nachtergaele}},\ and\ \bibinfo {author} {\bibfnamefont {R.~F.}\ \bibnamefont
  {Werner}},\ }\bibfield  {title} {\bibinfo {title} {Finitely correlated states
  on quantum spin chains},\ }\href {https://doi.org/10.1007/BF02099178}
  {\bibfield  {journal} {\bibinfo  {journal} {Communications in Mathematical
  Physics}\ }\textbf {\bibinfo {volume} {144}},\ \bibinfo {pages} {443}
  (\bibinfo {year} {1992})}\BibitemShut {NoStop}%
\bibitem [{\citenamefont {Nachtergaele}(1996)}]{nachtergaele1996spectral}%
  \BibitemOpen
  \bibfield  {author} {\bibinfo {author} {\bibfnamefont {B.}~\bibnamefont
  {Nachtergaele}},\ }\bibfield  {title} {\bibinfo {title} {The spectral gap for
  some spin chains with discrete symmetry breaking},\ }\href
  {https://doi.org/10.1007/BF02099509} {\bibfield  {journal} {\bibinfo
  {journal} {Communications in Mathematical Physics}\ }\textbf {\bibinfo
  {volume} {175}},\ \bibinfo {pages} {565} (\bibinfo {year}
  {1996})}\BibitemShut {NoStop}%
\bibitem [{\citenamefont {Wolf}\ \emph {et~al.}(2006)\citenamefont {Wolf},
  \citenamefont {Ortiz}, \citenamefont {Verstraete},\ and\ \citenamefont
  {Cirac}}]{wolf2006quantum}%
  \BibitemOpen
  \bibfield  {author} {\bibinfo {author} {\bibfnamefont {M.~M.}\ \bibnamefont
  {Wolf}}, \bibinfo {author} {\bibfnamefont {G.}~\bibnamefont {Ortiz}},
  \bibinfo {author} {\bibfnamefont {F.}~\bibnamefont {Verstraete}},\ and\
  \bibinfo {author} {\bibfnamefont {J.~I.}\ \bibnamefont {Cirac}},\ }\bibfield
  {title} {\bibinfo {title} {{Quantum Phase Transitions in Matrix Product
  Systems}},\ }\href {https://doi.org/10.1103/PhysRevLett.97.110403} {\bibfield
   {journal} {\bibinfo  {journal} {Phys. Rev. Lett.}\ }\textbf {\bibinfo
  {volume} {97}},\ \bibinfo {pages} {110403} (\bibinfo {year}
  {2006})}\BibitemShut {NoStop}%
\bibitem [{\citenamefont {Page}(1993)}]{page1993}%
  \BibitemOpen
  \bibfield  {author} {\bibinfo {author} {\bibfnamefont {D.~N.}\ \bibnamefont
  {Page}},\ }\bibfield  {title} {\bibinfo {title} {{Average entropy of a
  subsystem}},\ }\href {https://doi.org/10.1103/PhysRevLett.71.1291} {\bibfield
   {journal} {\bibinfo  {journal} {Phys. Rev. Lett.}\ }\textbf {\bibinfo
  {volume} {71}},\ \bibinfo {pages} {1291} (\bibinfo {year}
  {1993})}\BibitemShut {NoStop}%
\bibitem [{\citenamefont {D'Alessio}\ \emph {et~al.}(2016)\citenamefont
  {D'Alessio}, \citenamefont {Kafri}, \citenamefont {Polkovnikov},\ and\
  \citenamefont {Rigol}}]{Alessio:2014_rev}%
  \BibitemOpen
  \bibfield  {author} {\bibinfo {author} {\bibfnamefont {L.}~\bibnamefont
  {D'Alessio}}, \bibinfo {author} {\bibfnamefont {Y.}~\bibnamefont {Kafri}},
  \bibinfo {author} {\bibfnamefont {A.}~\bibnamefont {Polkovnikov}},\ and\
  \bibinfo {author} {\bibfnamefont {M.}~\bibnamefont {Rigol}},\ }\bibfield
  {title} {\bibinfo {title} {From quantum chaos and eigenstate thermalization
  to statistical mechanics and thermodynamics},\ }\href
  {https://doi.org/10.1080/00018732.2016.1198134} {\bibfield  {journal}
  {\bibinfo  {journal} {Advances in Physics}\ }\textbf {\bibinfo {volume}
  {65}},\ \bibinfo {pages} {239} (\bibinfo {year} {2016})}\BibitemShut
  {NoStop}%
\bibitem [{\citenamefont {Poilblanc}\ \emph {et~al.}(1993)\citenamefont
  {Poilblanc}, \citenamefont {Ziman}, \citenamefont {Bellissard}, \citenamefont
  {Mila},\ and\ \citenamefont {Montambaux}}]{Poilblanc:1993}%
  \BibitemOpen
  \bibfield  {author} {\bibinfo {author} {\bibfnamefont {D.}~\bibnamefont
  {Poilblanc}}, \bibinfo {author} {\bibfnamefont {T.}~\bibnamefont {Ziman}},
  \bibinfo {author} {\bibfnamefont {J.}~\bibnamefont {Bellissard}}, \bibinfo
  {author} {\bibfnamefont {F.}~\bibnamefont {Mila}},\ and\ \bibinfo {author}
  {\bibfnamefont {G.}~\bibnamefont {Montambaux}},\ }\bibfield  {title}
  {\bibinfo {title} {Poisson vs. {GOE} statistics in integrable and
  non-integrable quantum {H}amiltonians},\ }\href
  {http://iopscience.iop.org/article/10.1209/0295-5075/22/7/010/meta}
  {\bibfield  {journal} {\bibinfo  {journal} {Europhys. Lett.}\ }\textbf
  {\bibinfo {volume} {22}},\ \bibinfo {pages} {537} (\bibinfo {year}
  {1993})}\BibitemShut {NoStop}%
\bibitem [{\citenamefont {Atas}\ \emph {et~al.}(2013)\citenamefont {Atas},
  \citenamefont {Bogomolny}, \citenamefont {Giraud},\ and\ \citenamefont
  {Roux}}]{Atas:2013}%
  \BibitemOpen
  \bibfield  {author} {\bibinfo {author} {\bibfnamefont {Y.~Y.}\ \bibnamefont
  {Atas}}, \bibinfo {author} {\bibfnamefont {E.}~\bibnamefont {Bogomolny}},
  \bibinfo {author} {\bibfnamefont {O.}~\bibnamefont {Giraud}},\ and\ \bibinfo
  {author} {\bibfnamefont {G.}~\bibnamefont {Roux}},\ }\bibfield  {title}
  {\bibinfo {title} {{Distribution of the Ratio of Consecutive Level Spacings
  in Random Matrix Ensembles}},\ }\href
  {https://doi.org/10.1103/PhysRevLett.110.084101} {\bibfield  {journal}
  {\bibinfo  {journal} {Phys. Rev. Lett.}\ }\textbf {\bibinfo {volume} {110}},\
  \bibinfo {pages} {084101} (\bibinfo {year} {2013})}\BibitemShut {NoStop}%
\bibitem [{\citenamefont {Anderson}(1967)}]{Anderson:1967}%
  \BibitemOpen
  \bibfield  {author} {\bibinfo {author} {\bibfnamefont {P.~W.}\ \bibnamefont
  {Anderson}},\ }\bibfield  {title} {\bibinfo {title} {{Infrared Catastrophe in
  Fermi Gases with Local Scattering Potentials}},\ }\href
  {https://doi.org/10.1103/PhysRevLett.18.1049} {\bibfield  {journal} {\bibinfo
   {journal} {Phys. Rev. Lett.}\ }\textbf {\bibinfo {volume} {18}},\ \bibinfo
  {pages} {1049} (\bibinfo {year} {1967})}\BibitemShut {NoStop}%
\bibitem [{\citenamefont {Khemani}\ \emph {et~al.}(2015)\citenamefont
  {Khemani}, \citenamefont {Nandkishore},\ and\ \citenamefont
  {Sondhi}}]{Khemani:2015}%
  \BibitemOpen
  \bibfield  {author} {\bibinfo {author} {\bibfnamefont {V.}~\bibnamefont
  {Khemani}}, \bibinfo {author} {\bibfnamefont {R.}~\bibnamefont
  {Nandkishore}},\ and\ \bibinfo {author} {\bibfnamefont {S.~L.}\ \bibnamefont
  {Sondhi}},\ }\bibfield  {title} {\bibinfo {title} {Nonlocal adiabatic
  response of a localized system to local manipulations},\ }\href
  {https://doi.org/10.1038/nphys3344} {\bibfield  {journal} {\bibinfo
  {journal} {Nat. Phys.}\ }\textbf {\bibinfo {volume} {11}},\ \bibinfo {pages}
  {560} (\bibinfo {year} {2015})}\BibitemShut {NoStop}%
\bibitem [{\citenamefont {Deng}\ \emph {et~al.}(2015)\citenamefont {Deng},
  \citenamefont {Pixley}, \citenamefont {Li},\ and\ \citenamefont
  {Das~Sarma}}]{Deng:2015}%
  \BibitemOpen
  \bibfield  {author} {\bibinfo {author} {\bibfnamefont {D.-L.}\ \bibnamefont
  {Deng}}, \bibinfo {author} {\bibfnamefont {J.~H.}\ \bibnamefont {Pixley}},
  \bibinfo {author} {\bibfnamefont {X.}~\bibnamefont {Li}},\ and\ \bibinfo
  {author} {\bibfnamefont {S.}~\bibnamefont {Das~Sarma}},\ }\bibfield  {title}
  {\bibinfo {title} {Exponential orthogonality catastrophe in single-particle
  and many-body localized systems},\ }\href
  {https://doi.org/10.1103/PhysRevB.92.220201} {\bibfield  {journal} {\bibinfo
  {journal} {Phys. Rev. B}\ }\textbf {\bibinfo {volume} {92}},\ \bibinfo
  {pages} {220201(R)} (\bibinfo {year} {2015})}\BibitemShut {NoStop}%
\bibitem [{\citenamefont {Michailidis}\ \emph {et~al.}(2020)\citenamefont
  {Michailidis}, \citenamefont {Turner}, \citenamefont
  {Papi\ifmmode~\acute{c}\else \'{c}\fi{}}, \citenamefont {Abanin},\ and\
  \citenamefont {Serbyn}}]{Michailidis:2020}%
  \BibitemOpen
  \bibfield  {author} {\bibinfo {author} {\bibfnamefont {A.~A.}\ \bibnamefont
  {Michailidis}}, \bibinfo {author} {\bibfnamefont {C.~J.}\ \bibnamefont
  {Turner}}, \bibinfo {author} {\bibfnamefont {Z.}~\bibnamefont
  {Papi\ifmmode~\acute{c}\else \'{c}\fi{}}}, \bibinfo {author} {\bibfnamefont
  {D.~A.}\ \bibnamefont {Abanin}},\ and\ \bibinfo {author} {\bibfnamefont
  {M.}~\bibnamefont {Serbyn}},\ }\bibfield  {title} {\bibinfo {title} {Slow
  quantum thermalization and many-body revivals from mixed phase space},\
  }\href {https://doi.org/10.1103/PhysRevX.10.011055} {\bibfield  {journal}
  {\bibinfo  {journal} {Phys. Rev. X}\ }\textbf {\bibinfo {volume} {10}},\
  \bibinfo {pages} {011055} (\bibinfo {year} {2020})}\BibitemShut {NoStop}%
\bibitem [{\citenamefont {Tal‐Ezer}\ and\ \citenamefont
  {Kosloff}(1984)}]{Ezer:1984}%
  \BibitemOpen
  \bibfield  {author} {\bibinfo {author} {\bibfnamefont {H.}~\bibnamefont
  {Tal‐Ezer}}\ and\ \bibinfo {author} {\bibfnamefont {R.}~\bibnamefont
  {Kosloff}},\ }\bibfield  {title} {\bibinfo {title} {{An accurate and
  efficient scheme for propagating the time dependent Schrödinger equation}},\
  }\href {https://doi.org/10.1063/1.448136} {\bibfield  {journal} {\bibinfo
  {journal} {The Journal of Chemical Physics}\ }\textbf {\bibinfo {volume}
  {81}},\ \bibinfo {pages} {3967} (\bibinfo {year} {1984})}\BibitemShut
  {NoStop}%
\bibitem [{\citenamefont {Schaefer}\ \emph {et~al.}(2017)\citenamefont
  {Schaefer}, \citenamefont {Tal-Ezer},\ and\ \citenamefont
  {Kosloff}}]{schaefer2017}%
  \BibitemOpen
  \bibfield  {author} {\bibinfo {author} {\bibfnamefont {I.}~\bibnamefont
  {Schaefer}}, \bibinfo {author} {\bibfnamefont {H.}~\bibnamefont {Tal-Ezer}},\
  and\ \bibinfo {author} {\bibfnamefont {R.}~\bibnamefont {Kosloff}},\
  }\bibfield  {title} {\bibinfo {title} {Semi-global approach for propagation
  of the time-dependent {S}chrödinger equation for time-dependent and
  nonlinear problems},\ }\href
  {https://doi.org/https://doi.org/10.1016/j.jcp.2017.04.017} {\bibfield
  {journal} {\bibinfo  {journal} {J. Comput. Phys.}\ }\textbf {\bibinfo
  {volume} {343}},\ \bibinfo {pages} {368} (\bibinfo {year}
  {2017})}\BibitemShut {NoStop}%
\bibitem [{\citenamefont {De~Grandi}\ and\ \citenamefont
  {Polkovnikov}(2010)}]{grandi2010}%
  \BibitemOpen
  \bibfield  {author} {\bibinfo {author} {\bibfnamefont {C.}~\bibnamefont
  {De~Grandi}}\ and\ \bibinfo {author} {\bibfnamefont {A.}~\bibnamefont
  {Polkovnikov}},\ }\bibinfo {title} {Adiabatic perturbation theory: From
  landau--zener problem to quenching through a quantum critical point},\ in\
  \href {https://doi.org/10.1007/978-3-642-11470-0_4} {\emph {\bibinfo
  {booktitle} {Quantum Quenching, Annealing and Computation}}}\ (\bibinfo
  {publisher} {Springer},\ \bibinfo {address} {Berlin, Heidelberg},\ \bibinfo
  {year} {2010})\ pp.\ \bibinfo {pages} {75--114}\BibitemShut {NoStop}%
\bibitem [{\citenamefont {Polkovnikov}\ and\ \citenamefont
  {Gritsev}(2008)}]{Polkovnikov:2008}%
  \BibitemOpen
  \bibfield  {author} {\bibinfo {author} {\bibfnamefont {A.}~\bibnamefont
  {Polkovnikov}}\ and\ \bibinfo {author} {\bibfnamefont {V.}~\bibnamefont
  {Gritsev}},\ }\bibfield  {title} {\bibinfo {title} {Breakdown of the
  adiabatic limit in low-dimensional gapless systems},\ }\href
  {https://doi.org/10.1038/nphys963} {\bibfield  {journal} {\bibinfo  {journal}
  {Nature Physics}\ }\textbf {\bibinfo {volume} {4}},\ \bibinfo {pages} {477}
  (\bibinfo {year} {2008})}\BibitemShut {NoStop}%
\bibitem [{\citenamefont {Zurek}\ \emph {et~al.}(2005)\citenamefont {Zurek},
  \citenamefont {Dorner},\ and\ \citenamefont {Zoller}}]{Zurek:2005}%
  \BibitemOpen
  \bibfield  {author} {\bibinfo {author} {\bibfnamefont {W.~H.}\ \bibnamefont
  {Zurek}}, \bibinfo {author} {\bibfnamefont {U.}~\bibnamefont {Dorner}},\ and\
  \bibinfo {author} {\bibfnamefont {P.}~\bibnamefont {Zoller}},\ }\bibfield
  {title} {\bibinfo {title} {{Dynamics of a Quantum Phase Transition}},\ }\href
  {https://doi.org/10.1103/PhysRevLett.95.105701} {\bibfield  {journal}
  {\bibinfo  {journal} {Phys. Rev. Lett.}\ }\textbf {\bibinfo {volume} {95}},\
  \bibinfo {pages} {105701} (\bibinfo {year} {2005})}\BibitemShut {NoStop}%
\bibitem [{\citenamefont {Polkovnikov}(2005)}]{Polkovnikov:2006}%
  \BibitemOpen
  \bibfield  {author} {\bibinfo {author} {\bibfnamefont {A.}~\bibnamefont
  {Polkovnikov}},\ }\bibfield  {title} {\bibinfo {title} {Universal adiabatic
  dynamics in the vicinity of a quantum critical point},\ }\href
  {https://doi.org/10.1103/PhysRevB.72.161201} {\bibfield  {journal} {\bibinfo
  {journal} {Phys. Rev. B}\ }\textbf {\bibinfo {volume} {72}},\ \bibinfo
  {pages} {161201(R)} (\bibinfo {year} {2005})}\BibitemShut {NoStop}%
\bibitem [{\citenamefont {De~Grandi}\ \emph {et~al.}(2008)\citenamefont
  {De~Grandi}, \citenamefont {Barankov},\ and\ \citenamefont
  {Polkovnikov}}]{Grandi:2008}%
  \BibitemOpen
  \bibfield  {author} {\bibinfo {author} {\bibfnamefont {C.}~\bibnamefont
  {De~Grandi}}, \bibinfo {author} {\bibfnamefont {R.~A.}\ \bibnamefont
  {Barankov}},\ and\ \bibinfo {author} {\bibfnamefont {A.}~\bibnamefont
  {Polkovnikov}},\ }\bibfield  {title} {\bibinfo {title} {{Adiabatic Nonlinear
  Probes of One-Dimensional Bose Gases}},\ }\href
  {https://doi.org/10.1103/PhysRevLett.101.230402} {\bibfield  {journal}
  {\bibinfo  {journal} {Phys. Rev. Lett.}\ }\textbf {\bibinfo {volume} {101}},\
  \bibinfo {pages} {230402} (\bibinfo {year} {2008})}\BibitemShut {NoStop}%
\bibitem [{\citenamefont {Polkovnikov}(2011)}]{Polkovnikov:2011}%
  \BibitemOpen
  \bibfield  {author} {\bibinfo {author} {\bibfnamefont {A.}~\bibnamefont
  {Polkovnikov}},\ }\bibfield  {title} {\bibinfo {title} {Microscopic diagonal
  entropy and its connection to basic thermodynamic relations},\ }\href
  {https://doi.org/https://doi.org/10.1016/j.aop.2010.08.004} {\bibfield
  {journal} {\bibinfo  {journal} {Annals of Physics}\ }\textbf {\bibinfo
  {volume} {326}},\ \bibinfo {pages} {486} (\bibinfo {year}
  {2011})}\BibitemShut {NoStop}%
\bibitem [{\citenamefont {Kolodrubetz}\ \emph {et~al.}(2017)\citenamefont
  {Kolodrubetz}, \citenamefont {Sels}, \citenamefont {Mehta},\ and\
  \citenamefont {Polkovnikov}}]{Kolodrubetz:2017}%
  \BibitemOpen
  \bibfield  {author} {\bibinfo {author} {\bibfnamefont {M.}~\bibnamefont
  {Kolodrubetz}}, \bibinfo {author} {\bibfnamefont {D.}~\bibnamefont {Sels}},
  \bibinfo {author} {\bibfnamefont {P.}~\bibnamefont {Mehta}},\ and\ \bibinfo
  {author} {\bibfnamefont {A.}~\bibnamefont {Polkovnikov}},\ }\bibfield
  {title} {\bibinfo {title} {Geometry and non-adiabatic response in quantum and
  classical systems},\ }\href
  {https://doi.org/https://doi.org/10.1016/j.physrep.2017.07.001} {\bibfield
  {journal} {\bibinfo  {journal} {Physics Reports}\ }\textbf {\bibinfo {volume}
  {697}},\ \bibinfo {pages} {1} (\bibinfo {year} {2017})}\BibitemShut {NoStop}%
\bibitem [{\citenamefont {You}\ \emph {et~al.}(2007)\citenamefont {You},
  \citenamefont {Li},\ and\ \citenamefont {Gu}}]{you2007}%
  \BibitemOpen
  \bibfield  {author} {\bibinfo {author} {\bibfnamefont {W.-L.}\ \bibnamefont
  {You}}, \bibinfo {author} {\bibfnamefont {Y.-W.}\ \bibnamefont {Li}},\ and\
  \bibinfo {author} {\bibfnamefont {S.-J.}\ \bibnamefont {Gu}},\ }\bibfield
  {title} {\bibinfo {title} {Fidelity, dynamic structure factor, and
  susceptibility in critical phenomena},\ }\href
  {https://doi.org/10.1103/PhysRevE.76.022101} {\bibfield  {journal} {\bibinfo
  {journal} {Phys. Rev. E}\ }\textbf {\bibinfo {volume} {76}},\ \bibinfo
  {pages} {022101} (\bibinfo {year} {2007})}\BibitemShut {NoStop}%
\bibitem [{\citenamefont {Zanardi}\ \emph {et~al.}(2007)\citenamefont
  {Zanardi}, \citenamefont {Giorda},\ and\ \citenamefont
  {Cozzini}}]{Zanardi:2007}%
  \BibitemOpen
  \bibfield  {author} {\bibinfo {author} {\bibfnamefont {P.}~\bibnamefont
  {Zanardi}}, \bibinfo {author} {\bibfnamefont {P.}~\bibnamefont {Giorda}},\
  and\ \bibinfo {author} {\bibfnamefont {M.}~\bibnamefont {Cozzini}},\
  }\bibfield  {title} {\bibinfo {title} {{Information-Theoretic Differential
  Geometry of Quantum Phase Transitions}},\ }\href
  {https://doi.org/10.1103/PhysRevLett.99.100603} {\bibfield  {journal}
  {\bibinfo  {journal} {Phys. Rev. Lett.}\ }\textbf {\bibinfo {volume} {99}},\
  \bibinfo {pages} {100603} (\bibinfo {year} {2007})}\BibitemShut {NoStop}%
\bibitem [{\citenamefont {Pandey}\ \emph {et~al.}(2020)\citenamefont {Pandey},
  \citenamefont {Claeys}, \citenamefont {Campbell}, \citenamefont
  {Polkovnikov},\ and\ \citenamefont {Sels}}]{pandey2020}%
  \BibitemOpen
  \bibfield  {author} {\bibinfo {author} {\bibfnamefont {M.}~\bibnamefont
  {Pandey}}, \bibinfo {author} {\bibfnamefont {P.~W.}\ \bibnamefont {Claeys}},
  \bibinfo {author} {\bibfnamefont {D.~K.}\ \bibnamefont {Campbell}}, \bibinfo
  {author} {\bibfnamefont {A.}~\bibnamefont {Polkovnikov}},\ and\ \bibinfo
  {author} {\bibfnamefont {D.}~\bibnamefont {Sels}},\ }\bibfield  {title}
  {\bibinfo {title} {Adiabatic eigenstate deformations as a sensitive probe for
  quantum chaos},\ }\href {https://doi.org/10.1103/PhysRevX.10.041017}
  {\bibfield  {journal} {\bibinfo  {journal} {Phys. Rev. X}\ }\textbf {\bibinfo
  {volume} {10}},\ \bibinfo {pages} {041017} (\bibinfo {year}
  {2020})}\BibitemShut {NoStop}%
\bibitem [{\citenamefont {LeBlond}\ \emph {et~al.}(2021)\citenamefont
  {LeBlond}, \citenamefont {Sels}, \citenamefont {Polkovnikov},\ and\
  \citenamefont {Rigol}}]{LeBlond:2021}%
  \BibitemOpen
  \bibfield  {author} {\bibinfo {author} {\bibfnamefont {T.}~\bibnamefont
  {LeBlond}}, \bibinfo {author} {\bibfnamefont {D.}~\bibnamefont {Sels}},
  \bibinfo {author} {\bibfnamefont {A.}~\bibnamefont {Polkovnikov}},\ and\
  \bibinfo {author} {\bibfnamefont {M.}~\bibnamefont {Rigol}},\ }\bibfield
  {title} {\bibinfo {title} {Universality in the onset of quantum chaos in
  many-body systems},\ }\href {https://doi.org/10.1103/PhysRevB.104.L201117}
  {\bibfield  {journal} {\bibinfo  {journal} {Phys. Rev. B}\ }\textbf {\bibinfo
  {volume} {104}},\ \bibinfo {pages} {L201117} (\bibinfo {year}
  {2021})}\BibitemShut {NoStop}%
\bibitem [{\citenamefont {Surace}\ \emph {et~al.}(2021)\citenamefont {Surace},
  \citenamefont {Votto}, \citenamefont {Lazo}, \citenamefont {Silva},
  \citenamefont {Dalmonte},\ and\ \citenamefont {Giudici}}]{surace2021}%
  \BibitemOpen
  \bibfield  {author} {\bibinfo {author} {\bibfnamefont {F.~M.}\ \bibnamefont
  {Surace}}, \bibinfo {author} {\bibfnamefont {M.}~\bibnamefont {Votto}},
  \bibinfo {author} {\bibfnamefont {E.~G.}\ \bibnamefont {Lazo}}, \bibinfo
  {author} {\bibfnamefont {A.}~\bibnamefont {Silva}}, \bibinfo {author}
  {\bibfnamefont {M.}~\bibnamefont {Dalmonte}},\ and\ \bibinfo {author}
  {\bibfnamefont {G.}~\bibnamefont {Giudici}},\ }\bibfield  {title} {\bibinfo
  {title} {Exact many-body scars and their stability in constrained quantum
  chains},\ }\href {https://doi.org/10.1103/PhysRevB.103.104302} {\bibfield
  {journal} {\bibinfo  {journal} {Phys. Rev. B}\ }\textbf {\bibinfo {volume}
  {103}},\ \bibinfo {pages} {104302} (\bibinfo {year} {2021})}\BibitemShut
  {NoStop}%
\bibitem [{\citenamefont {Abramowitz}\ \emph {et~al.}(1988)\citenamefont
  {Abramowitz}, \citenamefont {Stegun},\ and\ \citenamefont
  {Romer}}]{abramowitz1988}%
  \BibitemOpen
  \bibfield  {author} {\bibinfo {author} {\bibfnamefont {M.}~\bibnamefont
  {Abramowitz}}, \bibinfo {author} {\bibfnamefont {I.~A.}\ \bibnamefont
  {Stegun}},\ and\ \bibinfo {author} {\bibfnamefont {R.~H.}\ \bibnamefont
  {Romer}},\ }\bibfield  {title} {\bibinfo {title} {Handbook of mathematical
  functions with formulas, graphs, and mathematical tables},\ }\href
  {https://doi.org/10.1119/1.15378} {\bibfield  {journal} {\bibinfo  {journal}
  {American Journal of Physics}\ }\textbf {\bibinfo {volume} {56}},\ \bibinfo
  {pages} {958} (\bibinfo {year} {1988})}\BibitemShut {NoStop}%
\bibitem [{\citenamefont {Altshuler}\ \emph {et~al.}(2010)\citenamefont
  {Altshuler}, \citenamefont {Krovi},\ and\ \citenamefont
  {Roland}}]{Altshuler:2010}%
  \BibitemOpen
  \bibfield  {author} {\bibinfo {author} {\bibfnamefont {B.}~\bibnamefont
  {Altshuler}}, \bibinfo {author} {\bibfnamefont {H.}~\bibnamefont {Krovi}},\
  and\ \bibinfo {author} {\bibfnamefont {J.}~\bibnamefont {Roland}},\
  }\bibfield  {title} {\bibinfo {title} {Anderson localization makes adiabatic
  quantum optimization fail},\ }\href {https://doi.org/10.1073/pnas.1002116107}
  {\bibfield  {journal} {\bibinfo  {journal} {Proceedings of the National
  Academy of Sciences}\ }\textbf {\bibinfo {volume} {107}},\ \bibinfo {pages}
  {12446} (\bibinfo {year} {2010})}\BibitemShut {NoStop}%
\bibitem [{\citenamefont {Haegeman}\ \emph {et~al.}(2013)\citenamefont
  {Haegeman}, \citenamefont {Osborne},\ and\ \citenamefont
  {Verstraete}}]{Haegeman:2013}%
  \BibitemOpen
  \bibfield  {author} {\bibinfo {author} {\bibfnamefont {J.}~\bibnamefont
  {Haegeman}}, \bibinfo {author} {\bibfnamefont {T.~J.}\ \bibnamefont
  {Osborne}},\ and\ \bibinfo {author} {\bibfnamefont {F.}~\bibnamefont
  {Verstraete}},\ }\bibfield  {title} {\bibinfo {title} {Post-matrix product
  state methods: {T}o tangent space and beyond},\ }\href
  {https://doi.org/10.1103/PhysRevB.88.075133} {\bibfield  {journal} {\bibinfo
  {journal} {Phys. Rev. B}\ }\textbf {\bibinfo {volume} {88}},\ \bibinfo
  {pages} {075133} (\bibinfo {year} {2013})}\BibitemShut {NoStop}%
\bibitem [{\citenamefont {Mark}\ \emph {et~al.}(2020)\citenamefont {Mark},
  \citenamefont {Lin},\ and\ \citenamefont {Motrunich}}]{Mark:2020}%
  \BibitemOpen
  \bibfield  {author} {\bibinfo {author} {\bibfnamefont {D.~K.}\ \bibnamefont
  {Mark}}, \bibinfo {author} {\bibfnamefont {C.-J.}\ \bibnamefont {Lin}},\ and\
  \bibinfo {author} {\bibfnamefont {O.~I.}\ \bibnamefont {Motrunich}},\
  }\bibfield  {title} {\bibinfo {title} {Unified structure for exact towers of
  scar states in the {A}ffleck-{K}ennedy-{L}ieb-{T}asaki and other models},\
  }\href {https://doi.org/10.1103/PhysRevB.101.195131} {\bibfield  {journal}
  {\bibinfo  {journal} {Phys. Rev. B}\ }\textbf {\bibinfo {volume} {101}},\
  \bibinfo {pages} {195131} (\bibinfo {year} {2020})}\BibitemShut {NoStop}%
\bibitem [{\citenamefont {Gustafson}\ \emph {et~al.}(2023)\citenamefont
  {Gustafson}, \citenamefont {Li}, \citenamefont {Khan}, \citenamefont {Kim},
  \citenamefont {Kurkcuoglu}, \citenamefont {Alam}, \citenamefont {Orth},
  \citenamefont {Rahmani},\ and\ \citenamefont {Iadecola}}]{Gustafson:2023}%
  \BibitemOpen
  \bibfield  {author} {\bibinfo {author} {\bibfnamefont {E.~J.}\ \bibnamefont
  {Gustafson}}, \bibinfo {author} {\bibfnamefont {A.~C.}\ \bibnamefont {Li}},
  \bibinfo {author} {\bibfnamefont {A.}~\bibnamefont {Khan}}, \bibinfo {author}
  {\bibfnamefont {J.}~\bibnamefont {Kim}}, \bibinfo {author} {\bibfnamefont
  {D.~M.}\ \bibnamefont {Kurkcuoglu}}, \bibinfo {author} {\bibfnamefont
  {M.~S.}\ \bibnamefont {Alam}}, \bibinfo {author} {\bibfnamefont {P.~P.}\
  \bibnamefont {Orth}}, \bibinfo {author} {\bibfnamefont {A.}~\bibnamefont
  {Rahmani}},\ and\ \bibinfo {author} {\bibfnamefont {T.}~\bibnamefont
  {Iadecola}},\ }\bibfield  {title} {\bibinfo {title} {Preparing quantum
  many-body scar states on quantum computers},\ }\href
  {https://doi.org/10.22331/q-2023-11-07-1171} {\bibfield  {journal} {\bibinfo
  {journal} {Quantum}\ }\textbf {\bibinfo {volume} {7}},\ \bibinfo {pages}
  {1171} (\bibinfo {year} {2023})}\BibitemShut {NoStop}%
\bibitem [{\citenamefont {Moore}\ and\ \citenamefont {Read}(1991)}]{moore1991}%
  \BibitemOpen
  \bibfield  {author} {\bibinfo {author} {\bibfnamefont {G.}~\bibnamefont
  {Moore}}\ and\ \bibinfo {author} {\bibfnamefont {N.}~\bibnamefont {Read}},\
  }\bibfield  {title} {\bibinfo {title} {Nonabelions in the fractional quantum
  {H}all effect},\ }\href
  {https://www.sciencedirect.com/science/article/pii/055032139190407O}
  {\bibfield  {journal} {\bibinfo  {journal} {Nucl. Phys. B}\ }\textbf
  {\bibinfo {volume} {360}},\ \bibinfo {pages} {362} (\bibinfo {year}
  {1991})}\BibitemShut {NoStop}%
\bibitem [{\citenamefont {Nielsen}\ \emph {et~al.}(2012)\citenamefont
  {Nielsen}, \citenamefont {Cirac},\ and\ \citenamefont
  {Sierra}}]{nielsen2012}%
  \BibitemOpen
  \bibfield  {author} {\bibinfo {author} {\bibfnamefont {A.~E.~B.}\
  \bibnamefont {Nielsen}}, \bibinfo {author} {\bibfnamefont {J.~I.}\
  \bibnamefont {Cirac}},\ and\ \bibinfo {author} {\bibfnamefont
  {G.}~\bibnamefont {Sierra}},\ }\bibfield  {title} {\bibinfo {title}
  {{Laughlin Spin-Liquid States on Lattices Obtained from Conformal Field
  Theory}},\ }\href {https://doi.org/10.1103/PhysRevLett.108.257206} {\bibfield
   {journal} {\bibinfo  {journal} {Phys. Rev. Lett.}\ }\textbf {\bibinfo
  {volume} {108}},\ \bibinfo {pages} {257206} (\bibinfo {year}
  {2012})}\BibitemShut {NoStop}%
\bibitem [{\citenamefont {Nielsen}(2015)}]{nielsen2015}%
  \BibitemOpen
  \bibfield  {author} {\bibinfo {author} {\bibfnamefont {A.~E.~B.}\
  \bibnamefont {Nielsen}},\ }\bibfield  {title} {\bibinfo {title} {Anyon
  braiding in semianalytical fractional quantum {H}all lattice models},\ }\href
  {https://doi.org/10.1103/PhysRevB.91.041106} {\bibfield  {journal} {\bibinfo
  {journal} {Phys. Rev. B}\ }\textbf {\bibinfo {volume} {91}},\ \bibinfo
  {pages} {041106(R)} (\bibinfo {year} {2015})}\BibitemShut {NoStop}%
\bibitem [{\citenamefont {Tu}\ \emph {et~al.}(2014)\citenamefont {Tu},
  \citenamefont {Nielsen}, \citenamefont {Cirac},\ and\ \citenamefont
  {Sierra}}]{tu2014a}%
  \BibitemOpen
  \bibfield  {author} {\bibinfo {author} {\bibfnamefont {H.-H.}\ \bibnamefont
  {Tu}}, \bibinfo {author} {\bibfnamefont {A.~E.~B.}\ \bibnamefont {Nielsen}},
  \bibinfo {author} {\bibfnamefont {J.~I.}\ \bibnamefont {Cirac}},\ and\
  \bibinfo {author} {\bibfnamefont {G.}~\bibnamefont {Sierra}},\ }\bibfield
  {title} {\bibinfo {title} {Lattice {L}aughlin states of bosons and fermions
  at filling fractions $1/q$},\ }\href
  {https://doi.org/10.1088/1367-2630/16/3/033025} {\bibfield  {journal}
  {\bibinfo  {journal} {New J. Phys.}\ }\textbf {\bibinfo {volume} {16}},\
  \bibinfo {pages} {033025} (\bibinfo {year} {2014})}\BibitemShut {NoStop}%
\bibitem [{\citenamefont {Srivatsa}\ \emph {et~al.}(2020)\citenamefont
  {Srivatsa}, \citenamefont {Wildeboer}, \citenamefont {Seidel},\ and\
  \citenamefont {Nielsen}}]{Srivatsa2020}%
  \BibitemOpen
  \bibfield  {author} {\bibinfo {author} {\bibfnamefont {N.~S.}\ \bibnamefont
  {Srivatsa}}, \bibinfo {author} {\bibfnamefont {J.}~\bibnamefont {Wildeboer}},
  \bibinfo {author} {\bibfnamefont {A.}~\bibnamefont {Seidel}},\ and\ \bibinfo
  {author} {\bibfnamefont {A.~E.~B.}\ \bibnamefont {Nielsen}},\ }\bibfield
  {title} {\bibinfo {title} {Quantum many-body scars with chiral topological
  order in two dimensions and critical properties in one dimension},\ }\href
  {https://doi.org/10.1103/PhysRevB.102.235106} {\bibfield  {journal} {\bibinfo
   {journal} {Phys. Rev. B}\ }\textbf {\bibinfo {volume} {102}},\ \bibinfo
  {pages} {235106} (\bibinfo {year} {2020})}\BibitemShut {NoStop}%
\bibitem [{\citenamefont {Robnik}\ and\ \citenamefont
  {Berry}(1986)}]{Robnik:1986}%
  \BibitemOpen
  \bibfield  {author} {\bibinfo {author} {\bibfnamefont {M.}~\bibnamefont
  {Robnik}}\ and\ \bibinfo {author} {\bibfnamefont {M.~V.}\ \bibnamefont
  {Berry}},\ }\bibfield  {title} {\bibinfo {title} {False time-reversal
  violation and energy level statistics: the role of anti-unitary symmetry},\
  }\href {https://doi.org/10.1088/0305-4470/19/5/020} {\bibfield  {journal}
  {\bibinfo  {journal} {Journal of Physics A: Mathematical and General}\
  }\textbf {\bibinfo {volume} {19}},\ \bibinfo {pages} {669} (\bibinfo {year}
  {1986})}\BibitemShut {NoStop}%
\bibitem [{\citenamefont {Fremling}\ \emph {et~al.}(2018)\citenamefont
  {Fremling}, \citenamefont {Repellin}, \citenamefont {Stéphan}, \citenamefont
  {Moran}, \citenamefont {Slingerland},\ and\ \citenamefont
  {Haque}}]{Fremling:2018}%
  \BibitemOpen
  \bibfield  {author} {\bibinfo {author} {\bibfnamefont {M.}~\bibnamefont
  {Fremling}}, \bibinfo {author} {\bibfnamefont {C.}~\bibnamefont {Repellin}},
  \bibinfo {author} {\bibfnamefont {J.-M.}\ \bibnamefont {Stéphan}}, \bibinfo
  {author} {\bibfnamefont {N.}~\bibnamefont {Moran}}, \bibinfo {author}
  {\bibfnamefont {J.~K.}\ \bibnamefont {Slingerland}},\ and\ \bibinfo {author}
  {\bibfnamefont {M.}~\bibnamefont {Haque}},\ }\bibfield  {title} {\bibinfo
  {title} {Dynamics and level statistics of interacting fermions in the lowest
  {L}andau level},\ }\href {https://doi.org/10.1088/1367-2630/aae73f}
  {\bibfield  {journal} {\bibinfo  {journal} {New Journal of Physics}\ }\textbf
  {\bibinfo {volume} {20}},\ \bibinfo {pages} {103036} (\bibinfo {year}
  {2018})}\BibitemShut {NoStop}%
\bibitem [{\citenamefont {Wei\ss{}e}\ \emph {et~al.}(2006)\citenamefont
  {Wei\ss{}e}, \citenamefont {Wellein}, \citenamefont {Alvermann},\ and\
  \citenamefont {Fehske}}]{Weise:2006}%
  \BibitemOpen
  \bibfield  {author} {\bibinfo {author} {\bibfnamefont {A.}~\bibnamefont
  {Wei\ss{}e}}, \bibinfo {author} {\bibfnamefont {G.}~\bibnamefont {Wellein}},
  \bibinfo {author} {\bibfnamefont {A.}~\bibnamefont {Alvermann}},\ and\
  \bibinfo {author} {\bibfnamefont {H.}~\bibnamefont {Fehske}},\ }\bibfield
  {title} {\bibinfo {title} {The kernel polynomial method},\ }\href
  {https://doi.org/10.1103/RevModPhys.78.275} {\bibfield  {journal} {\bibinfo
  {journal} {Rev. Mod. Phys.}\ }\textbf {\bibinfo {volume} {78}},\ \bibinfo
  {pages} {275} (\bibinfo {year} {2006})}\BibitemShut {NoStop}%
\bibitem [{\citenamefont {Torrontegui}\ \emph {et~al.}(2013)\citenamefont
  {Torrontegui}, \citenamefont {Ibáñez}, \citenamefont {Martínez-Garaot},
  \citenamefont {Modugno}, \citenamefont {{del Campo}}, \citenamefont
  {Guéry-Odelin}, \citenamefont {Ruschhaupt}, \citenamefont {Chen},\ and\
  \citenamefont {Muga}}]{Torrontegui2013}%
  \BibitemOpen
  \bibfield  {author} {\bibinfo {author} {\bibfnamefont {E.}~\bibnamefont
  {Torrontegui}}, \bibinfo {author} {\bibfnamefont {S.}~\bibnamefont
  {Ibáñez}}, \bibinfo {author} {\bibfnamefont {S.}~\bibnamefont
  {Martínez-Garaot}}, \bibinfo {author} {\bibfnamefont {M.}~\bibnamefont
  {Modugno}}, \bibinfo {author} {\bibfnamefont {A.}~\bibnamefont {{del
  Campo}}}, \bibinfo {author} {\bibfnamefont {D.}~\bibnamefont
  {Guéry-Odelin}}, \bibinfo {author} {\bibfnamefont {A.}~\bibnamefont
  {Ruschhaupt}}, \bibinfo {author} {\bibfnamefont {X.}~\bibnamefont {Chen}},\
  and\ \bibinfo {author} {\bibfnamefont {J.~G.}\ \bibnamefont {Muga}},\
  }\bibfield  {title} {\bibinfo {title} {Chapter 2 - shortcuts to
  adiabaticity},\ }in\ \href
  {https://doi.org/https://doi.org/10.1016/B978-0-12-408090-4.00002-5} {\emph
  {\bibinfo {booktitle} {Advances in Atomic, Molecular, and Optical
  Physics}}},\ \bibinfo {series} {Advances In Atomic, Molecular, and Optical
  Physics}, Vol.~\bibinfo {volume} {62},\ \bibinfo {editor} {edited by\
  \bibinfo {editor} {\bibfnamefont {E.}~\bibnamefont {Arimondo}}, \bibinfo
  {editor} {\bibfnamefont {P.~R.}\ \bibnamefont {Berman}},\ and\ \bibinfo
  {editor} {\bibfnamefont {C.~C.}\ \bibnamefont {Lin}}}\ (\bibinfo  {publisher}
  {Academic Press},\ \bibinfo {year} {2013})\ pp.\ \bibinfo {pages}
  {117--169}\BibitemShut {NoStop}%
\bibitem [{\citenamefont {Gu\'ery-Odelin}\ \emph {et~al.}(2019)\citenamefont
  {Gu\'ery-Odelin}, \citenamefont {Ruschhaupt}, \citenamefont {Kiely},
  \citenamefont {Torrontegui}, \citenamefont {Mart\'{\i}nez-Garaot},\ and\
  \citenamefont {Muga}}]{Odelin2019}%
  \BibitemOpen
  \bibfield  {author} {\bibinfo {author} {\bibfnamefont {D.}~\bibnamefont
  {Gu\'ery-Odelin}}, \bibinfo {author} {\bibfnamefont {A.}~\bibnamefont
  {Ruschhaupt}}, \bibinfo {author} {\bibfnamefont {A.}~\bibnamefont {Kiely}},
  \bibinfo {author} {\bibfnamefont {E.}~\bibnamefont {Torrontegui}}, \bibinfo
  {author} {\bibfnamefont {S.}~\bibnamefont {Mart\'{\i}nez-Garaot}},\ and\
  \bibinfo {author} {\bibfnamefont {J.~G.}\ \bibnamefont {Muga}},\ }\bibfield
  {title} {\bibinfo {title} {Shortcuts to adiabaticity: {C}oncepts, methods,
  and applications},\ }\href {https://doi.org/10.1103/RevModPhys.91.045001}
  {\bibfield  {journal} {\bibinfo  {journal} {Rev. Mod. Phys.}\ }\textbf
  {\bibinfo {volume} {91}},\ \bibinfo {pages} {045001} (\bibinfo {year}
  {2019})}\BibitemShut {NoStop}%
\bibitem [{\citenamefont {Ljubotina}\ \emph {et~al.}(2022)\citenamefont
  {Ljubotina}, \citenamefont {Roos}, \citenamefont {Abanin},\ and\
  \citenamefont {Serbyn}}]{Ljubotina:2020}%
  \BibitemOpen
  \bibfield  {author} {\bibinfo {author} {\bibfnamefont {M.}~\bibnamefont
  {Ljubotina}}, \bibinfo {author} {\bibfnamefont {B.}~\bibnamefont {Roos}},
  \bibinfo {author} {\bibfnamefont {D.~A.}\ \bibnamefont {Abanin}},\ and\
  \bibinfo {author} {\bibfnamefont {M.}~\bibnamefont {Serbyn}},\ }\bibfield
  {title} {\bibinfo {title} {Optimal steering of matrix product states and
  quantum many-body scars},\ }\href
  {https://doi.org/10.1103/PRXQuantum.3.030343} {\bibfield  {journal} {\bibinfo
   {journal} {PRX Quantum}\ }\textbf {\bibinfo {volume} {3}},\ \bibinfo {pages}
  {030343} (\bibinfo {year} {2022})}\BibitemShut {NoStop}%
\bibitem [{\citenamefont {Benseny}\ and\ \citenamefont
  {M\o{}lmer}(2021)}]{Benseny2021}%
  \BibitemOpen
  \bibfield  {author} {\bibinfo {author} {\bibfnamefont {A.}~\bibnamefont
  {Benseny}}\ and\ \bibinfo {author} {\bibfnamefont {K.}~\bibnamefont
  {M\o{}lmer}},\ }\bibfield  {title} {\bibinfo {title} {Adiabatic theorem
  revisited: {T}he unexpectedly good performance of adiabatic passage},\ }\href
  {https://doi.org/10.1103/PhysRevA.103.062215} {\bibfield  {journal} {\bibinfo
   {journal} {Phys. Rev. A}\ }\textbf {\bibinfo {volume} {103}},\ \bibinfo
  {pages} {062215} (\bibinfo {year} {2021})}\BibitemShut {NoStop}%
\bibitem [{\citenamefont {Kolb}\ and\ \citenamefont
  {Pakrouski}(2023)}]{Kolb2023}%
  \BibitemOpen
  \bibfield  {author} {\bibinfo {author} {\bibfnamefont {P.}~\bibnamefont
  {Kolb}}\ and\ \bibinfo {author} {\bibfnamefont {K.}~\bibnamefont
  {Pakrouski}},\ }\bibfield  {title} {\bibinfo {title} {Stability of the
  many-body scars in fermionic spin-1/2 models},\ }\href
  {https://doi.org/10.1103/PRXQuantum.4.040348} {\bibfield  {journal} {\bibinfo
   {journal} {PRX Quantum}\ }\textbf {\bibinfo {volume} {4}},\ \bibinfo {pages}
  {040348} (\bibinfo {year} {2023})}\BibitemShut {NoStop}%
\end{thebibliography}%

\end{document}